\documentclass[12pt,a4paper]{article}
\usepackage{elliptic}
\usepackage[dvipsnames]{xcolor}
\usepackage{amscd,amsmath,amssymb,amsfonts,xspace,mathrsfs,mathtools,scalefnt}
\usepackage[utf8]{inputenc}
\usepackage{bbm}
\usepackage{graphicx} 
\usepackage{tabu}
\usepackage[color,matrix,arrow]{xy}
\usepackage{wasysym}

\newcommand{\tq}{\mathtt{q}} 
\newcommand{\be}{\begin{equation}} 
\newcommand{\ee}{\end{equation}} 
\newcommand{\bes}{\begin{equation*}}
\newcommand{\ees}{\end{equation*}}

\newcommand{\im}{\text{Im}}
\newcommand{\pd}[2]{\frac{\partial #1}{\partial #2}} 
  
\newcommand{\CA}{\mathcal{A}} 
  
\newcommand{\CC}{\mathcal{C}}   
  
\newcommand{\CE}{\mathcal{E}}  
\newcommand{\CF}{\mathcal{F}}

\newcommand{\CL}{\mathcal{L}} 
\newcommand{\CM}{\mathcal{M}}  
\newcommand{\CN}{\mathcal{N}}
\newcommand{\CO}{\mathcal{O}} 
\newcommand{\CP}{\mathcal{P}}

\newcommand{\CR}{\mathcal{R}}
\newcommand{\CS}{\mathcal{S}}
\newcommand{\CT}{\mathcal{T}} 
\newcommand{\CU}{\mathcal{U}}

\newcommand{\BR}{\mathbb{R}}

\newcommand{\BZ}{\mathbb{Z}}
\newcommand{\BQ}{\mathbb{Q}}

\newcommand{\Tr}{\text{Tr}}

\title{Elliptic loci of $SU(3)$ vacua}

\abstract{ 
	
\vspace{30pt}\noindent\today
 
}

\author{Johannes Aspman$^a$, Elias Furrer$^b$, Jan Manschot$^c$ \\
{\it School of Mathematics, Trinity College, Dublin 2, Ireland\\
\it Hamilton Mathematical Institute, Trinity College, Dublin 2 \vspace{20pt}

$^a$\href{mailto:aspmanj@maths.tcd.ie}{aspmanj@maths.tcd.ie}\\
$^b$\href{mailto:furrere@maths.tcd.ie}{furrere@maths.tcd.ie}\\
$^c$\href{mailto:manschot@maths.tcd.ie}{manschot@maths.tcd.ie}

}}

\abstract{The space of vacua of many four-dimensional, $\CN=2$ supersymmetric gauge
  theories can famously be identified with a family of complex curves. For
  gauge group $SU(2)$, this gives a fully explicit description of the
  low-energy effective theory in terms of an elliptic curve and
  associated modular fundamental domain. The two-dimensional space of
  vacua for gauge group $SU(3)$ parametrizes an intricate family of
  genus two curves. We analyze this family using the so-called  
  Rosenhain form for these curves. We demonstrate that two natural
  one-dimensional subloci of the space of $SU(3)$ vacua,
  $\mathcal{E}_u$ and $\mathcal{E}_v$, each parametrize a family of elliptic curves. For these
  {\it elliptic loci}, we describe the order parameters and fundamental
  domains explicitly. The locus $\mathcal{E}_u$ contains the points where mutually local dyons
  become massless, and is a fundamental domain for a classical congruence
  subgroup. Moreover, the locus $\mathcal{E}_v$ contains the superconformal
  Argyres-Douglas points, and is a fundamental domain for a Fricke group.

\vspace{30pt}\noindent

}

\preprint{}

\begin{document}
\maketitle

\section{Introduction}

Supersymmetric field theories provide a rich ground for qualitative and quantitative analyses in quantum field
theory \cite{Seiberg:1994rs, Seiberg:1994aj, Vafa:1994tf, Ne}. Many
observables have been determined non-perturbatively in terms of
hypergeometric, modular or other special functions.
The best understood example is $\CN=2$ supersymmetric Yang-Mills theory with gauge 
group $SU(2)$ \cite{Seiberg:1994rs, Seiberg:1994aj}. 
Its space of vacua is parametrized by the vacuum expectation value (vev)
$u=\frac 12\langle \text{Tr} \,\phi^2\rangle$, where $\phi$ is the
complex scalar in the $\CN=2$ vector multiplet. The renormalisation group flow generates a quantum scale $\Lambda$, at which the gauge coupling becomes strong.
 In the weak-coupling  
region $|u|\gg \Lambda^2$, the semi-classical BPS spectrum consists of massive
monopoles and dyons. 
The theory can be solved non-perturbatively in  
terms of the Seiberg-Witten (SW) curve
\cite{Seiberg:1994rs}. This solution demonstrates that the effective
abelian gauge theory breaks down at two special points, $u=\pm
\Lambda^2$. The electric-magnetic duality group is generated by the
monodromies around these singular points. It is  a 
subgroup of $SL(2,\mathbb{Z})$, which acts by linear fractional
transformations on the effective coupling constant $\tau$. With the SW
solution, various physical quantities can be exactly determined as functions of
$\tau$ using modular functions \cite{Nahm:1996di, Moore:1997pc, Losev:1997tp,
  Aganagic:2006wq, Huang:2009md}.   
 
Similar non-perturbative solutions have been developed for gauge theories with matter
multiplets \cite{Seiberg:1994aj} and theories with other gauge groups \cite{Klemm:1994qs, Klemm:1994qj, Klemm:1995wp,
  Danielsson:1995is, Masuda:1997nv, suzuki1, suzuki2}. In pure
Yang-Mills theory with compact gauge group $G$, the Coulomb branch has
complex dimension $r=\mathrm{rank}(G)$. Classically, the moduli space 
is parametrized by the vevs $u_{I+1}\sim  \langle
\text{Tr} \,\phi^{I+1}\rangle$, $I=1,\dots,r$. The $r(r+1)/2$
couplings $\tau_{IJ}$ are determined by the $r$ order parameters $u_I$. The
electric-magnetic duality group is a subgroup of $Sp(2r,\mathbb Z)$,
generated by monodromies around singular loci. While this also
demonstrates a link to modularity, the connection has
remained more elusive, and the connection is best established for the superconformal theories
\cite{Minahan:1995er, Minahan:1996ws, Argyres:1998bn, Ashok:2015cba, Ashok:2016oyh}. 

One complication for asymptotically free theories is that the structure of the singular loci is in general quite intricate. 
This article focuses on the asymptotically free $SU(3)$ theory without
hypermultiplets, whose singular loci have a rich
structure \cite{Argyres:1995jj, ARGYRES199671,
  EGUCHI1996430, Galakhov:2013oja, wang:2019}. There are six singular (complex) lines which intersect in five
points. A particularly interesting phenomenon occurs at two of these
five vacua, namely those where three mutually non-local dyons
become massless, such that there is no duality frame in which all of these
states only carry electric charge. This indicates that the system is
in a critical phase, which led to the discovery
of new superconformal theories \cite{Argyres:1995jj, ARGYRES199671,
  EGUCHI1996430}. 

Another complication for $SU(N>2)$ is that the number of couplings exceeds the dimension of the
Coulomb branch. The observables are therefore defined on a subspace of
the genus $N-1$ Siegel upper half-space $\mathbb{H}_{N-1}$. For the $SU(3)$
theory, the Coulomb branch is parametrized by
two order parameters which determine three coupling constants, $\tau_{11}$, $\tau_{12}$ and $\tau_{22}$.  The curve and the SW differential for pure $SU(N)$ gauge theory have
first been proposed in \cite{Klemm:1994qs}.
As a first step to explore the modularity of the $SU(3)$ theory, we relate the hyperelliptic Seiberg-Witten
curve to the Rosenhain form, which is an algebraic expression in terms of Siegel theta series. To exactly match the Rosenhain curve
and Seiberg-Witten curve, we use the fact that the complexified
masses $a_I$ and  $a_{D,I}=\frac{\partial \CF}{\partial a_I}$ are solutions of second order partial differential equations of Picard-Fuchs (PF)
type. The solutions to such equations can be expressed in terms of the generalized
hypergeometric function $F_4$ of Appell \cite{Klemm:1995wp}. The
Siegel theta series and their modular transformations can provide insights
for the analytic continuation and monodromies of the solution in terms
of $F_4$.

The Rosenhain curve allows us to characterize the $SU(3)$ Coulomb branch, parametrized by the two
Casimirs $u=u_2$ and $v=u_3$, as the zero-locus of three equations
inside a five-dimensional space. The structure of these equations simplifies on one-dimensional loci of
the Coulomb branch. We study two of these loci in detail, namely 
$\CE_u$ where $v=0$ and  $\CE_v$ where $u=0$. On each of these loci, the equations reduce to
two algebraic relations of Siegel theta functions, relating the couplings
$\tau_{IJ}$ to a single independent one. Interestingly, each of
these loci in the space of genus two curves also parametrizes a
family of (genus 1) elliptic curves. Both loci interpolate between a weak-coupling regime with large order parameters and a
strong-coupling regime where $u/\Lambda^2$ and $v/\Lambda^3$ are $\CO(1)$. Locus $\CE_u$
contains three cusps where mutually local dyons become massless, while locus $\CE_v$
contains two special points where mutually non-local
dyons becomes massless. The latter are the superconformal
Argyres-Douglas points.  

Since an elliptic locus parametrizes a family of elliptic curves, there must be a coupling $\tau$ 
valued in a fundamental domain (or modular curve) for a discrete group in the
upper half-plane $\mathbb{H}$. We derive the
generators of the discrete subgroup from the monodromies of the
$SU(3)$ theory. We provide two solutions for the locus $\CE_u$. The
coupling for the first solution is $\tau_-=\tau_{11}-\tau_{12}$, while
$\tau_{22}=\tau_{11}$. The order parameter $u$ equals a modular form $u_-$ for the congruence
subgroup $\Gamma^0(9)\subset SL(2,\mathbb Z)$ (\ref{largeuv=0}),
\be
\label{uMinIntro}
u=u_-(\tau_-).
\ee 
The cusps of the fundamental domain of $\Gamma^0(9)$ map exactly to
the singular points on this locus. The coupling for the second
solution is $\tau_+=\tau_{11}+\tau_{12}$. In terms of this coupling,
Equation (\ref{u_thetas}) expresses $u$ as
\be 
\label{uMaxIntro}
u=u_+(\tau_+), 
\ee
where $u_+$ is expressed in terms of roots of modular forms, while it is not
a modular function for a congruence subgroup of $SL(2,\mathbb Z)$. We call it a
{\it sextic modular function} since it is a solution to a sextic equation. The
inverses of the identities (\ref{uMinIntro}) and (\ref{uMaxIntro}) provide
all order $u$-expansions for $\tau_{11}=\tau_{22}$ and $\tau_{12}$ on
this locus. The function $u_+$ appeared earlier as the solutions for the order
parameter on the Coulomb branch of the $\CN=2$, $SU(2)$ theory with
one massless hypermultiplet \cite{Nahm:1996di}. While this Coulomb branch and $\CE_u$ are
isomorphic as four punctured spheres, it is striking that the
solutions of the order parameters are identical.

We find another intriguing structure for the second locus
$\CE_v$ where $u=0$. We are
able to demonstrate for this locus that $v$ is left invariant by
the action of the principal congruence subgroup $\Gamma(6)\subset SL(2,\BZ)$. The
fundamental domain $\Gamma(6)\backslash\mathbb H$ has 12 cusps, where $v$ diverges. Surprisingly, this appears to imply the existence of strongly
coupled vacua in the region where $v$ is large, which is 
unexpected since large $v$ is known to correspond to weak coupling. The paradox is
resolved by realizing that $v$ is invariant under a transformation
which is not contained in $SL(2,\mathbb{Z})$, namely a \emph{Fricke
  involution} $\tau\mapsto -1/n\tau$ for integer $n\geq2$. This transformation maps the putative cusps to $i\infty$. The result is that
$v$ is a modular function for a discrete subgroup $ \Gamma_v \subset SL(2,\BR)$ of Atkin-Lehner type, and we show that the non-trivial
monodromies on this locus do generate this group. 
 
We demonstrate furthermore that the elliptic curves underlying the two loci $\CE_u$ and $\CE_v$ are
related to the genus two curve in a precise way. For a genus two
curve $\Sigma_2$, a holomorphic map $\varphi:\Sigma_2\to \Sigma_1$ to
an elliptic curve $\Sigma_1$ may
exist. Such
maps were studied in the classic works by Legendre and Jacobi, and
more recently in \cite{shaska2001, shaska2012genus}. The existence of the map 
$\varphi$ depends on the complex structure moduli
$\tau_{IJ}$. The family of such curves spans a complex co-dimension one
locus $\mathcal{L}_2$ in the complex three-dimensional space of genus two 
curves. At the elliptic loci of the Coulomb branch of the $SU(3)$ theory
mentioned above, $\mathcal{L}_2$ intersects the $SU(3)$ Coulomb branch, such that for any point on the elliptic loci, there is a
degree two map from the genus two curve to an elliptic curve, or in
other words the genus two curve is a double cover of the elliptic
curve. Besides $\CE_u$ and $\CE_v$, $\CL_2$ also includes a third elliptic locus,
$\CE_3$ (\ref{l2locus}), which does not contain any of the singular points of
the Coulomb branch. 
   
Our work motivates a similar analysis for $SU(N)$ gauge theories, whose Coulomb branch
parametrizes a curve of genus $N-1$. The order parameters $u_I$, $I=2,\dots,N$,
are expected to be given by higher genus modular functions of the coupling
matrix $\tau_{IJ}$. They should furthermore be invariant under a
subgroup of $Sp(2r,\mathbb Z)$ generated by the monodromies. The
existence of maps to elliptic or lower genus curves is however more
subtle for such theories \cite{shaska2006, gutierrez2012}.

The outline of the paper is as follows. In Section \ref{sec:rank1} we
give an overview of the $SU(2)$ theory. In Section \ref{sec:higher_rank} we
review the geometry of the $SU(3)$ theory and write down asymptotic
expansions of the periods which we later use. In Section
\ref{sec:Rosenhain} we discuss the Seiberg-Witten curve in Rosenhain
form to match cross-ratios of the hyperelliptic
curve with the theta constants. Sections \ref{sec:Rosenhain} and \ref{sectionu=0} are
devoted to studying these equations on the loci  $v=0$ and $u=0$
respectively, which allows us to express $u$ and $v$ on these loci
in terms of modular functions. In Section \ref{sec:pfsolution}, we
study the global symmetries of the moduli space and calculate the
corresponding monodromies, along with the BPS spectrum at strong
coupling. In Section \ref{discussion}, we comment on further
directions.  

We have included three appendices. Appendix \ref{sec:modularforms}
gives an overview of classical and Siegel modular forms, Appendix
\ref{sec:pfsolutionappend} discusses the Picard-Fuchs solutions, and Appendix
\ref{sec:variouscomputations} concludes with proofs of modular
identities.

\section{Review of the $SU(2)$ theory}\label{sec:rank1}

We will begin our discussion by reviewing some of the features of pure $\CN=2$ Yang-Mills theory with gauge group $SU(2)$, in order to  familiarize the reader with the concepts that are going to be expanded to higher rank in the following. For a more extensive review see \cite{AlvarezGaume,Tachikawa13,Bilal:1995hc,Laba05,Klemm:1997gg}. The $\CN=2$ vector multiplet consists of a gauge field $A$, complex scalar field $\phi$ and Weyl fermions $\lambda$ and $\psi$. They are all in the adjoint representation of the gauge group.

One of the important insights of Seiberg and Witten was that the quantum moduli space of pure $SU(2)$ super-Yang-Mills (SYM) coincides with that of an elliptic curve, parametrized by the quadratic Casimir $u=\frac{1}{2}\langle\Tr\phi^2\rangle$  \cite{Seiberg:1994rs}. The complex structure of the curve is then identified with the complexified effective gauge coupling $\tau=\frac{\theta}{\pi }+\frac{8\pi i}{g^2}$,  where $\theta$ is the vacuum angle. The elliptic curve can be written in a few different ways depending on conventions. For example, in \cite{Seiberg:1994aj, Klemm:1995wp} we find two different descriptions in terms of a cubic and a quartic polynomial, respectively. They do, however, correspond to isomorphic curves.  Let us denote the two curves by $\CC_1$ and $\CC_2$,
\be
\begin{aligned}   
	&\text{$\CC_1$}: \hspace{5pt} y^2 = x^3-u\,x^2+\frac{1}{4}\Lambda_1^4x, \\
	& \text{$\CC_2$}: \hspace{5pt} y^2 = (x^2-u)^2-\Lambda_2^4. \\
\end{aligned}
\ee
We will henceforth work in units where the dynamical scales are set to
1, $\Lambda_1=\Lambda_2=1$. We can better understand the relation between
these curves by studying the cross-ratios of the roots of the
polynomials on the right hand side. The rank three polynomial can be
considered as a rank four polynomial with one root at infinity, and
with a certain choice of numbering the roots are\footnote{the
  numbering is of course arbitrary, in the sense that the relations
  between the curves will continue to hold for other choices} 
\be
\begin{aligned}
	&\text{$\CC_1$}:\hspace{5pt} r_1=\infty,\hspace{5pt} r_2=\frac{1}{2}(u+\sqrt{u^2-1}),\hspace{5pt} r_3=0,\hspace{5pt} r_4=\frac{1}{2}(u-\sqrt{u^2-1}), \\
	&\text{$\CC_2$}: \hspace{5pt} r_1=\sqrt{u+1},\hspace{5pt} r_2=\sqrt{u-1},\hspace{5pt} r_3=-\sqrt{u+1},\hspace{5pt} r_4=-\sqrt{u-1}.
\end{aligned}
\ee
The cross-ratios $C = \frac{(r_3-r_1)(r_4-r_2)}{(r_3-r_2)(r_4-r_1)}$ for both curves then become 
\be
\begin{aligned}
	&\text{$\CC_1$,  $\CC_2$ }:\hspace{5pt}C =  \frac{2\sqrt{u^2-1}}{u+\sqrt{u^2-1}},\end{aligned}
\ee
which demonstrates that the two curves are isomorphic for fixed $u$, $\tau_1=\tau_2$.

\begin{figure}[h]\centering
	\includegraphics[scale=1]{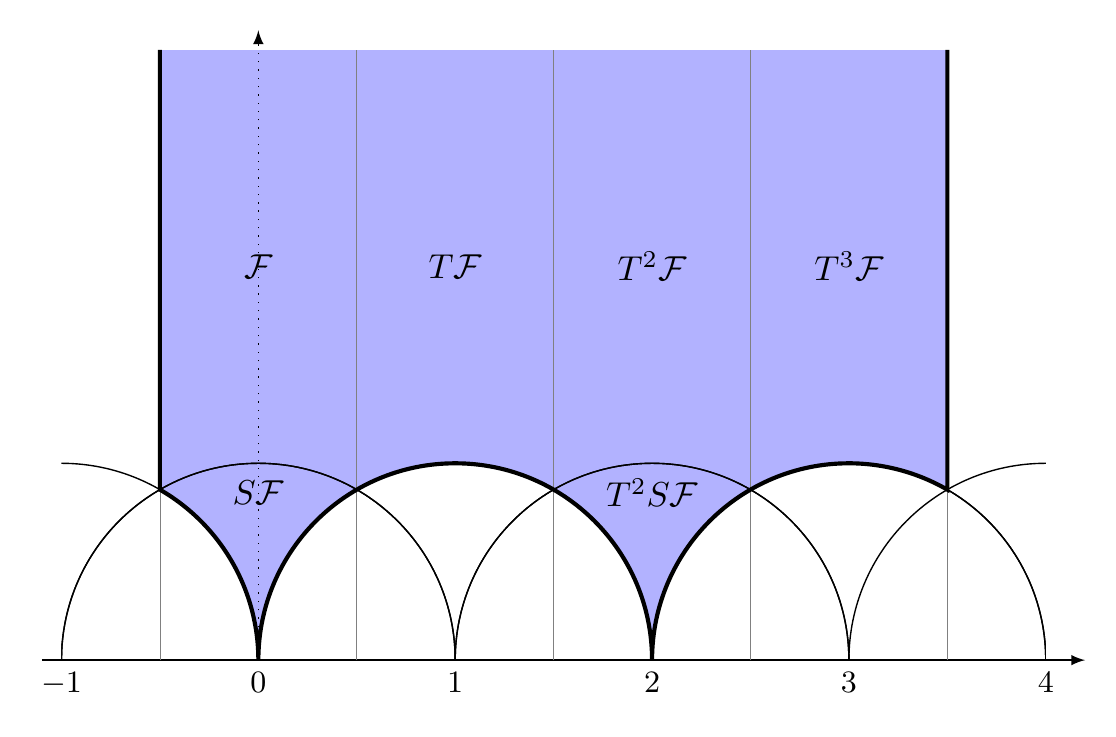} 
	\caption{Fundamental domain $\Gamma^0(4)\backslash \mathbb{H}$
          of the congruence subgroup $\Gamma^0(4)$. It consists of
          six images of the key-hole fundamental domain $\CF$. }\label{fig:fund_dom_gamma04}
\end{figure}

A striking feature of  Seiberg-Witten theory is modularity. To see this, we first note that every genus one curve can be written in  Weierstra{\ss} form
\begin{equation}
y^2=4x^3-g_2x-g_3,
\end{equation}
where the coefficients $g_i$ depend on the complex structure. These
curves have been studied extensively  and the coefficients have been
shown to be modular forms (see \cite{Diamond} for a pedagogical
review). Moreover, the roots of the polynomial are
functions of modular forms. Using this, one can relate the roots of
the SW curve to elliptic theta functions and then further use the
cross-ratios to derive an expression for $u$ in terms of these theta
functions, 
\be
\label{utau2}
u=\frac{\vartheta_2^4(\tau)+\vartheta_3^4(\tau)}{2 \vartheta_2^2(\tau) \vartheta_3^2(\tau)}=\frac18\left(q^{-\frac14}+20\,q^{\frac14}-62\,q^{\frac34}+216\,q^{\frac54}+\CO(q^{\frac74})\right),
\ee
with $q=e^{2\pi i \tau}$. The  $q$-expansion of $8\, u$ is
known in the mathematics literature as the McKay-Thompson series of 
class 4C for the  Monster group \cite{Conway:1979qga, Alexander:1992, ford1994, Ferenbaugh1993}. The order
parameter $u$ is a modular function of the effective coupling $\tau$
for the congruence subgroup $\Gamma^0(4)$ of $ SL(2,\mathbb Z)$ (see
Appendix \ref{appendA} for the definition of the Jacobi theta
functions and congruence subgroups). The fundamental domain of
$\Gamma^0(4)$ is  given by the image of $\CF = SL(2,\mathbb
Z)\backslash \mathbb H$ under six elements in $SL(2,\mathbb Z)$, see
Figure \ref{fig:fund_dom_gamma04}. An equivalent way to derive
\eqref{utau2} is to calculate the $j$-invariant of the SW curve and
equate it with the $j$-invariant of the Weierstra{\ss} curve, which has
a known expression in terms of Jacobi theta functions.  

Singularities appear in the quantum moduli space when two branch points coincide, or equivalently when the discriminant 
\begin{equation}
\Delta = \prod_{i<j}(r_i-r_j)^2
\end{equation}
of the curve vanishes. It is proportional to $u^2-1$ and we thus find three singularities, $u=\pm 1$ and $u\to\infty$. Using \eqref{utau2} it is easily shown that this corresponds to $\tau\to0$, $2$ and $i\infty$ respectively. The strong coupling points $\tau\to 0$ and $\tau\to 2$ correspond to the rational cusps of the fundamental domain of $\Gamma^0(4)$ (see Figure \ref{fig:fund_dom_gamma04}). These are the points where the monopole and the dyon become massless, respectively.

The complex masses $a_D$ and $a$ are given by period integrals of a meromorphic $1$-form over the elliptic curve. In order to determine these, one can use the fact that they form a system of solutions to a set of Picard-Fuchs equations. This allows to express the periods in terms of hypergeometric functions \cite{Ferrari:1996sv},
\be
\label{su2periods}  
\begin{split}
a_D(u)&= \tfrac{i}{2} (u-1)\,_2F_1(\tfrac{1}{2},\tfrac{1}{2},2;\tfrac{1-u}{2}),\\
a(u)&=\sqrt{\frac{(u+1)}{2}}\,_2F_1\!\left(-\tfrac{1}{2},\tfrac{1}{2},1;\tfrac{2}{1+u} \right).
\end{split}
\ee
At the strong coupling points, the periods $\pi(u)=(a_D(u),a(u))$ become $\pi(1)=(0,\frac2\pi)$ and $\pi(-1)=(-\frac{4i}{\pi},-\frac{2i}{\pi})$. According to the central charge formula $Z_\gamma=\gamma\cdot \pi$, these values confirm that for $u=1$, the monopole $\gamma=(1,0)$ becomes massless while for $u=-1$ the dyon $\gamma=(-1,2)$ becomes massless. The limits $\lim_{u\to \pm1}a(u)$ depend on the direction from which $\pm1$ are approached, which is due to the branch cut in the hypergeometric function \cite{Ferrari:1996sv}. We choose to take the limit from the lower half $u$-plane.  

To demonstrate that the periods \eqref{su2periods} are indeed correct, define the effective coupling as $\tau= \frac{\partial a_D}{\partial a}$. Using the chain rule, one can then compute $\tau$ as a function of $u$. The quantity $q=e^{2\pi i \tau}$ can be expanded for large $u$ and the resulting relation can be inverted order by order, and one finds the same $q$-expansion as in \eqref{utau2}. 

The $u$-plane has a spontaneously broken global $\mathbb Z_2$ symmetry, acting by $u\mapsto e^{\pi i} u$. It acts on the periods $\pi$ as $\rho= i\left(\begin{smallmatrix}1&-2\\0&1\end{smallmatrix}\right)$ and sends $\tau\mapsto \tau-2$, thus mapping the dyonic cusp, $u=-1$, to the monopole cusp, $u=1$. Its square is the monodromy at infinity, 
\begin{equation}\label{minf}
\rho^2=M_\infty=\begin{pmatrix} -1&4\\0&-1\end{pmatrix}.
\end{equation}
It  can be directly checked that \eqref{utau2} picks up a minus sign under $T^2$, $u(\tau-2)=-u(\tau)$, and therefore is invariant under \eqref{minf}. In fact, $u$ also picks up a minus sign under $TST^{-1}=\left(\begin{smallmatrix}1&-2\\1&-1\end{smallmatrix}\right)$, which together with $T^2$ generates $\Gamma^0(2)$. One could therefore consider $u$ to be a modular function for the congruence group $\Gamma^0(2)$ with multipliers $\pm1$. 

The zeros of \eqref{utau2} are given by the $\Gamma^0(4)$ orbit of $\tau_0=1+i$, lying on the boundary of the  fundamental domain (proof in Appendix \ref{zerosu2}). This can be understood as follows. The  origin $u=0$ is invariant under the global $\mathbb Z_2$ symmetry $\rho$, which acts as $\tau\mapsto \tau-2$. The boundary arcs near the cusps $0$ and $2$ are identified, and the origin is the symmetric point  $\tau_0$ where the arcs from both cusps meet. The two points $\tau_0$ and $\tau_0+2$ in Fig. \ref{fig:fund_dom_gamma04} with this property are identified under $\left(\begin{smallmatrix}-3&4\\-1&1\end{smallmatrix}\right)\in\Gamma^0(4)$.

\section{The Coulomb branch of the $SU(3)$ theory}
\label{sec:higher_rank}
We study in this Section the $SU(3)$ Coulomb branch. We first recall
the Seiberg-Witten geometry in Section \ref{SU3SW} following \cite{Klemm:1994qs, Klemm:1995wp, Argyres:1994xh}. Section
\ref{PFsolution} reviews the Picard-Fuchs solution for the 
complexified masses and couplings. Section \ref{SWRHform} uses those
results to write the curve in Rosenhain form.

\subsection{Seiberg-Witten geometry} 
\label{SU3SW}
The vector multiplet scalar $\phi$ can  be gauge rotated into the
Cartan subalgebra of $SU(3)$.  Then, $\phi$ can be expanded in terms
of the two Cartan generators 
$H_I$, $I=1,2$, as
\be
\phi =a_1H_1+a_2H_2.
\ee
Non-vanishing vevs of $\phi$ break the gauge group in general to
$U(1)^{2}$. The central charges of the gauge
bosons are then given by
\be 
\label{ccharges}
\begin{split} 
&Z_1=2a_1-a_2,\\  
&Z_2=2a_2-a_1,\\
&Z_3=a_1+a_2.
\end{split}
\ee
We denote electric-magnetic charges under $U(1)^2$ as
$\gamma=(m_1,m_2,n_1,n_2)$, where $m_i$ are the magnetic and $n_i$ the
electric charges respectively, and the period vector as
$\pi=(a_{D,1},a_{D,2},a_1,a_2)^T$.  The central charge for a generic
$\gamma$ is then given
by $Z_\gamma=\gamma\cdot \pi$, where $\cdot$ is the standard scalar
product.

The Coulomb branch is parametrized by vevs of Casimirs of $\phi$, $u_I\sim \langle\Tr\phi^I\rangle$, $I=2,3$. 
Gauge invariant combinations for $SU(3)$ are 
\be
\begin{split}\label{casimirs}
u=u_2&=\frac 12\langle \mathrm{Tr}(\phi^2)\rangle_{\mathbb{R}^4}=a_1^2+a_2^2-a_1a_2,\\
v=u_3&=\frac 13\langle \mathrm{Tr}(\phi^3)\rangle_{\mathbb{R}^4}=a_1a_2(a_1-a_2).
\end{split}
\ee
These relations can be rewritten in terms of two cubic equations for $a_1$ and $a_2$ as
\be
\begin{split}\label{aIcubic}
	a_1^3-ua_1-v=0,\\
	a_2^3-ua_2+v=0.
\end{split}
\ee
There is a spontaneously broken global $\mathbb{Z}_6$ symmetry   acting on $u$ and $v$ by
$u\mapsto \alpha\,u$ and $v\mapsto -v$, with $\alpha=e^{2\pi i/3}$. Classically, the discriminant is the  determinant $\Delta_{\rm classical}$ of the matrix $B_{IJ} =\pd{u_{I+1}}{a_J}$. It reads
\be 
\Delta_{\rm classical}=\det B_{IJ} = (a_1-2a_2)(2a_1-a_2)(a_1+a_2),
\ee
and vanishes when one of the gauge bosons \eqref{ccharges} becomes massless. 

Let us denote the space parametrized by $u$ and $v$ by
$\CU$. We  parametrize points on this space by $(\underline u, v)\in
\CU$, where $\underline u$ is the normalized parameter, 
$\underline u=\sqrt[3]{\frac{4}{27}}\, u$. The moduli space $\CU$ parametrizes a
complex  two-dimensional family of 
hyperelliptic curves of genus two \cite{Klemm:1994qs, Argyres:1994xh},
\be
\label{g2curve}
y^2=(x^3-u\,x-v)^2-\Lambda^6,
\ee 
which has  discriminant 
\be \label{discr} 
\Delta_\Lambda = \Lambda^{18}(4u^3-27(v+\Lambda^3)^2)(4u^3-27(v-\Lambda^3)^2).\ee
This can be viewed as a product of the discriminants of two  elliptic curves whose $v$
parameters are separated by $2\Lambda^3$. 
Note that the $\mathbb{Z}_6$ global symmetry leaves the discriminant
invariant. It vanishes if and only if $\underline u^3 = (v\pm
\Lambda^3)^2$. We will frequently use units where the dynamical scale
$\Lambda =1$ and we note that  it can always be restored from
dimensional analysis.

\begin{figure}[h]\centering
	\includegraphics[scale=0.9,trim={6.5cm 7.5cm 6.5cm 7cm},clip]{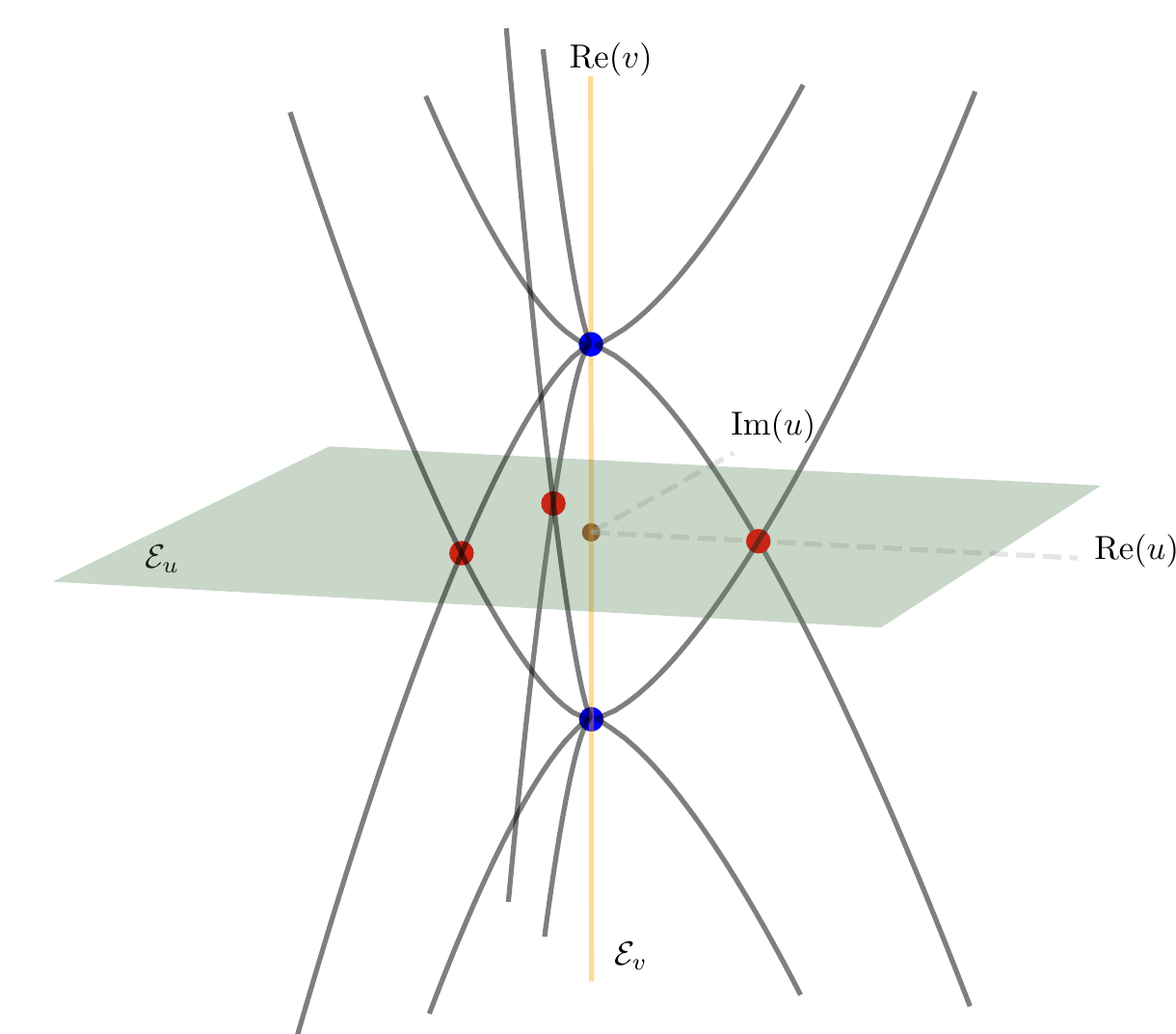}
	\caption{Singular lines $\Delta(u,v)=0$ in the $SU(3)$ moduli space with $\text{Im } v=0$, associated to massless dyons \cite{Klemm:1995wp}. The red dots represent the strong coupling points $(\underline u,v)=(1,0)$, $(\alpha,0)$ and $(\alpha^2,0)$ on the $v=0$ plane $\CE_u$, where two singular lines intersect. The blue dots represent the AD points $(\underline u,v)=(0,1)$ and $(0,-1)$ respectively, where three singular lines intersect. They lie on $\CE_v$, which is represented by the $\text{Re}\, v$  axis here. The two loci $\CE_u$ and $\CE_v$ intersect in the origin $(u,v)=(0,0)$ (brown).}\label{fig:singularlines}
\end{figure}

If we restrict to $\text{Im}\, v= 0$,  the zero locus
of the discriminant describes six singular curves which intersect in
the following points. On the $v=0$ plane, there are four
singularities, namely $\underline u\in
\{\infty,1,\alpha,\alpha^2\}$. On the other hand for $\underline u=0$,
there are two singularities at $v=\pm 1$. These are the
Argyres-Douglas points, where mutually non-local BPS states become
massless and the theory becomes superconformal
\cite{Argyres:1995jj}. Figure \ref{fig:singularlines} sketches the
singular lines on the subset of $\CU$ where $\text{Im}\, v= 0$. The
singular lines represent regions in $\CU$ where the effective action
of the pure $\CN=2$ theory becomes singular, and they are associated
with vacua where hypermultiplets become massless.  

\label{sec:semi_class_analysis}
Similarly to the $SU(2)$ case, the periods transform under monodromies which generate the duality group of the theory. The classical part of the monodromy group is given by the Weyl group of the $SU(3)$ root lattice, which acts as reflections on lines perpendicular to the positive roots. The perturbative quantum correction comes from the one-loop effective action. It contributes to the prepotential as
\begin{equation}\label{f1l}
\mathcal F_{1-\text{loop}} = \frac{i}{2\pi} \sum_\alpha Z_\alpha^2 \log Z_\alpha,
\end{equation}
where the sum runs over all positive roots $\alpha_1$, $\alpha_2$ and $\alpha_3=\alpha_1+\alpha_2$. Here, $Z_\alpha$ are the central charges \eqref{ccharges} of the gauge bosons.

The semi-classical monodromies  can be derived in the following way. The Weyl group of the root lattice $A_2$ is generated by  two reflections, $r_1$ and $r_2$. The element $r_k$ reflects the root lattice on the line perpendicular to $\alpha_k$ . For instance, $r_2$ induces the map $\alpha_2\mapsto -\alpha_2$, $\alpha_1\mapsto \alpha_1+\alpha_2$. Using \eqref{ccharges}, we find that $a_1\mapsto a_1$ and $a_2 \mapsto a_1-a_2$. The semi-classical transformation of the dual variables can be obtained using \eqref{f1l} and the fact that, semi-classically, $a_{D,I} = \frac{\partial \mathcal F_{1-\text{loop}}}{\partial a_I}$ holds.
The crucial insight is that  $Z_2\mapsto -Z_2$ induces a shift of $\pi i$ due to the logarithm, and the result can be written as an integer linear combination of the periods. The other two Weyl elements transform $a_1$ and  $a_2$ in the following way,
\begin{equation}\begin{aligned}\label{weylr123}
r_1: (a_1,a_2)&\mapsto (a_2-a_1,a_2), \\ 
r_2: (a_1,a_2)&\mapsto (a_1,a_1-a_2), \\ 
r_3: (a_1,a_2)&\mapsto (-a_2,-a_1) .
\end{aligned}\end{equation}
 The corresponding monodromies can be obtained in a similar fashion, the result is
\be\label{semiclassicalmono}\scalefont{0.75}
\CM^{(r_1)}=\begin{pmatrix}-1&0&4&-2\\1&1&-2&1\\0&0&-1&1\\0&0&0&1\end{pmatrix},
\hspace{15pt} 
\CM^{(r_2)}= \begin{pmatrix}1&1&1&-2\\0&-1&-2&4\\0&0&1&0\\0&0&1&-1\end{pmatrix},
\hspace{15pt}
\CM^{(r_3)}=\begin{pmatrix}0&-1&1&-2\\-1&0&4&1\\0&0&0&-1\\0&0&-1&0\end{pmatrix}, 
\ee
which satisfy $\CM^{(r_3)}=\CM^{(r_2)}\CM^{(r_1)}(\CM^{(r_2)})^{-1}$.

\subsection{Picard-Fuchs solution}
\label{PFsolution}
One way to find the non-perturbative solution is to notice that the
periods satisfy second order partial differential equations of 
Picard-Fuchs (PF) type, whose solution space is spanned by the
generalized hypergeometric function $F_4$ of Appell
\cite{Klemm:1995wp}. We review some aspects of the PF solution in the
following, and left further details for Appendix \ref{sec:pfsolutionappend}. We study two interesting regions, one where $u$ is large and $v$ small, and the other one where $v$ is large and $u$ is small. 

The non-perturbative effective action is characterized by the holomorphic prepotential $\CF$, which allows to define the dual periods $a_{D,I}=\frac{\partial \CF}{\partial a_I}$.
Both periods $a_I$ and $a_{D,I}$ are given by linear combinations of
Appell functions. The large $u$ expansion reads \cite{Klemm:1995wp} 
\be 
\label{PFEu}
\begin{split}
	a_{D,1}(u,v)&=-\frac{i}{2\pi}
	\left(\sqrt{u}+\frac{3}{2}\frac{v}{u}\right)\log\! \left(
	\frac{27}{4u^3} \right)-\frac{1}{\pi} \left(\frac{i}{2}+2\alpha_1\right)\sqrt{u}+\dots,\\
	a_1(u,v) &= \sqrt{u} +\frac{1}{2} \frac{v}{u}+\dots,
\end{split}
\ee
with $a_{D,2}(u,v) =a_{D,1}(u,-v)$, $a_2(u,v) =a_1(u,-v)$ and $\alpha_1\in \mathbb C$ a constant (see Appendix \ref{pfuappendix}). The coupling constants $\tau_{IJ}=\frac{\partial a_{D,I}}{\partial a_J}$  are determined using the chain rule,
\begin{equation}\label{eq:tau_nonzero_v}
\tau_{11}(u,v)=\tau_{22}(u,-v)= \frac{i}{\pi}\log(8u^3)+\frac{9iv}{2\pi
}u^{-3/2}-\left(\frac{129i}{32\pi}+\frac{63
	iv^2}{8\pi}\right)u^{-3}+\dots,
\end{equation}
The off-diagonal $\tau_{12}$ is given by the series 
\be\label{tau12largeu}
\tau_{12}(u,v)=-\frac{\tau_{11}(u,v)+\tau_{22}(u,v)}{4}-\frac{1}{2\pi i}\log(8)+\frac{1}{2\pi i}\frac{27}{4}f(u,v),
\ee
where
\be\label{eq:fuv}
f(u,v) = \frac{(1-4v^2)}{8}u^{-3}+\left(\frac{453}{1024}-3v^2-\frac{31}{16}v^4\right)u^{-6}+\dots.
\ee

Similarily, we find that the  large $v$ expansion of the coupling matrix reads (see Appendix \ref{pfderivation_vlarge} for details, $\omega=e^{\pi i/6}$)
\be\label{taulargev}
\tau_{11} \sim \frac{i}{\pi}\log(108v^2)-1+\frac{\omega}{\pi}uv^{-2/3}+\frac{\omega^5}{6\pi}u^2v^{-4/3}-\left(\frac{11i}{18\pi}+\frac{4i}{27\pi}u^3\right)v^{-2}+\dots,
\ee
and $\tau_{12}$ and $\tau_{22}$ are given by similar series. At $u=0$ we have $\tau_{11}=\tau_{22}+1$ and $\tau_{12}=-\frac{\tau_{11}}{2}+1$. 

For pure $SU(N)$ supersymmetric gauge theory, the periods  satisfy the following interesting relation \cite{Matone:1995rx,Eguchi:1995jh},
\begin{equation}\label{eq:matone}
\sum_{I=1}^{N-1} a_I a_{D,I}-2\mathcal F = \frac{Ni}{\pi}u,
\end{equation}
with $u=u_2 =\frac 12 \langle \Tr(\phi^2)\rangle$. Here, $\CF$ is the prepotential of the pure $SU(N)$ theory. 
For  $N=3$, we  can differentiate \eqref{eq:matone} with respect to $u$ and $v$ to get
\begin{equation}\begin{aligned}\label{firstderiv}
\frac{3i}{\pi} &= a_1a_{D,1}'-a_1'a_{D,1}+a_2a_{D,2}'-a_2'a_{D,2},\\
0&= a_1\dot a_{D,1}-\dot a_1a_{D,1}+a_2\dot a_{D,2}-\dot a_2a_{D,2},
\end{aligned}\end{equation}
where $'$ ( $\dot {}$ ) denotes $\pd{}{u}$ ( $\pd{}{v}$ ). Both relations  serve as useful checks of our solutions.

\subsection{Seiberg-Witten curve in Rosenhain form}\label{SWRHform}
In this section, we will relate the $SU(3)$ Seiberg-Witten curve to
the curve in Rosenhain form, which is a degree 5 equation. Every genus
two hyperelliptic curve can be brought to the Rosenhain form \cite{Rosenhain:1851}
\be
\label{curveRosenhain}
y^2 = x(x-1)(x-\lambda_1)(x-\lambda_2)(x-\lambda_3).
\ee 
The three roots $\lambda_i$  of the polynomial are also referred to as
\emph{Rosenhain invariants}. These invariants are complementary to
the Igusa invariants \cite{igusa1967,igusa1962}.

By a lemma of Picard, the Rosenhain invariants can be expressed in terms of even theta constants as
\be \label{thetaconstants}
\lambda_1 = \frac{\Theta_1^2\,\Theta_3^2}{\Theta_2^2\,\Theta_4^2},
\hspace{10pt} \lambda_2 =
\frac{\Theta_3^2\,\Theta_8^2}{\Theta_4^2\,\Theta_{10}^2},
\hspace{10pt}\lambda_3 =
\frac{\Theta_1^2\,\Theta_8^2}{\Theta_2^2\,\Theta_{10}^2}. 
\ee
The functions $\Theta_j$ are instances of  genus two Siegel modular forms,
\be
\Theta\begin{bmatrix}a\\b\end{bmatrix}(\Omega) = \sum_{k\in \BZ^2}\exp\left(\pi i(k+a)^T\Omega(k+a)+2\pi i(k+a)^T\, b  \right),
\ee
where the entries of the column vectors $a$ and $b$ take values in the
set $\{0,\frac{1}{2} \}$. The argument $\Omega$ is a $2\times2$-matrix 
\begin{equation}\label{Omega}
\Omega=\begin{pmatrix}\tau_{11} & \tau_{12}\\ \tau_{12} & \tau_{22}\end{pmatrix},
\end{equation} 
valued in the Siegel upper half-plane $\mathbb{H}_2$.  We refer to Appendix
\ref{sec:SiModForms} for a precise
definition and references. The moduli space of genus two curves $\CM_2$ is complex
 three-dimensional. Since the SW order parameters $u$ and $v$ are two complex parameters, the
 $SU(3)$ Coulomb branch maps out a complex two-dimensional space
 $\CU\subset \CM_2$ in the moduli space of genus two curves. In other words, $\CU$ is a
 divisor of $\CM_2$.

To relate the Rosenhain curve (\ref{curveRosenhain}) to the
Seiberg-Witten curve (\ref{g2curve}), note that a degree 5 polynomial as in
(\ref{curveRosenhain}) can be obtained by a linear fractional
transformation of a degree 6 hyperelliptic equation
$y^2=\prod_{j=1}^6(x-r_j)$, which maps three of
the roots to $\infty$, $0$ and 1. Linear fractional maps leave
cross-ratios invariant, which is a convenient way to relate the
$\lambda_j$ to $u$ and $v$. Let us define the cross-ratio of four
points $z_i\in \mathbb{CP}^1$ as
\be
\label{crossratio}
C(z_1,z_2,z_3,z_j)=\frac{(z_1-z_3)(z_2-z_j)}{(z_1-z_j)(z_2-z_3)},
\ee
such that $C(\{\infty,0,1,\lambda_j\})=\lambda_j$.

Note that we have 120 different possibilities to map three
roots among the $\{r_j\}$ to ${0,1,\infty}$, and another $3!$ possibilities
to identify the three cross-ratios in the hyperelliptic setting with
the $\lambda_j$. By studying the large $u$
expansions of these for non-zero $v$, one can easily
identify which cross-ratios, in terms of the $r_i$, correspond to which
$\lambda_j$. To this end, let $\alpha=e^{2\pi i /3}$ as before. The roots of the rhs of
(\ref{g2curve}) are then given by (with $\Lambda=1$)
\be\label{r1-r6}
\begin{alignedat}{2}
	r_1&=s_+(u,v+1)+s_-(u,v+1), \hspace{15pt} &&r_4=s_+(u,v-1)+s_-(u,v-1),\\ 
	r_2&=\alpha\,s_+(u,v+1)+\alpha^2\,s_-(u,v+1),\hspace{15pt}  &&r_5=\alpha\,s_+(u,v-1)+\alpha^2\,s_-(u,v-1),\\
	r_3&=\alpha^2\,s_+(u,v+1)+\alpha\,s_-(u,v+1),\hspace{15pt} 	&&r_6=\alpha^2\,s_+(u,v-1)+\alpha\,s_-(u,v-1),
\end{alignedat}
\ee
where 
\begin{equation}
s_\pm(u,v)=\sqrt[3]{\frac{v}{2}\pm \sqrt{\frac{v^2}{4}-\frac{u^3}{27}}}.
\end{equation}
To simplify notation, let us set $s_{\pm\pm}\coloneqq s_\pm(u,v\pm 1)$. The large $u$, small $v$ expansions for the roots are
\be
\label{asymp_roots}
\begin{alignedat}{2}
r_1&=\sqrt{u}+\frac{1+v}{2u}+\dots,\hspace{15pt} &&r_4=\sqrt{u}-\frac{1-v}{2u}+\dots,\\
r_2&=-\sqrt{u}+\frac{1+v}{2u}+\dots,\hspace{15pt} &&r_5=-\sqrt{u}-\frac{1-v}{2u}+\dots,\\
r_3&=-\frac{1+v}{u}+\dots,\hspace{15pt} &&r_6=\frac{1-v}{u}+\dots.
\end{alignedat}
\ee
Plugging the weak-coupling expansions (\ref{eq:tau_nonzero_v}) into
the Rosenhain invariants gives the leading behaviour for the
$\lambda_j$. From this we can see that each invariant $\lambda_j$
approaches 1 in the large $u$ limit. 

We continue by determining which of the 720 possible sets of cross-ratios matches with the
theta constants. We have to determine which roots correspond to the
first three points $z_i$, $i=1,2,3$, in the cross-ratio
(\ref{crossratio}). Since the three theta constants approach 1 in the
large $u$ limit, we should take for $\{z_1, z_2\}$ in
(\ref{crossratio}) the roots which vanish in this limit, thus
$\{r_3,r_6\}$. Together with the choice of $z_2$, this reduces to 8
possible triplets. From a further comparison between the Rosenhain invariants
and the cross-ratios, we determine that 
$z_1=r_6$, $z_2=r_3$ and $z_3=r_2$. With $C_j\coloneqq C({r_6,r_3,r_2,r_j})$
for $j=1$, $4$ and $5$, we arrive at 
\be 
\label{lambdaC}
\begin{split}
\lambda_1&=C_5,\qquad \lambda_2=C_1,\qquad \lambda_3=C_4.
\end{split} 
\ee
These are three equations for five unknowns, namely $\tau_{11}$,
$\tau_{12}$, $\tau_{22}$, $u$ and $v$. To make it more manifest that
the right hand side depends on only two variables, let
us express the cross-ratios $C_j$ in terms of $s_{\pm\pm}$,
\be\label{eq:cross_ratios}
\begin{split}
	&C_1=\alpha^2\frac{[\alpha\,s_{+-}+s_{--}- s_{++}-\alpha\,s_{-+}]\,[s_{++}-\alpha\,s_{-+}]}{[\alpha^2s_{+-}+\alpha\,s_{--}-s_{++}-s_{-+}]\,[s_{-+}-\,s_{++}]},\\
	&C_4=-\frac{[\alpha\,s_{+-}+\,s_{--}-s_{++}-\alpha\,s_{-+}]\,[\alpha^2\,s_{++}+\alpha\,s_{-+}-s_{+-}-\,s_{--}]}{3[s_{+-}-\alpha\,s_{--}]\,[s_{-+}-s_{++}]},\\
	&C_5=-\alpha^2\frac{[\alpha\,s_{+-}+s_{--}-s_{++}-\alpha\,s_{-+}]\,[\alpha\,s_{++}+s_{-+}-s_{+-}-\alpha\,s_{--}]}{3[s_{--}-\,s_{+-}]\,[s_{-+}-s_{++}]}.
\end{split}
\ee

Note that these expressions are true on the full moduli space.
For $u\neq 0$, we can define
\be
\label{XYs}
X=\frac{s_{++}}{\sqrt{u/3}},\qquad Y=\frac{s_{+-}}{\sqrt{u/3}},
\ee
such that $X^{-1}=s_{-+}/\sqrt{u/3}$ and
$Y^{-1}=s_{--}/\sqrt{u/3}$, since $s_{+\pm}\,s_{-\pm}=u/3$. The
cross-ratios can then be expressed as
\be \label{generaluv}
\begin{aligned}
	C_1 =& -\alpha^2\frac{X(X-\alpha Y)(X-Y^{-1})(X-\alpha X^{-1})}{(X^2-1)(X-\alpha^2 Y)(X-\alpha Y^{-1})}, \\
	C_4 =&-\frac13\alpha^2\frac{(X-\alpha Y)^2(X-Y^{-1})(X-\alpha Y^{-1})}{X(X^2-1)(Y-\alpha Y^{-1})} ,\\
	C_5 =&\enspace \enspace\; \frac13\frac{(X-\alpha Y)(X-Y^{-1})^2(X-\alpha^2 Y)}{X(X^2-1)(Y-Y^{-1})}.
\end{aligned}
\ee 
We thus see that the Coulomb branch can be identified with the
zero-locus of the three equations \eqref{generaluv} inside the space
$(\lambda_1,\lambda_2,\lambda_3, X,Y)$. One may in principle eliminate
$X$ and $Y$ to arrive at a single equation in terms of the
$\lambda_j$. In the following two sections, we will restrict to the
two one-dimensional sub-loci $\CE_u$ and $\CE_v$ of the solution space of \eqref{lambdaC}, where $v=0$ and $u=0$ respectively.

\section{Locus $\CE_u$: $v=0$}\label{sec:Rosenhain}
In this section we analyse the locus $v=0$. We will demonstrate
that the order parameter $u$ can be expressed in terms of classical modular
forms on this locus. In fact, we will arrive at two distinct expressions depending
on a choice of effective coupling. In Section
\ref{EllipticLoci}, we will discuss these aspects from the geometric
point of view.

\subsection{Algebraic relations}\label{v=0,ularge}
On the locus $v=0$ we have that $\tau_{11}(u,0)=\tau_{22}(u,0)$ and $\tau_{12}(u,0)$ is given by \eqref{tau12largeu}. Let us analyse these coupling constants, now from the perspective of
Section \ref{SWRHform}. For $u$ large and positive,
$s_{+\pm}$ has a large magnitude and phase $e^{\pi i /6}$. Similarly, the phase of $s_{-\pm}$ is approximately given by $e^{-\pi i /6}$ (see Appendix \ref{sec:roots} for a discussion on the subtlety of the cubic root). This means that
\be
\label{s--s++}
s_{--}=-\alpha\, s_{++},\qquad s_{+-}=-\alpha^2\, s_{-+}, \qquad X=-\alpha^2\,Y^{-1}.
\ee
Using this and  \eqref{XYs}, we find that  \eqref{generaluv} now turns into
\be 
\label{Cjv0}
\begin{aligned}
	C_1 =& -\frac{(X+X^{-1})\,(X-\alpha X^{-1})}{(X-X^{-1})\,(X+\alpha X^{-1})}, \\
	C_4 =& -\frac{1}{3}\frac{(X+X^{-1})^2}{(X-X^{-1})^2}, \\
	C_5 =&+ \frac{1}{3}\frac{(X+X^{-1})\,(X+\alpha X^{-1})}{(X-\alpha X^{-1})\,(X-X^{-1})}.
\end{aligned}
\ee
Since the rhs of (\ref{Cjv0}) depends only on one variable $X$, the cross-ratios $C_j$  satisfy two algebraic equations, which can be determined by solving the equations for $X^2$. One finds
\be\label{algrelv=0}
\begin{split}
&C_1\,C_5-C_4=0,\\
& (3C_4-C_1)^2-C_4(C_1+1)^2=0.
\end{split}
\ee
Using \eqref{lambdaC} and \eqref{thetaconstants}, the cross-ratios are identified with quotients of Siegel theta functions (see Appendix \ref{app:thetaconstants}), and the above equations take the form
\begin{align}\label{thetarelv=0} 
0&=\Theta _3^4-\Theta _4^4,  \\
0&=\Theta _1^2 \Theta _2^2 \Theta _8^4 \Theta _3^4-\Theta _2^4 \Theta _8^2 \Theta _{10}^2
   \Theta _3^4+8 \,\Theta _1^2 \Theta _2^2 \Theta _4^2 \Theta _8^2 \Theta _{10}^2 \Theta
   _3^2+\Theta _1^2 \Theta _2^2 \Theta _4^4 \Theta _{10}^4-9\, \Theta _1^4 \Theta _4^4
   \Theta _8^2 \Theta _{10}^2.\nonumber
\end{align}
The two systems of equations above are equivalent given that none of the $\lambda_j$ vanish or are infinite, which is an assumption of Picard's lemma \eqref{thetaconstants}.  We can use the second relation of \eqref{Cjv0} to  solve for $u$, 
\be
\label{u3v0}
u^3=\frac{\sqrt{27}}{2}\frac{(3C_4+1)^3}{\sqrt{C_4}(C_4-1)},
\ee
 and in terms of  theta constants  this gives
 \begin{equation}
u^3=\frac{\sqrt{27}}{2} \frac{(3\Theta_1^2 \Theta_8^2+\Theta_2^2 \Theta_{10}^2)^3}{\Theta_1 \Theta_2^3 \Theta_8 \Theta_{10}^3(\Theta_1^2 \Theta_8^2-\Theta_2^2 \Theta_{10}^2)}.
\end{equation}
This  can be viewed as a generalisation of the rank 1 result \eqref{utau2}, in the sense that we can write the  parameter $u$ as a rational function of theta series.  It follows naively that $u$ transforms as a weight 0 function under a subgroup of $Sp(4,\mathbb Z)$.

\subsection{A modular expression for $u$}
\label{umodexp}
The solutions to the algebraic relations \eqref{thetarelv=0} are not unique due to the periodicity in the $\tau_{IJ}$. The first equation implies $\tau_{11}-\tau_{22}= 2k$ with $k\in \mathbb Z$, but we know from \eqref{eq:tau_nonzero_v} that $k=0$. 
From \eqref{tau12largeu} we can make a power series expansion for
$\tau_{12}$ in terms of $p=e^{2\pi i \tau_{11}}$. One  finds 
\be\label{eq:tau_12_from_cs}
\tau_{12}=-\frac{1}{2}\tau_{11}-\frac{1}{2\pi i}\log(8)+\frac{1}{2\pi i} \frac{27}{4}h(p),
\ee
with   
\be 
h(p) = p^{\frac{1}{2}}-\frac{63}{16}\,p +\frac{1447}{64}p^{\frac 32}-\frac{307679}{2048}p^2+\CO(p^{\frac{5}{2}}),
\ee
by satisfying the second relation in \eqref{thetarelv=0} order by order. 
Substitution of (\ref{eq:tau_12_from_cs}) in (\ref{u3v0}) gives the following $p$-expansion for $u$,
\be \label{u(p)_ularge}
u=\frac{1}{2}\,p^{-\frac{1}{6}}+\frac{43}{8}\,p^{\frac{1}{3}}-\frac{2923}{128}\,p^{\frac{5}{6}}+\frac{1713}{16}\,p^{\frac{4}{3}}+\CO(p^{\frac{11}{6}}).
\ee 
One can verify agreement with the  Picard-Fuchs approach by
substituting this expansion in Eq. \eqref{eq:tau_nonzero_v}. As this
series is only an expansion for small $p$, it is not very
elucidating. To arrive at a closed expression, we aim to express $u$ 
as a function of a ``coupling constant'' which transforms well under
the duality transformations. This is not the case for $\tau_{11}$.
 
However when $\tau_{11}=\tau_{22}$,  the inversion $\CS= \left(\begin{smallmatrix}0& -\mathbbm
    1\\\mathbbm 1&0\end{smallmatrix} \right)\in Sp(4,\mathbb Z)$
 acts naturally on the linear combinations
$\tau_\pm=\tau_{11}\pm \tau_{12}$, which are in one-to-one correspondence
with $\tau_{11}$ and $\tau_{12}$. From \eqref{symplectictransf}, we
deduce for the action of $\CS$ on $\tau_\pm$
\be 
\label{Staupm} 
\CS: \tau_{11}\pm \tau_{12}\mapsto -\frac{1}{\tau_{11}\pm \tau_{12}}.
\ee
That is to say, it reduces to the ordinary $S$-transformation
$\tau_\pm \mapsto - 1/\tau_\pm$. Moreover, $\tau_\pm \in \mathbb{H}$
for both $\pm$. To see this note that since $\im(\Omega)$ is positive
definite, we have that $y_{11}>0$ and $y_{11}y_{22}-y_{12}^2>0$, where
$y_{IJ}=\im(\tau_{IJ})$. Whenever $y_{11}=y_{22}$,  the latter inequality implies that
$y_{11}^2>y_{12}^2$. Since $y_{11}>0$, it implies $y_{11}>y_{12}$ and
$y_{11}>-y_{12}$ simultaneously. From this we learn that
$y_{11}-y_{12}$ and $y_{11}+y_{12}$ are both positive and therefore
$\tau_\pm\coloneqq \tau_{11}\pm\tau_{12}\in\mathbb H$.

We will proceed by considering $\tau_-\eqqcolon\tau$, leaving the discussion
on $\tau_+$ for Section \ref{SubSecu+}. To determine $u$ as function
of $\tau$, one can first find the series expansion for $\tau$ in
terms of $p$, invert and substitute $p(\tau)$ in (\ref{u(p)_ularge}). Alternatively, one can revert to
the Picard-Fuchs solution,
by inverting the series \eqref{eq:tau_nonzero_v} for $v=0$, 
\begin{equation} 
q=e^{2\pi
  i\left(\tau_{11}(u)-\tau_{12}(u)\right)}=U^{3}+45U^{4}+1512U^{5}+45672U^{6}+\dots,
\qquad U=\frac{1}{4u^3}.
\end{equation}
Either method gives us the following series for $u$,
\be \label{v0solution}
\sqrt[3]{4}\, u = q^{-\frac{1}{9}}+5\,q^{\frac{2}{9}}-7\,q^{\frac{5}{9}}+3\,q^{\frac{8}{9}}+15\,q^{\frac{11}{9}}-32\, q^{\frac{14}{9}}+\CO(q^{\frac{17}{9}}).
\ee
This expansion is also known as the McKay-Thompson
series of class 9B for the Monster group \cite{Conway:1979qga, Alexander:1992, ford1994, Ferenbaugh1993}. Thus
similarly to the $u$ for rank 1 (\ref{utau2}), we find a
McKay-Thompson series. 
We then have
\be  \label{largeuv=0}
 u = u_-(\tau)= \sqrt[3]{\tfrac{27}{4}}\, \frac{b_{3,0}\left(\frac \tau3\right)}{b_{3,1}\left( \frac\tau3\right)},
\ee
where $b_{3,j}$ are
theta series for the $A_2$ root lattice,
\be\label{b3j}
b_{3,j}(\tau) = \sum_{k_1, k_2\in\mathbb Z+\frac j3}q^{k_1^2+k_2^2+k_1k_2}, \quad j\in\{-1,0,1\}.
\ee
The theta series $b_{3,j}$ transform under the generators of $SL(2,\mathbb Z)$ as ($\alpha=e^{2\pi i/3}$)
\be 
\label{b3trafos}
\begin{aligned}
S: \quad b_{3,j}\left(-\frac1\tau\right)&=-\frac{i\tau}{\sqrt3}\sum_{l\mod 3}\alpha^{2jl}\, b_{3,l}(\tau), \\
T: \quad b_{3,j}(\tau+1)&=\alpha^{j^2} b_{3,j}(\tau).
\end{aligned} \ee
The solution $u_-$ can also be expressed in terms of the Dedekind $\eta$-function (\ref{etaf}) as
\begin{equation}\label{newformulau}
u_-(\tau) = \sqrt[3]{\tfrac{27}{4}} \left(1+\frac 13 \frac{\eta\left(\tfrac{\tau}{9}\right)^3}{\eta(\tau)^3}\right).
\end{equation}

Using Theorem 1 in Appendix \ref{appendA}, one finds that $u_-(9\tau)$ is a modular function for the congruence subgroup
$\Gamma_0(9)$ (also defined in  Appendix \ref{appendA}). This implies
that $ u$ is a modular function for $\Gamma^0(9)$, which is
generated by the matrices $T^9$, $STS$ and $(T^3S)T(T^3S)^{-1}$. In
fact, it is easy to see from \eqref{b3trafos} that $u_-(\tau-3)=\alpha
u_-(\tau)$ for all $\tau\in \mathbb H$. Furthermore, $u$ rotates as well
under $TST^{-2}$, $u_-\!\left(\frac{\tau-3}{\tau-2}\right)=\alpha
u_-(\tau)$. The two elements $T^3$ and $TST^{-2}$ generate $\Gamma^0(3)$
and $u$ can therefore be interpreted as a modular function for
$\Gamma^0(3)$ with multipliers $\alpha^k$, analogous to the
discussion on rank 1 in Section \ref{sec:rank1}.

Let us analyse the strong coupling singularities $u^3=\tfrac{27}{4}$ for $v=0$ in terms of the variable $\tau$. We will demonstrate that these correspond to
$\tau\to 0, 3$ and $-3$. Using \eqref{b3trafos}, one finds that the  expansion around $ 0$ takes the form
\be\begin{aligned}\label{udtaud}
\sqrt[3]{\tfrac{4}{27}}\, u_{-,D}(\tau_D)& = \frac{b_{3,0}(3\tau_D)+2b_{3,1}(3\tau_D)}{b_{3,0}(3\tau_D)-b_{3,1}(3\tau_D)}\\
&=1+9\, q_D+27 \,q_D^2+81 \,q_D^3+198 \,q_D^4+\CO(q_D^5),
\end{aligned}\ee
with $\tau_D=-1/\tau$, $q_D=e^{2\pi i \tau_D}$ and $  u_{-,D}(\tau_D)\coloneqq  u_-(-1/\tau_D)$. In the same notation we can invert the series to find 
\be\begin{aligned}
q_D&=\chi -3 \chi ^2+9 \chi ^3-22 \chi ^4+21 \chi ^5+207 \chi ^6+\CO(\chi^7),
\end{aligned}\ee
where $\chi:= (\sqrt[3]{4/27}\, u -1)/9$. It follows that $q_D\to 0$
for $\sqrt[3]{4/27}\, u\to 1$ or $\chi\to 0$. This can be directly
confirmed by analytically continuing the Picard-Fuchs expansion around
$u=\sqrt[3]{27/4}$.

The expansion around $\pm 3$ can then be obtained from the one around
$0$ by shifting the argument $\tau_{D,\pm}=-\frac1\tau \pm 3$, and one
finds using  the $T$-transformation \eqref{b3trafos} that  
\begin{equation}
 u_{-,D}(\tau_{D,\pm})=\alpha^{\mp 1} \sqrt[3]{\tfrac{27}{4}}\, \frac{b_{3,0}(3\tau_{D})+2b_{3,1}(3\tau_{D})}{b_{3,0}(3\tau_{D})-b_{3,1}(3\tau_{D})}
\end{equation}
The expansions around the points $3$ and $-3$ differ from the one around 0 only by the phases $\alpha^{-1}=\alpha^2$ and $\alpha$. 
Together with \eqref{udtaud}, this proves that indeed $\tau\to \{0,-3,3\}$ corresponds to the three singularities $\underline u\to \{1,\alpha,\alpha^2\}$.  Due to the $T^9$-invariance of the solution \eqref{largeuv=0}, there is an ambiguity in identifying the  $\tau$-parameter  with $\tau+9\, \mathbb Z$. 
These $\mathbb Z_2$ points are studied in detail in
\cite{Argyres:1994xh, Douglas:1995nw}. They correspond to the 3 vacua
of the $\CN=1$ theory after deforming the $\CN=2$ theory by relevant
or marginal terms. 
 
\begin{figure}[h]\centering
	\includegraphics[scale=0.75]{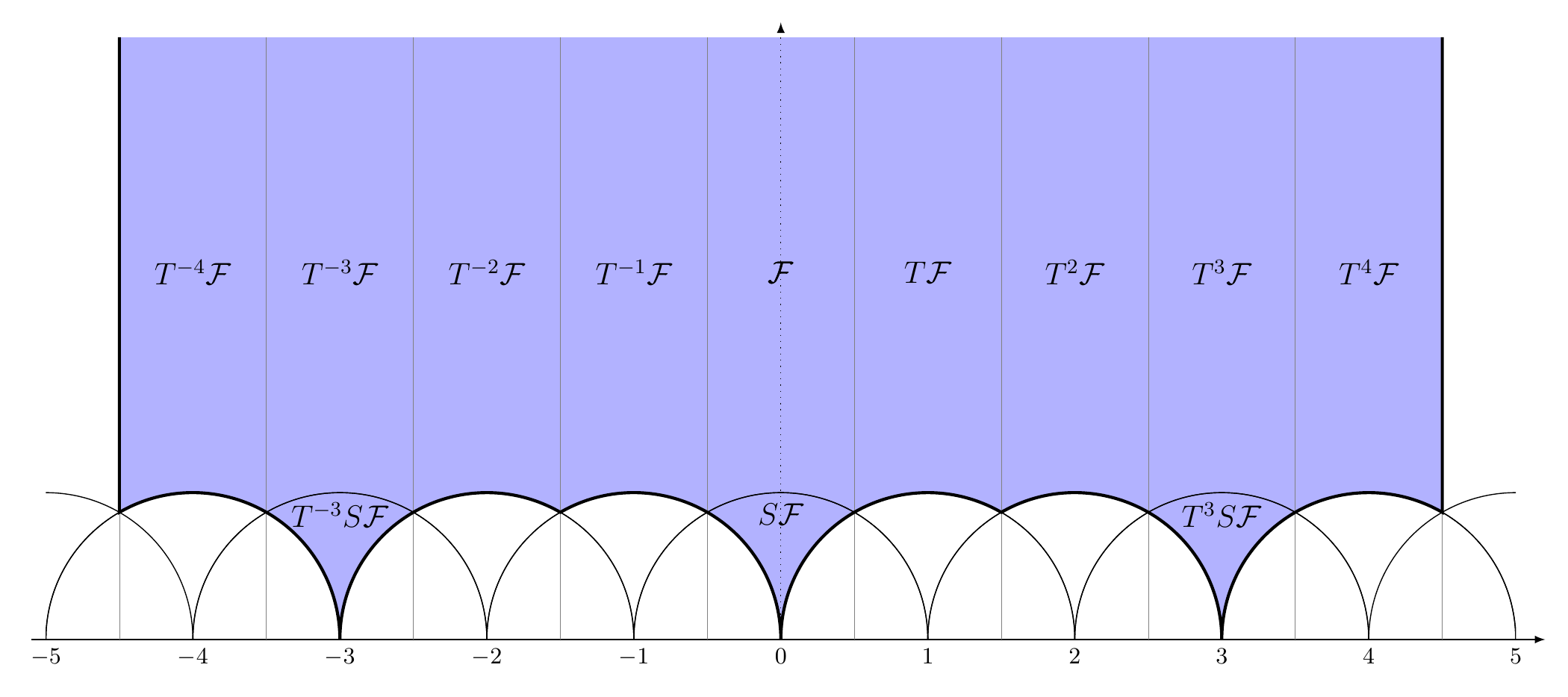}
	\caption{Fundamental domain $\Gamma^0(9)\backslash \mathbb{H}$
          of the congruence subgroup $\Gamma^0(9)$. It consists of
          12 images of the key-hole fundamental domain $\CF$. }\label{fig:fund_dom_gamma09}
\end{figure}

The modular analysis is completely analogous to the $SU(2)$ theory, as
reviewed in Section \ref{sec:rank1}: The cusps of $\Gamma^0(9)$ are
$\{ 0,-3,3,i\infty\}$, which is exactly where  $u$ assumes the
$\mathbb Z_2$ vacua and the semi-classical limit.  The fundamental
domain of $\Gamma^0(9)$ is given in Figure \ref{fig:fund_dom_gamma09}
and is
the union of 12 images of the $SL(2,\mathbb Z)$ fundamental domain
$\CF = SL(2,\mathbb Z)\backslash \mathbb H$,  
\begin{equation}
\Gamma^0(9)\backslash \mathbb H = \bigcup_{\ell=-4}^{4} T^\ell \CF\,\cup\, S\CF\cup\,T^3S\CF \cup\, T^{-3}S\CF.
\end{equation}
 
Using \eqref{newformulau}, we can  find the exact coupling at the origin of the moduli space. We have that $u(\tau_0)=0$ for the $\Gamma^0(9)$ orbit of 
\begin{equation}\label{tauorigin}
\tau_0 = \sqrt3\, \omega = \frac 32 +\frac{\sqrt 3}{2}i,
\end{equation}
with $\omega=e^{\pi i/6}$. This is rigorously proven in Appendix
\ref{zerosu3}.  The point $\tau_0$ lies on the boundary of the
fundamental domain, on the  point where the boundary arcs from
different cusps meet. The elements $(STS)^k\in \Gamma^0(9)$ map
$\tau_0\mapsto \tau_0-3k$ for integer $k$, which identifies the
``corners'' in Figure \ref{fig:fund_dom_gamma09}. This is compatible
with the global $\mathbb Z_3$ symmetry, which also acts by $T^{-3}$
and leaves the origin invariant. 
 It is in complete analogy to the $SU(2)$ picture, see Section \ref{sec:rank1}: 
We find the nice picture that the cusps of $\Gamma^0(9)\backslash
\mathbb H$ are in one-to-one correspondence with the singularities
$u^3=\frac{27}{4}$ and $ u=\infty$ and the origin is the symmetric point where the boundary arcs meet. 

We will derive the modular expression for $u$ from the SW geometry in Section
\ref{EllipticLoci}. Section \ref{sec:strongmon} will discuss how the action
of the $SU(3)$ monodromies reduce to the generators of $\Gamma^0(9)$ for
the action on $\tau_-$.

The connection between elliptic curves and theta constants furthermore
allows  to express the periods $\frac{\partial a_I}{\partial  u_J}$ as
modular forms. Indeed, the period matrix  $\frac{\partial a_I}{\partial
  u_J}$ can be written as a combination of even, odd and
differentiated theta constants \cite{Eilers:2017}. By substituting the
solution for $u$ and $v$ into the asymptotic expansion of the periods,
we can confirm this for some cases. Recall
that in the $SU(2)$ theory, $a$ is a quasi-modular form and
$\frac{da}{du}$ is a  modular form of $\Gamma^0(4)$ with non-trivial multipliers, both of
weight $1$ \cite{Laba05}. For rank 2, one finds that on $v=0$,  
\begin{equation}
\frac{\partial a_1}{\partial v}(\tau)=-\frac{\partial a_2}{\partial v}(\tau)=\frac{1}{3\sqrt[3]{2}}b_{3,1}(\tfrac\tau3)=\frac{1}{\sqrt[3]{2}}\frac{\eta(\tau)^3}{\eta(\tfrac\tau3)}.
\end{equation}
Theorem 1 in Appendix \ref{appendA} confirms that it is a modular form
of weight $1$ on $\Gamma^0(9)$, which is the same modular
group as for $u$.

\subsection{$u$ as a sextic modular function} 
\label{SubSecu+} 
While we chose in the above the modular parameter
$\tau_-=\tau_{11}-\tau_{12}$, Equation (\ref{Staupm}) shows that we
could equally well consider $\tau_+=\tau_{11}+\tau_{12}$. We will
consider the variable $\tau\coloneqq \tau_+$ in this subsection.  
We can determine the first terms in the $q$-expansion of $u$, which results in 
\begin{equation} 
\label{uuPlus}
u= u_+(\tau)=\frac{1}{4}\left(q^{-1/3}+104\,q^{2/3}-7396\,q^{5/3}+\CO(q^{8/3})\right).
\end{equation}
This series can be recognized as the  $q$-expansion of
\begin{equation}\label{u_thetas}
u_+(\tau)= \sqrt[3]{\tfrac{27}{2}}\,\frac{E_4(\tau)^{1/2}}{(E_4(\tau)^{3/2}-E_6(\tau))^{1/3}},
\end{equation}
where $E_4$ and $E_6$ are the Eisenstein series (\ref{Ek}). We will
derive this explicitly in Section \ref{EllipticLoci}.
The function $u_+$ is a root of the sextic equation
\be 
\label{sextic}
X^6- \frac{j(\tau)}{64}\,X^3+\frac{27\,j(\tau)}{256}=0,
\ee
where $j$ is the $j-$invariant (\ref{jfunction}). Since the
coefficients of this sextic equation are modular functions for $SL(2,\mathbb{Z})$, we call $u_+$ a {\it sextic
  modular function}. Due to the fractional powers in (\ref{u_thetas}), $u_+$ is not a
modular function for $SL(2,\mathbb{Z})$. In fact, $E_4^{1/2}$ and $u_+$ are
not invariant under {\it any} subgroup of $SL(2,\mathbb{Z})$. One way to see this is that $E_4$
has a simple zero for $\tau=\alpha$, such that the square root
introduces a branch cut. While the family of sextic modular functions
thus includes functions which are not modular for $SL(2,\mathbb{Z})$,
this family also includes functions which are modular for an index
6 congruence subgroup of $SL(2,\mathbb{Z})$. The order parameter for
$SU(2)$ (\ref{utau2}) is an example
of the latter. One can thus view the family of sextic modular
functions as an extension of the family of modular functions for 
index 6 congruence subgroups. We will discuss the modular properties
of \eqref{u_thetas} in more detail in a future
work \cite{AFMtoappear}.

Interestingly, $u_+$ is up to an overall factor the same function as the order parameter of the
massless $N_f=1$ theory with gauge group $SU(2)$ \cite{Nahm:1996di,
  Huang:2009md, Magro:1997qs}.  This aspect distinguishes massless $N_f=1$ from
  $N_f=0,2,3$, since for the latter theories the order parameters are
  modular functions for congruence subgroups isomorphic to
  $\Gamma^0(4)$ \cite{Nahm:1996di}. On the other hand, it is known
since the time of Fricke and Klein that similar
fractional powers of modular forms as in $u_+$ do appear in the context 
of Picard-Fuchs equations and hypergeometric functions
\cite{Lian:1995js, Alim:2013eja}.

\begin{figure}[h]\centering 
	\includegraphics[height=10cm]{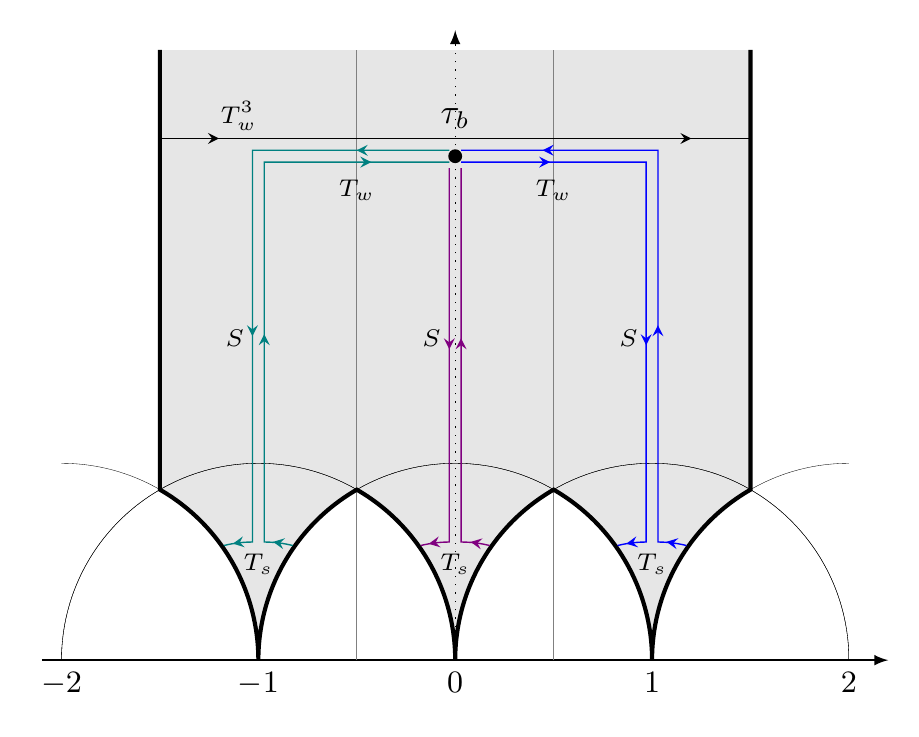}
	\caption{Fundamental domain for $u_+$. The vertical lines at
          $\tau=\pm 3/2$ are identified, as well as each pair of the
          two arcs meeting at a cusp $-1,0$ or 1. The point $\tau_b$
          is the base point for the monodromies, which are compositions of
 $T_w$, $T_s$ and $S$. $T_w$ is a shift
          $\tau\mapsto \tau+1$ at weak coupling, $T_s$ circles around
          a strong coupling cusp, and $S$ maps $\tau$ from weak to
          strong coupling.}\label{fig:funduPlus}
\end{figure}

As mentioned before, the fractional powers in (\ref{u_thetas}) are
incompatable with any subgroup of $SL(2,\mathbb{Z})$. 
Nevertheless, if we choose a basepoint, we can show that
$u_+$ is invariant under transformations of $\tau$, which combine to a closed trajectory 
with starting and endpoint equal to the base point. We choose the base
point $\tau_b$ with $\mathrm{Re}(\tau_b)=0$ and
$\mathrm{Im}(\tau_b)\gg 1$. First, using the modular transformation of $E_4$ and $E_6$,
we find for the expansion of $\tau$ near 0,
\be
\tau\to 0:\qquad u_+(\tau)=u_{+,D}(-1/\tau),
\ee
with  
\be
\begin{split}
u_{+,D}(\tau_D)&= \sqrt[3]{\tfrac{27}{2}}\, \frac{E_4(\tau_D)^{1/2}}{(E_4(\tau_D)^{3/2}+E_6(\tau_D))^{1/3}}\\
&=\sqrt[3]{\tfrac{27}{4}}\,\left( 1+144\,q_D-3456\,q_D^2+596160\,q_D^3+\dots\right).
\end{split}
\ee 
The $S$-transform $u_{+,D}$ is also a solution to (\ref{sextic}) and thus also a sextic
modular function. From Eq. \eqref{uuPlus} we see that $u_+$ is invariant under $\tau\mapsto \tau+3$ at weak coupling,
$\mathrm{Im}(\tau)\gg 1$. Let us introduce $T_w$ for the translation
at weak coupling. Moreover at strong coupling, $0<\mathrm{Im}(\tau)\ll
1$, $u_+$ is invariant under $\tau_D=-1/\tau \mapsto \tau_D+1$. Let us
introduce $T_s$ for the translation at strong coupling. 
We can get the monodromies around 
the other cusps, $\tau=\pm 1$ from conjugation with $T_w$.  We then find
that $u_+$ is left invariant by
\begin{equation}
\label{uPtrafos1}
T_w^{3n},\qquad (T_w^\ell S)T_s(T_w^\ell S)^{-1},\qquad \ell,n\in \mathbb{Z}, 
\end{equation}
where $S$ is the usual inversion $\tau\mapsto -1/\tau$, mapping $\tau$ from
weak to strong coupling. These transformations are sketched in Figure
\ref{fig:funduPlus} for $n=1$ and $\ell=0,\pm 1$. 

We denote the invariance group of $u_+$ by $\Gamma_{u_+}$. It is
generated by the elements in (\ref{uPtrafos1}) with $n=1$, and
$\ell=0,1$. From the invariance under (\ref{uPtrafos1}), one derives that a
fundamental domain is given by
\be
\label{Gammau+}
\Gamma_{u_+}\backslash\mathbb{H}=\bigcup_{\ell=-1}^{1} T^\ell \CF\,\cup T^\ell S\CF.
\ee
It consists of six copies of $\CF$, which is directly related to $u_+$
being a sextic modular function. This fundamental domain is the grey area in Figure
\ref{fig:funduPlus}. The domain is clearly topologically equivalent to the fundamental
domain in Figure \ref{fig:fund_dom_gamma09}.
The expansions of $u_+$ and $u_{+,D}$ demonstrate that $u_+(i\infty)=\infty$, $u_+(0)=\sqrt[3]{\tfrac{27}{4}}$  and
$u_+(\pm 1)=\alpha^\mp \sqrt[3]{\tfrac{27}{4}}$. We will derive $u_+$ from the SW geometry in Section \ref{EllipticLoci}, and the
transformations (\ref{uPtrafos1}) in Section \ref{sec:strongmon} from the $SU(3)$ monodromies around the 
strong coupling cusps.      
 
Because $u_+$ is not a weakly holomorphic modular form, but involves
fractional powers of modular forms, it is problematic to identify the transformations (\ref{uPtrafos1}) with elements of
$SL(2,\mathbb{Z})$. One way to see that this identification is
problematic is that the composition of $S$, $T_w$ and $T_s$
does not satisfy the relation $(ST)^3=-\mathbbm 1$, if we
identify $T_w=T_s=T=\left( 
\begin{array}{cc}
 1 & 1 \\
 0 & 1 \\
\end{array}\right)$. To further study this aspect, let us list the $SL(2,\mathbb{Z})$ matrices corresponding to (\ref{uPtrafos1}),
\begin{equation}
\label{uPtrafos}
\begin{split}
T^3&=\left(
\begin{array}{cc}
 1 & 3 \\
 0 & 1 \\
\end{array}\right), \\
STS^{-1}&=\left(\begin{array}{cc}
 1 & 0 \\
 -1 & 1 \\
\end{array}\right),\\
 (TS)T(TS)^{-1}&=\left(
\begin{array}{cc}
 0 & 1 \\
 -1 & 2 \\
\end{array}\right), \\
(T^{-1}S)T(T^{-1}S)^{-1}&=\left(
\begin{array}{cc}
 2 & 1 \\
 -1 & 0 \\
\end{array}\right).
\end{split}
\end{equation}
These matrices fix each of the cusps $\{\infty, 0,1,-1\}$. On the other hand,
$u_+$ is not invariant under the modular action of the matrices on $\tau$,
$\tau\mapsto (a\tau+b)/(c\tau+d)$ except for $T^{3n}$. For example,
$STS^{-1}$ would map $\tau=i\infty$ to $-1$. The values of $u_+$ are
however different for these two arguments: $u_+(i\infty)=\infty$ and
$u_+(-1)=\alpha \sqrt[3]{\tfrac{27}{4}}$. Furthermore, the matrices
(\ref{uPtrafos}) generate the full modular group $SL(2,\mathbb{Z})$.

The origin $u_+(\tau_0)=0$ of the moduli space is again given by the
points where the boundary arcs meet: At $\tau_0=\alpha$ we have that
$E_4$ vanishes but $E_6$ does not. From  \eqref{u_thetas} it is then
clear that $\tau_0+\mathbb Z$ are indeed the zeros of $u_+$. This is
also compatible with the $\mathbb Z_3$ global symmetry, which
according to \eqref{uuPlus} acts as $T^{-1}$ and leaves the
origin invariant.

\section{Locus $\CE_v$:  $u=0$} \label{sectionu=0}
We will now consider the second elliptic locus, namely where $u=0$. 
By doing a similar analysis as in Section \ref{sec:Rosenhain} but now
for large $v$, we find that the correct matching between the
cross-ratios and the Rosenhain invariants for this limit is 
\begin{equation} 
\label{Identu0}
\lambda_1 = C_5, \hspace{20pt}\lambda_2 = C_4,\hspace{20pt}\lambda_3 = C_1.
\end{equation}
Note that the only difference from before is that the r\^{o}les of
$\lambda_2$ and $\lambda_3$ have been interchanged. One could perform
a change of symplectic basis to have the same matching as
\eqref{lambdaC}. This can be be done by acting on the periods with the
matrix $\CT_\theta=\left(\begin{smallmatrix}\mathbbm{1} &\theta \\ 0 &
    \mathbbm{1} \end{smallmatrix}\right)\in Sp(4,\BZ)$ with
$\theta=\left(\begin{smallmatrix}-1&2\\2&-4\end{smallmatrix}\right)$.\footnote{Note that there is an ambiguity in the
  choice of $\CT_\theta$. The $\lambda_j$ are invariant under a subgroup of $Sp(4,\BZ)$. Multiplying $\CT_\theta$ with an element of this group thus gives the same result.} 
This changes the  $\omega_1$, $\omega_2$ prefactors  of $a_{D,1}$ in
\eqref{ajlargev}. This would however also change the Rosenhain form, and we
therefore prefer to continue with the identification in
(\ref{Identu0}).  

We will proceed by deriving the relations satisfied by the couplings
$\tau_{IJ}$ on the locus $u=0$. 

\subsection{Algebraic relations} 
To determine the algebraic relations among the theta constants, we
assume that $v$ is real, large and positive. In this limit we find
that $s_{+\pm} = \sqrt[3]{v\pm 1}$ and $s_{-\pm}=0$. The cross-ratios
\eqref{eq:cross_ratios} simplify to 
\be
\label{Ciu0}
\begin{aligned}
	C_1 =& -\alpha^2\frac{s_{++}-\alpha s_{+-}}{s_{++}-\alpha^2s_{+-}}, \\
	C_4=& -\frac{\alpha^2}{3}\frac{(s_{++}-\alpha s_{+-})^2}{s_{++}s_{+-}}, \\
	C_5=&+\frac{1}{3}\frac{(s_{++}-\alpha s_{+-})\,(s_{++}-\alpha^2s_{+-})}{s_{++}s_{+-}} . 
\end{aligned}
\ee
From this we find two algebraic relations between the
cross-ratios, namely
\be\label{eq:C_rel_v_large}
\begin{split}
& C_1C_5-C_4=0, \\
&C_5^2+C_4^2-C_5C_4-C_4=0.
\end{split}
\ee
Writing these in terms of the theta constants, we have
\be \begin{aligned}\label{moalgrel}
0&=\Theta_1^4-\Theta_2^4, \\
0&=\Theta_2^4\Theta_3^2\Theta_8^4+\Theta_1^4\Theta_3^2\Theta_{10}^4-\Theta_1^2\Theta_2^2\Theta_3^2\Theta_8^2\Theta_{10}^2-\Theta_2^4\Theta_4^2\Theta_8^2\Theta_{10}^2.
\end{aligned}\ee 

\subsection{Modular expression for $v$}\label{monstroussolution}
Our next aim is to determine a modular expression for $v$ on this
elliptic locus. The first relation in (\ref{moalgrel}) implies
$\tau_{11}=\tau_{22}+2\mathbb Z+1$, while the second one implies $
\tau_{12}=\pm\frac 12 \tau_{11} +\mathbb Z$. We claim that these are
all the solutions. As in the case $v=0$, the PF solution
\eqref{taulargev} fixes these relations,  
\begin{equation}\label{Ttauu=0}
\tau_{11}=\tau_{22}+1, \quad  \tau_{12}=-\frac{\tau_{11}}{2}+1.
\end{equation}
In contrast to the locus $\CE_u$, these linear relations between the $\tau_{11}$, $\tau_{22}$ and
$\tau_{12}$ are exact on $\CE_v$. Using the first equation in \eqref{Ciu0}, we can solve for $v$,
\be\label{eq:v_exp_c1}
v=-\frac{i}{\sqrt{27}}\frac{(C_1-2)(C_1+1)(2C_1-1)}{C_1(C_1-1)}.
\ee
This can again  be written  as a rational function of Siegel theta functions,
\begin{equation}\label{u=0solutiontheta}
v= -\frac{i}{\sqrt{27}}\frac{(\Theta_8^2-2\Theta_{10}^2)(\Theta_8^2+\Theta_{10}^2)(2\Theta_8^2-\Theta_{10}^2)}{\Theta_8^2 \Theta_{10}^2(\Theta_8^2-\Theta_{10}^2)}.
\end{equation}
As a function of $\tau_-=\tau_{11}-\tau_{12}$, one finds ($q_- = e^{2\pi i \tau_-}$)
\be\label{vexpansion}
v= \frac{i}{2 \sqrt{27}} \left(\alpha\, q_-^{-\frac 16}-33\,\alpha^2\, q_-^{\frac 16}-153\,q_-^{\frac 12}-713\,\alpha\, q_-^{\frac 56}+\CO(q_-^{\frac 76})\right).
\ee
The expansion in terms of $\tau_+=\tau_{11}+\tau_{12}$ is very
similar. One can recognize these series as  
\begin{equation}\label{mfromcurve}
\begin{aligned}
 v&=\tfrac{i}{2\sqrt{27}}\,m(\tfrac{\tau_+}{2}),\\
 v&=\tfrac{i}{2\sqrt{27}}\,m(\tfrac{\tau_-}{6}+\tfrac23),
\end{aligned}
\end{equation}
where
\be \begin{aligned} \label{mo}
	m(\tau) &= \left(\frac{\eta\left(2\tau\right)}{\eta\left(6\tau\right)}\right)^6-27 \left(\frac{\eta\left(6\tau\right)}{\eta\left(2\tau\right)}\right)^6  \\ 
	&= q^{-1} - 33\, q - 153\, q^3 - 713\, q^5 - 2550\, q^7 - 7479\, q^{9}+\CO(q^{11}).
\end{aligned} \ee
The function $m$ is known  in the literature as the completely
replicable function of class 6a
\cite{Alexander:1992,ford1994,Ferenbaugh1993}.  Since the relations
between the $\tau_{IJ}$ are linear in this case, one can prove
the step from \eqref{u=0solutiontheta} to \eqref{mfromcurve}. The
details of the proof are given in Appendix \ref{proofmo}.  
The perturbative expansion \eqref{vexpansion} can also be verified from the Picard-Fuchs
solution by starting from Eq. \eqref{taulargev} and setting
$u=0$. Then, expand $q=e^{2\pi i(\tau_{11}(v)-\tau_{12}(v))}$ as a
series in $v$ and invert it to find \eqref{vexpansion}.

\subsection{The $\mathbb Z_3$ vacua}\label{sec:z3vacua}

Let us study the solution \eqref{mfromcurve} near the strong coupling
vacua. To this end, we eliminate the phases in (\ref{vexpansion}) by
substitution of $\tau\coloneqq\tau_-+1$ in (\ref{mfromcurve}). In the new variable $\tau$, the solution reads
\be \label{monstersol}  
v=-\tfrac{i}{ 2\sqrt{27}}\,  m\!\left(\tfrac {\tau}6\right).
\ee
It can be shown that the values of $\tau$ at the Argyres-Douglas (AD)
vacua $v_{\text{AD},1}= 1$ and $v_{\text{AD},2}=-1$ are ($\omega=e^{\pi i/6}$)
\begin{equation}\label{tildetauad}
\begin{split}
\tau_{\text{AD},1}&=-\frac{3}{2}+\frac{\sqrt{3}i}{2} =\sqrt 3\,
\omega^5,\\
\tau_{\text{AD},2}&=+\frac{3}{2}+\frac{\sqrt{3}i}{2} =\sqrt
3\,\omega  ,
\end{split}
\end{equation}
and the origin $(u,v)=(0,0)$ is located at  $ \tau_0=\sqrt3i$. This is rigorously proven in Appendix \ref{app:vtau=pm1} using
the properties of $m$. Note that these values lie in the interior of
the upper half-plane, rather than at the boundary. Section
\ref{sec:pflargev} will demonstrate that these values of $\tau$ also
match perfectly with the PF solutions.

The modular group of $v$ is closely related to the duality group of
the $SU(3)$ theory on this locus. It can be  shown that $v$ is a modular form for the principal
congruence subgroup $\Gamma(6)$, as defined in Appendix
\ref{appendA}. However, the fundamental domain of this group has
twelve cusps, and $v$ diverges at all of them. This suggests that we
found strongly coupled vacua in the region of the moduli space where
$v$ is large. But from the discriminant $\Delta_\Lambda|_{\CE_v}= v^2-1$
we expect the only singularities to be at  $v\in\{ 1,-1,\infty\}$.

 To resolve this problem, let us study the function
$m$ in more detail. It is a linear combination of eta quotients, whose modular
properties have been studied extensively
\cite{gordon1993,ono2004}. Applying  Theorem 1 in Appendix
\ref{appendA}, one finds that $m$ is a modular function for the Hecke
congruence subgroup $\Gamma_0(12)$. In addition, it satisfies the
following non-$SL(2,\mathbb{Z})$ transformations
\begin{subequations}\begin{eqnarray}\label{propm1}
m\left(\tau-\tfrac12\right)&=-m(\tau),\\ \label{propm2}
 m\left(-\tfrac{1}{12\tau}\right)&=-m(\tau),
\end{eqnarray}\end{subequations}
and further properties  are given in Appendix \ref{app:vtau=pm1}. The transformation \eqref{propm2} is also known as a \emph{Fricke involution}. Translating both equations to the argument  of $v$, we find that $v$ picks up a minus sign under both $T^{-3}$ and $F=\left(\begin{smallmatrix}0&-3\\1&0\end{smallmatrix}\right)$. Taking products, we find that $v$ is properly invariant under  $FT^{-3}=\left(\begin{smallmatrix}0&-3\\1&-3\end{smallmatrix}\right)$ and $T^{-6}$. Let us normalize the former to $X=\frac{1}{\sqrt3}\left(\begin{smallmatrix}0&-3\\1&-3\end{smallmatrix}\right)$, and denote the subgroup of $PSL(2,\mathbb R)$ generated by these two elements as 
\begin{equation}\label{G}
\Gamma_v=\left\langle X, T^{-6}\right\rangle.
\end{equation}
 This group is a proper subgroup of the modular group $\Gamma^0(6|2)+3$ of Atkin-Lehner type, in the notation of
 \cite{Ferenbaugh1993}. This Atkin-Lehner group extends the ordinary congruence
 subgroup $\Gamma^0(\tfrac 62)$ by elements in $PSL(2,\mathbb R)$. See Appendix \ref{appendA}
for the precise definition. If we allow for a non-trivial multiplier
system, the modular group associated
 with $m$ is $\Gamma^0(6|2)+3$ \cite{Ferenbaugh1993} . The latter
 contains for example $T^{-3}$, under which we have shown that $v$ is
 anti-invariant. We can write a similar set of
matrices as (\ref{uPtrafos}),
\begin{equation}\label{gengamma62}
M_1=\left(\begin{array}{cc}
 -3 & -3 \\
 1 & 0 \\
\end{array}
\right),\quad
M_2=\left(\begin{array}{cc}
 0 & 3 \\
 -1 & 3 \\
\end{array}
\right),\quad
M_\infty=\left(\begin{array}{cc}
 1& -6 \\
 0 & 1 \\
\end{array}
\right)=T^{-6}, 
\end{equation}
under which $ v\sim m(\tau/6)$ is  invariant. If we consider
their normalisation to unit determinant, $\Pi(M_j)\coloneqq |{\rm det}(M_j)|^{-1/2}\,M_j$, they lie in the group $\Gamma_v$ \eqref{G}, and  furthermore satisfy
\begin{equation}
\Pi(M_1)\Pi(M_2)=M_\infty.
\end{equation}
We will show in Section \ref{sec:strongmon} that
 these generators match with the monodromies.

\begin{figure}\centering
	\includegraphics[scale=0.85]{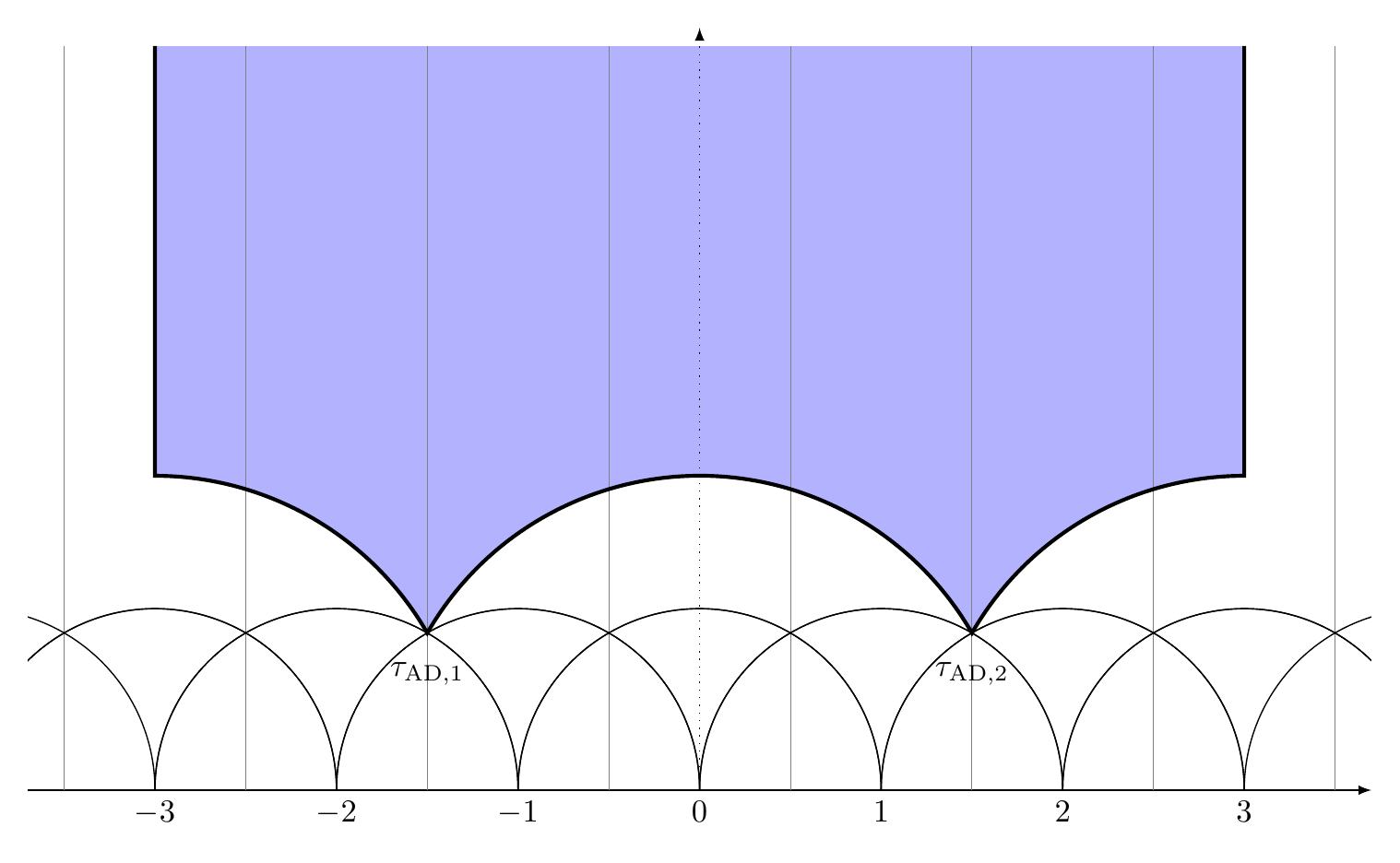}
	\caption{Fundamental domain $\Gamma_v\backslash \mathbb{H}$
          for the group $\Gamma_v$. The values of the special points
          are: $\tau_{{\rm AD},1}=\sqrt{3}\,\omega^5$ and $\tau_{{\rm AD},2}=\sqrt{3}\,\omega$.}\label{fig:fund_dom_v}
\end{figure}

A fundamental domain for $\Gamma_v$ can be drawn using the algorithm given in \cite{Ferenbaugh1993}, and it  is shown in Figure
\ref{fig:fund_dom_v}. The element $T^6$ contains the
domain to $|\text{Re }\tau|<3$. $X$ identifies the interior of the circle with radius
$\sqrt 3$ centered at 0, with a region inside the blue
domain in Figure \ref{fig:fund_dom_v}. Similarly, the  interior of the circles centered at $\pm
3$ is identified with a region of the blue domain.
We conclude, 
\begin{equation}
\Gamma_v \backslash \mathbb H=\left\{z\in\mathbb H\mid |\text{Re } z|<3\right\}\backslash \bigcup_{\ell=-1}^1 \overline D_{\sqrt3}(3\ell).
\end{equation}
where $\overline D_{r}(c)$ is the closed disc of radius $r$ and center $c$.

The Argyres-Douglas vacua $v=1$ and $v=-1$ correspond to the special
points $\tau_{\text{AD},j}$ (\ref{tildetauad}). They are stabilized by $M_1$  and $M_2$,
respectively. 
This makes the AD vacua elliptic points of $\Gamma_v$. They
are in fact expected to \emph{not} get mapped to cusps of $v$, since
their coupling matrix \eqref{ADcouplingT} lies inside the Siegel upper
half-space $\mathbb{H}_2$. This is a familiar property of superconformal points \cite{Argyres:1995jj,Argyres:2015ffa}. It is different from the $\mathbb Z_2$ points
where the coupling matrices \eqref{z2coupling} are located on the
boundary $\partial \mathbb{H}_2$ and therefore mapped to the real line
$\partial \mathbb H_1$.  The origin $\tau_0=\sqrt 3i $ is mapped under
$FT^{-3}$ to $\tau_0-3$, which is identified with $\tau_0$ since $v=0$
is a fixed point under $T^{-3}:v\mapsto -v$. The anti-invariance under
$T^{-3}$ is in fact directly derived from the $\mathbb Z_2$ symmetry
$\rho: v\mapsto e^{\pi i}v$  computed in \eqref{z2sym}. The large $v$
monodromy $\rho^2$ acts on $\tau$ as $T^{-6}$, under which $v$ is
invariant. The origin of the Fricke involution can therefore be
understood from the global structure on the $u=0$ plane. 

Similarly to Section \ref{umodexp}, we can express periods in terms of
modular forms. We have in terms of $\tau=\tau_{11}-\tau_{12}+1$,
\begin{equation} 
\label{pa1pu}
\frac{\partial a_1}{\partial u}(\tau)=\frac{\partial a_2}{\partial u}(\tau)=\frac{\sqrt[3]{2}\omega}{\sqrt3}\eta(\tfrac\tau3)\eta(\tau).
\end{equation}

The discussion is similar for the parameter
$\tau_+=\tau_{11}+\tau_{12}$. If we introduce here $\tau=\tau_+-1$,
$v$ equals $-\frac{i}{2\sqrt{27}}\,m(\tau/2)$, which is again invariant
under $\Gamma(6)$. It is multiplied by a sign under $T$ as well as under the
Fricke involution $\tilde
F=\left(\begin{smallmatrix}0& -1\\3&0\end{smallmatrix}\right)$. This
means that it is invariant under $T^2$ together with the involution
$\tilde X :=\tilde F
T^{-1}=\left(\begin{smallmatrix}0&-1\\3&-3\end{smallmatrix}\right)$,
which again generate an Atkin-Lehner type group. The fundamental
domain of this group equals that in Figure \ref{fig:fund_dom_v}, but
with all points divided by $3$.

\section{Elliptic curves for the two loci}
\label{EllipticLoci}
It is natural to expect that the complexified couplings $\tau_\pm$ for
both loci $\CE_u$ and $\CE_v$ have an interpretation as complex
structures of elliptic curves. Moreover, these elliptic curves are
expected to be related to the geometry of the genus two Seiberg-Witten
curve (\ref{g2curve}). We will make these expectations precise in this section.
 
Recall that the moduli space $\CM_2$ of genus two curves is complex
three-dimensional.  The moduli space $\CM_2$ contains two-dimensional
loci $\CL_2\subset \CM_2$, for which the genus two curves can be
mapped to genus one with a map of degree 2 \cite{igusa1960}. 
 The map can be lifted to a map of the Jacobians of the curves. The Jacobian of
the genus two curve is a four-torus, while the Jacobian of a genus one
curve is a two-torus. For the curves contained in $\CL_2$, there is a
degree two map from the genus two Jacobian to the genus one
Jacobian. The Jacobian of a curve in $\CL_2$ factors, $T^4 \equiv
T^2\times T^2$, which demonstrates that for a generic curve in $\CL_2$, there
are two distinct maps $\varphi_j: \Sigma_2 \to \Sigma_{1,j}$, $j=1,2$ to two elliptic
curves $\Sigma_{1,j}$. We will see in this section that these elliptic
curves $\Sigma_{1,j}$ have precisely the complex structures $\tau_\pm$ introduced above.  
 
The locus $\CL_2$ can be
characterized  as the zero locus 
of a weight 30 polynomial in the genus two Igusa invariants $J_2$, $J_4$, $J_6$, 
$J_{10}$ \cite[Theorem 3]{shaska2001}.\footnote{We found two small typos in
  Theorem 3 of \cite{shaska2001}: For the   coefficient of $J_{10}^2J_4^2J_2$, we find
  +507384000; and $-6912$ for the coefficient of $J_4^3J_6^3$.} See \cite{Klemm:2015iya} for a definition of the Igusa invariants. 
Additionally, the $SU(3)$ vacuum moduli space also corresponds to a
two-dimensional locus $\CU$ in $\CM_2$. On $\CU$ the weight
30 polynomial factors in three terms, such that $\CU$ and $\CL_2$
 intersect in three one-dimensional loci:       
\begin{equation}\label{l2locus}   
\begin{aligned}   
\CE_1=\CE_u:\qquad &v=0,\\ 
\CE_2=\CE_v:\qquad &u=0,\\
\CE_3\qquad \,\,:\qquad &784 u^9-24 u^6 \left(297 v^2+553\right)-15 u^3
\left(729 v^4+5454 v^2-4775\right)\\
& +8 
   \left(27 v^2-25\right)^3=0.
\end{aligned}
\end{equation}  
Not suprisingly, we have seen the first two of these loci before.
The latter is a cubic equation in $v^2$ as well as in $u^3$, which 
does not reduce further. It does not include special points of the
$SU(3)$ theory. For $v=0$, the equation reduces to the points $u^3=8$
and $u^3=\tfrac{125}{28}$ in the $u$-plane,
and for $u=0$ it intersects in $v^2=\tfrac{25}{27}$ on the $v$-plane. 

The locus $\CL_2$ can also be characterized in terms of Rosenhain
invariants of the curve \cite[Equation (18)]{shaska2001}. By plugging in the
cross-ratios we can  check that the $SU(3)$ Seiberg-Witten curve is not in $\CL_2$ for
generic $u,v$. For $v=0$ we rediscover the first algebraic relation
\eqref{algrelv=0}, while for $u=0$ we find both relations
\eqref{eq:C_rel_v_large}. This arises from an additional symmetry of
the $u=0$ curve, which we will comment on below. 

\subsection{Elliptic curves for locus $\CE_u$}
In this subsection we will establish two elliptic curves corresponding to the two
modular parameters $\tau_\pm$ in Section \ref{sec:Rosenhain}. 
The curves described by the locus $\CL_2$ can be written in the form \cite{shaska2001}
\begin{equation}
\label{genJacobi}
Y^2=X^6-s_1X^4+s_2X^2-1,  
\end{equation} 
with $s_1$ and $s_2$ complex coordinates for $\CL_2$. This family of
curves is left invariant by a non-trivial automorphism group, which
contains the Klein four-group $V_4$  \cite{shaska2003}. 
Namely, the curve \eqref{genJacobi} is left invariant by 
$(X,Y)\mapsto (-X,Y)$ and $(X,Y)\mapsto (X,-Y)$, which generate the
dihedral group $D_4\cong V_4\cong \mathbb{Z}_2\times \mathbb{Z}_2$. We
interpret this group as the symmetry group of BPS/anti-BPS spectrum,
and more precisely the central charges of
the W-bosons $Z_j$ (\ref{ccharges}) and their charge conjugates. For $v=0$,
Eq. (\ref{PFEu}) shows that $a_1=a_2=a$, such that $Z_1=Z_2=a$, and
$Z_3=2a$. One $\mathbb{Z}_2 \subset D_4$ corresponds to the charge conjugation
symmetry, while the other $\mathbb{Z}_2$ corresponds to the $a_1
\leftrightarrow a_2$ symmetry on $\CE_u$. Note that the automorphism group of a generic
genus two curve is $\mathbb{Z}_2$, which is consistent with the charge
conjugation symmetry for arbitrary $(u,v)$.

For $v=0$, the Seiberg-Witten curve $Y^2=(X^3-uX)^2-1$ is of the
form (\ref{genJacobi}), with $s_1=2u$ and $s_2=u^2$.
The degree two map to an elliptic curve is
\be
(x,y)=(X^2,Y), 
\ee
which maps the algebraic equation (\ref{genJacobi}) to
\begin{equation}
\label{v0Curve1}
y^2=x(x-u)^2-1.
\end{equation}
We can determine $u$ in terms of
the complex structure $\tau$ of the curve from the $j$-invariant,
$j=256u^6/(4u^3-27)$. This gives
\begin{equation}\label{uplus}
4u(\tau)=q^{-1/3}+104\,q^{2/3}-7396\,q^{5/3}+\CO(q^{8/3}).
\end{equation}
We immediately recognize this function as the function $u_+$
(\ref{uuPlus}), which was obtained from the Picard-Fuchs solution for the modular parameter
$\tau_+=\tau_{11}+\tau_{12}$. The curve (\ref{v0Curve1}) is exactly the Seiberg-Witten curve for the
$SU(2)$ theory with one massless hypermultiplet in the fundamental
representation and scales related by $\Lambda_{SU(2)}=2\Lambda_{SU(3)}$ \cite{Seiberg:1994aj}, which clarifies the observation
in Section \ref{SubSecu+}.

The curve that corresponds to $\tau_-=\tau_{11}-\tau_{12}$ can be
constructed as follows. On the curve \eqref{genJacobi}, the transformation $(X,Y)\mapsto (\tfrac 1X,\tfrac{iY}{X^3})$
interchanges $s_1$ and $s_2$. Interchanging those coefficients,
$s_1=u^2$ and $s_2=2u$, and setting again $(x,y)=(X^2,Y)$, we obtain  
\begin{equation}\label{uminuscurvenorm}
y^2=x(x^2-u^2x+2u)-1.
\end{equation}
One finds $j=256u^3(u^3-6)^3/(4u^3-27)$, which reproduces the solution
$u_-$ for the $\Gamma^0(9)$ curve \eqref{largeuv=0}. Note that the
equation for $j$ shows that $u_-$ is the root of a degree 12
polynomial, which matches with the number of copies of $\CF$ in Figure
\ref{fig:fund_dom_gamma09}. Another way to obtain this curve is to set
$x=X^2$ and $y=XY$, from which one gets a quartic curve with the same
$j$-invariant. 

We have thus demonstrated that the two natural choices
$\tau_\pm$ of the modular parameter indeed correspond to the complex
structures of two elliptic curves covering the hyperelliptic
curve. The physical $u$ is given in terms of two different functions
$u_\pm:\mathbb{H}\to \mathbb{C}$ with arguments
$\tau_\pm$.

\subsection{Elliptic curves for locus $\CE_v$}
The Seiberg-Witten curve $Y^2=(X^3-v)^2-1$ for $u=0$ is not in form
\eqref{genJacobi} for a curve of $\CL_2$. However, the discussion
around \eqref{l2locus} suggests that it can be written in this
form. We can achieve this by comparing the invariants of the $u=0$
hyperelliptic curve and \eqref{genJacobi}, and solving for $s_1,
s_2$. Just as two elliptic curves are isomorphic if and only if their
$j$-functions are equal, higher genus curves are isomorphic if and only if their
absolute invariants are equal \cite{shaska2012genus,igusa1967,igusa1962}. On $\CL_2$,
there are only two independent invariants. Comparing the absolute invariants
of \eqref{genJacobi}  and the $SU(3)$ curve for $u=0$, we arrive at
\begin{equation}\label{abv}
s_1s_2=9 \left(25-24 v^2\right), \qquad s_1^3+s_2^3 = 54 \left(216 v^4-340 v^2+125\right).
\end{equation}
These combinations of $s_1$ and $s_2$ are known as the ``dihedral''
constants, since they are left invariant by the action of the dihedral group $D_6$ on (\ref{genJacobi}).
To solve the two equations in (\ref{abv}), let us denote 
\begin{equation}
\mathcal Q^\pm(v) = 27\left(216 v^4-340 v^2+125\pm 8 v \left(27 v^2-25\right)\sqrt{ v^2-1}\right).
\end{equation}
Then, one of the six solutions is given by
\begin{equation}
\begin{aligned}\label{s12pm}
s_1^\pm =  \sqrt[3]{\mathcal Q^\mp(v)}, \quad s_2^\pm =9 \frac{25-24v^2}{\sqrt[3]{\mathcal Q^\mp(v)}}.
\end{aligned}
\end{equation}
In order to get an elliptic curve, we again take the map $(x,y)=(X^2, Y)$. This gives us the two curves 
\begin{equation}
\label{s12pm}
y^2=x^3-s_1^\pm x^2+s_2^\pm x-1
\end{equation}
with  $j$-function
\begin{equation}\label{jpm}
j^\pm = -432 \left(1458 v^6-2673 v^4+1340 v^2-125\mp 2 v\left(729 v^4-972 v^2+275\right)\sqrt{ v^2-1 }\right)
\end{equation}
 and  discriminant $\Delta=v^2-1$. 
 By inverting \eqref{jpm}, the resulting function $v$ matches precisely with \eqref{mfromcurve} in Section \ref{monstroussolution}.  Note that $j^\pm$
 vanish at the AD points  $v=\pm 1$ and the curve \eqref{s12pm}
 becomes a cusp $y^2=x^3$. This implies that the AD points are
 elliptic fixed points and are in the $SL(2,\mathbb Z)$ orbit of
 $\alpha$, which is easy to check from \eqref{tildetauad}: We have
 that $\tau_{\text{AD},1}=\alpha-1$ and
 $\tau_{\text{AD},2}=\alpha+2$. See also Figure
 \ref{fig:fund_dom_v}. They do however not fall into the (classical)
 Kodaira classification of singular fibers, since the Weierstra{\ss}
 invariants of \eqref{s12pm} are not polynomials in $v$ and their order
 of vanishing is half-integer rather than integer.

The curve  $Y^2=X^6-2vX^3+v^2-1$ for $u=0$ has enhanced symmetry
compared to the Klein four-group for (\ref{genJacobi}). Since $v^2-1$ is the discriminant, we can divide and rescale $X$ to find 
\begin{equation}
Y^2=X^6-\frac{2v}{\sqrt{v^2-1}}X^3+1.
\end{equation}
 It is easy to show that any curve of the form $Y^2=X^6-a X^3+1$ is
 invariant under $(X,Y)\mapsto (\tfrac 1X,\tfrac{Y}{X^3})$ and
 $(X,Y)\mapsto (\alpha X,-Y)$, where again $\alpha=e^{2\pi
   i/3}$. These order 2 and 6 elements generate the dihedral group
 $D_{12}$. Similarly to the enhanced automorphism group for $\CE_u$,
 we interpret this group as a symmetry group of the BPS/anti-BPS spectrum. From
 Appendix \ref{pfderivation_vlarge}, we know that $a_2=-\alpha
 a_1$ on $\CE_v$. The central charges $Z_j$ (\ref{ccharges}) of the W-bosons,
 together with their charge conjugates, span therefore a regular
 6-gon, whose symmetry group is $D_{12}$.

Hyperelliptic curves $C\in\CL_2$ with $\text{Aut}(C)\cong
 D_{12}$ satisfy an additional constraint, it is given by the zero
 locus of a weight $20$ polynomial in the  Igusa invariants
 \cite[Eq. (24)]{Shaska:2009}. Moreover, the elliptic subcovers
 of hyperelliptic curves with $\text{Aut}(C)\cong D_{12}$ are
 isogenous \cite{shaska2001}. We can check explicitly that the $u=0$
 curve is of this form. This explains why the elliptic curves for the
 two complex structures produce a single modular function
 \eqref{mfromcurve}, rather than the two independent functions $u_\pm$ for $\CE_u$. On $\CE_u$ the  first algebraic relation in
\eqref{algrelv=0} holds and places the curve in $\CL_2$. On $\CE_v$
both relations \eqref{eq:C_rel_v_large} hold, where the first one
projects into $\CL_2$ and the second one gives the augmented $D_{12}$
symmetry. This is consistent with the argument of Section
\ref{umodexp} that the maps $\varphi_j$ should exist as long as
$\text{Im}(\tau_{11})=\text{Im}(\tau_{22})$, such that it is possible
to define $\tau_\pm=\tau_{11}\pm\tau_{12}\in\mathbb H$. The first
relations in both \eqref{algrelv=0} and \eqref{eq:C_rel_v_large} are
equivalent to this condition.

\section{Monodromies}\label{sec:pfsolution}
We study the weak and strong coupling monodromies in this section. In this way,
we are able to derive the modular groups of the order parameters in
Section \ref{sec:Rosenhain}, which parametrize the elliptic loci. As before, we are interested in studying the two patches of the moduli space where one of the
parameters $u$ and $v$ is large compared to the other.

\subsection{Weak coupling monodromies}
The spontaneously broken global $\BZ_3$ and $\BZ_2$ symmetries are generated by $\sigma: u\mapsto \alpha u$ and $\rho: v\mapsto e^{\pi i}v$, respectively. Using the explicit Picard-Fuchs solutions \eqref{weakcouplingperiods} and \eqref{ajlargev}, we can determine how these symmetries act on the periods in the weak coupling region of the Coulomb branch.

\subsubsection*{Weak coupling in locus $\CE_u$}\label{sec:largeusym}
In the large $u$ regime we are interested in the action of $\sigma$ on the PF solutions in \eqref{weakcouplingperiods}. We can readily determine that it acts on the periods as the matrix
\be\label{sigma}
\sigma_u=\alpha^2\CP\left( \begin{array}{cccc} 0 & 1 & 1 & -2 \\ 1 & 0 & -2 & 1 \\ 0 & 0 & 0 & 1 \\ 0 & 0 & 1 & 0 \end{array} \right),
\ee
where the subscript $u$ indicates that the base point is at large $u$, and $\CP=\left(\begin{smallmatrix}-\mathbbm 1&0\\0&-\mathbbm 1\end{smallmatrix}\right)$ is the central element of $Sp(4,\BZ)$.  The matrix $\sigma_u$ conjugates the semi-classical monodromies \eqref{semiclassicalmono} to each other,
\be\label{monod_conj_sigma}
\begin{split}
\sigma^{-1}_u \CM^{(r_1)}\sigma_u &= \CM^{(r_2)},\\
\sigma^{-1}_u \CM^{(r_2)}\sigma_u &= \CM^{(r_1)},\\
\sigma^{-1}_u \CM^{(r_3)}\sigma_u&=\CM^{(r_1)}\CM^{(r_2)}(\CM^{(r_1)})^{-1}.
\end{split}
\ee
It holds that $\bar \sigma_u=\alpha\sigma_u\in Sp(4,\mathbb Z)$. We
introduce moreover the translation of $\tau_{IJ}$ at weak coupling,
\be
\label{defTw}
\CT_{w,u}= \left(
\begin{array}{cccc}
 0 & 1 & -1 & 2 \\
 1 & 0 & 2 & -1 \\
 0 & 0 & 0 & 1 \\
 0 & 0 & 1 & 0 \\
\end{array}
\right) = \alpha^2\CP \sigma_u^{-1}\in Sp(4,\mathbb{Z}),
\ee
which maps
\begin{equation}
\CT_{w,u}:\quad \left(
\begin{array}{cc}
 \tau_{11} & \tau_{12} \\
 \tau_{12} & \tau_{22} \\
\end{array}
\right)\mapsto \left(
\begin{array}{cc}
 \tau _{22}+2 & \tau _{12}-1 \\
 \tau _{12}-1 & \tau _{11}+2 \\
\end{array}
\right).
\end{equation}

Using \eqref{casimirs}, one checks that $\sigma_u$ maps  $u\mapsto \alpha u$, while $v\mapsto v$ is left invariant.
Moreover, $\sigma^3_u:u\mapsto e^{2\pi i}u$ leaves $u$ invariant, but acts as a monodromy on
the periods,
\be\label{sigma3}
\sigma^3_u=\CP\CT_{w,u}^{-3}=\CM^{(r_2)}\CM^{(r_1)}\CM^{(r_2)}=\left(
\begin{array}{cccc}
 0 & -1 & -3 & 6 \\
 -1 & 0 & 6 & -3 \\
 0 & 0 & 0 & -1 \\
 0 & 0 & -1 & 0 \\
\end{array} 
\right). 
\ee
This corresponds to the monodromy around  $u= \infty$ by construction.
In a similar way, we can determine the action of the $\BZ_2$ symmetry generated by $\rho:v\mapsto e^{\pi i}v$. Here, one finds the matrix representation
\begin{equation}\label{eq:z2_gen_large_u}
\rho_u = \begin{pmatrix}0&1&0&0\\1&0&0&0\\0&0&0&1\\0&0&1&0\end{pmatrix}\in Sp(4,\BZ).
\end{equation}
This matrix conjugates the semi-classical monodromies analogous to \eqref{monod_conj_sigma}, with $\sigma_u$
replaced by $\rho_u$. The large $u$ monodromy for $v$ is trivial,
$\rho_u^2=\mathbbm{1}$. We will see later that $\sigma_u$ and $\rho_u$
have a natural action on the charge vectors of the dyons that become
massless at the various strongly coupled singular vacua.  
The full $\mathbb Z_6$ symmetry can now be represented as 
\begin{equation}\label{quantummonodromy}
\rho_u^{-1}\,\CT_{w,u}=\CT_{\text q},
\end{equation}
with
$\CT_{\text q}=\left(\begin{smallmatrix}\mathbbm{1}&C\\0&\mathbbm{1}\end{smallmatrix}\right)$,
where  $C=\left(\begin{smallmatrix}2&-1\\-1&2\end{smallmatrix}\right)$
is the Cartan matrix of $SU(3)$. This represents the quantum monodromy
corresponding to a rotation of the  scale
$\Lambda^6\to e^{2 \pi i}\Lambda^6$ \cite{Klemm:1994qj}.

\subsubsection*{Weak coupling in locus $\CE_v$}\label{sec:pflargev}
We now turn to the patch with $v$ large and perform the analogous analysis as in the above. The action of $\sigma: u\mapsto \alpha u$ on the solution (\ref{PFP2}--\ref{ajlargev}) can be represented by the matrix
\begin{equation}\label{tildesigmav}
 \sigma_v =\alpha^2 \left(
\begin{array}{cccc}
 -1 & -1 & 2 & -1 \\
 1 & 0 & -1 & -1 \\
 0 & 0 & 0 & -1 \\
 0 & 0 & 1 & -1 \\
\end{array}
\right),
\end{equation}
where now the subscript $v$ indicates that we are in the large $v$ regime. It satisfies $ \sigma_v^3=\mathbbm 1$ and the large $v$ rotation is therefore a trivial monodromy. 
On this patch, the generator of the $\BZ_2$ symmetry $\rho_v: v\mapsto e^{\pi i} v$ is more interesting. Here, instead of \eqref{eq:z2_gen_large_u}, we now find
\begin{equation}\label{z2sym}
\rho_v=\left(
\begin{array}{cccc}
 0 & -1 & 1 & 1 \\
 1 & 1 & -2 & -2 \\
 0 & 0 & 1 & -1 \\
 0 & 0 & 1 & 0 \\
\end{array}
\right).
\end{equation}
Since $ \rho_v^2\neq \mathbbm 1$, $v\mapsto e^{2\pi i} v$ acts on the
periods as a monodromy, while  leaving $v$ invariant. The full $\BZ_6$ symmetry is again given by $\CP\alpha^2\rho_v^{-1}\sigma_v^{-1}= \CT_{\text q}$, as in \eqref{quantummonodromy}.

\subsection{Strong coupling monodromies}\label{sec:strongmon}

Analytically continuing the PF solution \eqref{weakcouplingperiods} to strong coupling, we can compute the periods near the singularities. 
At the $\mathbb Z_2$ point $(\underline
u,v)=(1,0)$, the coupling matrix can be computed explicitly and we can then use $\sigma_u$ to rotate to the other $\mathbb
Z_2$ points $\underline u= \alpha,\alpha^2$ by  means of the action
\eqref{symplectictransf}. The coupling matrices at these points evaluate to   
\be\label{z2coupling}
\begin{aligned} 
	\Omega(1,0)= \begin{pmatrix}0&0\\0&0\end{pmatrix}, \quad
	\Omega(\alpha,0)=  \begin{pmatrix} -2&1\\1&-2\end{pmatrix}, \quad 
	\Omega(\alpha^2,0)=  \begin{pmatrix} 2&-1\\-1&2\end{pmatrix}.
\end{aligned}
\ee
The above matrices lie on the boundary $\partial \mathbb{H}_2$ of the Siegel
upper half-plane. The relations among the entries are consistent with
the results from Section \ref{sec:Rosenhain}.

The coupling matrices at the $\mathbb{Z}_3$ (AD) points $(\underline u,v ) =(0,\pm1 )$ are
\begin{equation}\label{ADcouplingT}
	\Omega(0,1)=\begin{pmatrix} -1+\frac{i}{\sqrt3}&\frac{9-\sqrt3 i}{6}\\ \frac{9-\sqrt3 i}{6}&-2+\frac{i}{\sqrt3}\end{pmatrix}, \quad \Omega(0,-1)=\begin{pmatrix}1+\frac{i}{\sqrt3}&\frac{3-\sqrt 3i}{6}\\ \frac{3-\sqrt 3i}{6}&\frac{i}{\sqrt3}\end{pmatrix}.
\end{equation}
They lie in the interior of the Siegel upper half-space $\mathbb{H}_2$.

To determine the monodromies around these singularities, 
we recall the formula from \cite{Klemm:1994qj, Klemm:1995wp}.
It gives the monodromy matrix in terms of the charge vector $\gamma$ of the BPS
state with vanishing mass. The  charge vector is a left eigenvector
with unit eigenvalue. The monodromy $\CM_\gamma$ reads  
\begin{equation}\label{strongmon}
\CM_\gamma=\begin{pmatrix}\mathbbm 1+ n\otimes m& n\otimes n\\-m\otimes m& \mathbbm 1-m\otimes n\end{pmatrix}
\end{equation}
for $\gamma=(m,n)$ with $m=(m_1,m_2)$ and $n=(n_1,n_2)$ the magnetic and electric charge vectors. In locus $\CE_u$  we have three singular points where two mutually local dyons become massless, respectively, while in locus $\CE_v$ three mutually non-local dyons become massless at each of the two singular points.

\subsubsection*{Strong coupling in locus $\CE_u$}
To calculate the monodromies using \eqref{strongmon}, we need to first
choose a symplectic basis for the homology cycles. In locus $\CE_u$ we
choose it such that two monopoles $\gamma_1=(1,0,0,0)$ and
$\gamma_2=(0,1,0,0)$ become massless at $(\underline u,v)=(1,0)$. For
gauge group $SU(N)$ this choice is always possible
\cite{Klemm:1995wp}. In this subsection, we will consider monodromies
in locus $\CE_u$, keeping $v=0$ fixed. Restricting to this locus,
a monodromy circles a point rather than a line. We denote the
monodromy around the point $(\underline u,0)$ in $\CE_u$ by $\CM_{(\underline u,0)}$. The
charges of the dyons that become massless at the singular points
$(\underline u,v)=(\alpha,0)$ and $(\underline u,v)=(\alpha^2,0)$  are
obtained by acting on the periods with $\sigma_u$ and $\sigma_u^{-1}$
from the left, it turns out that this corresponds to acting on the
charges $\gamma_{1,2}$ from the right with $-\CT_{w,u}$ and its inverse. We
find   
\begin{alignat}{3}\nonumber
\gamma_1&=(1,0,0,0),\qquad &&\gamma_2 &&=\, (0,1,0,0), \\  
 \gamma_3&=-\gamma_1\CT_{w,u}= (0,-1,1,-2), \qquad&&\gamma_4 &&=-\gamma_2 \CT_{w,u}= \,(-1,0,-2,1), \\ 
\gamma_5 &=-\gamma_1 \CT_{w,u}^{-1}= (0,-1,-1,2),\qquad &&\gamma_6&&=-\gamma_2\CT_{w,u}^{-1}= \,(-1,0,2,-1),\nonumber
\end{alignat}
where each row corresponds to the charges of the mutually local states becoming massless at the respective points.  

We will first derive the four-dimensional monodromy matrices, and then
determine their action on the effective couplings constants $\tau_\pm$.
The monodromy around $(\underline u,v)=(1,0)$ can be computed
from the PF solution, it is 
\begin{equation}\label{m10}
\CM_{(1,0)}=\CM_{\gamma_1}\CM_{\gamma_2}=\left(
\begin{array}{cccc}
 1 & 0 & 0 & 0 \\
 0 & 1 & 0 & 0 \\
 -1 & 0 & 1 & 0 \\
 0 & -1 & 0 & 1 \\
\end{array}
\right)
\end{equation}
and agrees with the product of the monodromies \eqref{strongmon} of
the singular lines associated with the massless states of charges
$\gamma_1$ and $\gamma_2$. This monodromy can be written as a
``trajectory'' in the space of coupling constants as
\be
\CM_{(1,0)}=\mathcal{S} \CT_{s,u}\CS^{-1},
\ee
where $\CS$ is the symplectic inversion and $\CT_{s,u}$ is the translation at
strong-coupling,
\be\label{csct}
\CS=\left(
\begin{array}{cccc}
 0 & 0 & -1 & 0 \\
 0 & 0 & 0 & -1 \\
 1 & 0 & 0 & 0 \\
 0 & 1 & 0 & 0 \\
\end{array}
\right),\qquad \CT_{s,u}=\left(
\begin{array}{cccc}
 1 & 0 & 1 & 0 \\
 0 & 1 & 0 & 1 \\
 0 & 0 & 1 & 0 \\
 0 & 0 & 0 & 1 \\
\end{array}
\right).
\ee

The monodromies around $\underline u=\alpha$ and $\underline
u=\alpha^2$ can be obtained from the charges of the corresponding
states that become massless at the different points. Alternatively,
we can write them as conjugations of $\CT_{s,u}$. We find 
\begin{equation}\label{strongcouplmonU}
\begin{aligned}
\CM_{(\alpha,0)}=&\CM_{\gamma_3}\CM_{\gamma_4}=(\CT_{w,u}^{-1}\CS)\CT_{s,u}(\CT_{w,u}^{-1}\CS)^{-1}=\left(
\begin{array}{cccc}
3 & -1 & 5 & -4 \\
-1 & 3 & -4 & 5 \\
-1 & 0 & -1 & 1 \\
0 & -1 & 1 & -1 \\
\end{array}
\right), \\
\CM_{(\alpha^2,0)}=&\CM_{\gamma_5}\CM_{\gamma_6}=(\CT_{w,u}\CS)\CT_{s,u}(\CT_{w,u}\CS)^{-1}=
\left(
\begin{array}{cccc}
-1 & 1 & 5 & -4 \\
1 & -1 & -4 & 5 \\
-1 & 0 & 3 & -1 \\
0 & -1 & -1 & 3 \\
\end{array}
\right).
\end{aligned}
\end{equation}
They satisfy the consistency condition
\begin{equation}\label{m1m2m3m4m5m6}
\CP\CT_{w,u}^{-3}=\CM_\infty=\CM_{(\alpha,0)}\CM_{(1,0)}\CM_{(\alpha^2,0)}=\left(
\begin{array}{cccc}
0 & -1 & -3 & 6 \\
-1 & 0 & 6 & -3 \\ 
0 & 0 & 0 & -1 \\
0 & 0 & -1 & 0 \\
\end{array}
\right).
\end{equation}
Due to the singularity structure, the matrices
\eqref{m10}-\eqref{m1m2m3m4m5m6} are all the monodromies in the region
where $v$ is small. They all lie in $Sp(4,\mathbb Z)$, since
\eqref{strongmon} do.

For the elliptic locus $v=0$, we analyzed the couplings
$\tau_\pm=\tau_{11}\pm\tau_{12}$ in Section \ref{sec:Rosenhain}. We
will study here the action of $\CM_\infty$ and $\CM_{(\alpha^j,0)}$ on
$\tau_\pm$. We will find for $\tau_-$ that the action of the
monodromies generate a proper congruence subgroup $\Gamma^0(9)\subset
SL(2,\mathbb{Z})$. Therefore, the action of $\CT_{w,u}$ and
$\CT_{s,u}$ can be represented in terms of the same two-dimensional matrix
$T=\left( \begin{smallmatrix} 1& 1\\0 & 1 \end{smallmatrix}\right)$. The weak coupling shift $\CT_{w,u}$ corresponds
to the two-dimensional matrix $T^3$ for $\tau_-$, while the strong
coupling shift $\CT_{s,u}$ corresponds to $T$. Moreover, the four-dimensional symplectic $\CS$ reduces to the two-dimensional modular inversion $S$. Since $\tau_{11}=\tau_{22}$ on $\CE_u$, it is easy to show that the four-dimensional monodromies reduce to the matrices  
\begin{equation}\begin{aligned}
\CM_{(\infty,0)} \mapsto M^-_{(\infty,0)}&=T^{-9}= \left(
\begin{array}{cc}
 1 & -9 \\
 0 & 1 \\
\end{array}
\right),\\
\CM_{(1,0)}\mapsto M^-_{(1,0)}&=STS^{-1}= \left(
\begin{array}{cc}
 1 & 0 \\
 -1 & 1 \\
\end{array}
\right), \\
\CM_{(\alpha,0)}\mapsto M^-_{(\alpha,0)}&=(T^{-3}S)T(T^{-3}S)^{-1}=\left(
\begin{array}{cc}
 4 & 9 \\
 -1 & -2 \\
\end{array}
\right), \\
\CM_{(\alpha^2,0)}\mapsto  M^-_{(\alpha^2,0)}&=(T^3S)T(T^3S)^{-1} =\left(
\begin{array}{cc}
- 2 & 9 \\
 -1 & 4 \\
\end{array}
\right),
\end{aligned}\end{equation}
for $\tau_-$. They all lie in $\Gamma^0(9)$ and do in fact generate $\Gamma^0(9)$, and furthermore satisfy
\begin{equation}
M^-_{(\alpha,0)}M^-_{(1,0)}M^-_{(\alpha^2,0)}=M^-_{(\infty,0)}.
\end{equation}
Note there is here no sign between $M^-_{(\infty,0)}$ and $T^{-9}$. Of
course, this sign is irrelevant for the action on $\tau_-$. A good consistency check is that these monodromies fix the $\tau_-$ at the cusps $\tau_-=\{-3,0,3\}$.

The weak coupling shift $\CT_{w,u}$ corresponds
to the two-dimensional matrix $T_{w,u}$ for $\tau_+$, while the strong
coupling shift is $T_{s,u}$. For the parameter $\tau_+$, the monodromies reduce to
\begin{equation}  
\label{PlusMs}
\begin{aligned}
M^+_{(\infty,0)}&=PT_{w,u}^{-3 },\\
M^+_{(1,0)}&=ST_{s,u}S^{-1},\\ 
M^+_{(\alpha,0)}&=(T_{w,u}^{-1}S)T_{s,u}(T_{w,u}^{-1}S)^{-1},\\
M^+_{(\alpha^2,0)}&=(T_{w,u}S)T_{s,u}(T_{w,u}S)^{-1},\\
\end{aligned}
\end{equation}
which satisfy
\begin{equation}
M^+_{(\alpha,0)}M^+_{(1,0)}M^+_{(\alpha^2,0)}=M^+_{(\infty,0)}.
\end{equation}
This precisely reduces to the group $\Gamma_{u_+}$ (\ref{Gammau+}), which leaves the
function $u_+$ invariant. As discussed in Section \ref{SubSecu+}, these monodromies do not generate a congruence subgroup of
$SL(2,\mathbb{Z})$ if we identify $T_{w,u}$ and $T_{s,u}$ with $T$.

\subsubsection*{Strong coupling in locus $\CE_v$}
We can perform a similar analysis in the region where $v$ is large and $u$ small. At each of the two singular points we find that three mutually non-local states become massless. The corresponding charges are
\begin{alignat}{3}\nonumber\label{largevcharges}
\nu_1&=(1,1,0,0),\qquad && \nu_2 &&=\, (0,1,0,0), \\  
\nu_3 &= \nu_1 \bar\sigma_v^{-1}= (-1,0,-1,2), \qquad 
&& \nu_4 &&= \nu_2\bar\sigma_v^{-1} = \,(-1,-1,1,1), \\ 
\nu_5&= \nu_1 \bar\sigma_v = (0,-1,1,-2),\qquad && \nu_6&&=\nu_2 \bar\sigma_v= \,(1,0,-1,-1),\nonumber
\end{alignat}
where the left column represents the states that becomes massless at
$(u,v)=(0,1)$ and the second column the ones for $(u,v)=(0,-1)$, and
$\bar\sigma_v=\alpha\sigma_v\in Sp(4,\mathbb{Z})$.

The monodromy around $v=\infty$ is given by 
\begin{equation}\label{M0inf}
\CM_{(0,\infty)}= \rho_v^2=\left(
\begin{array}{cccc}
 -1 & -1 & 4 & 1  \\
 1 & 0 & -5 & 1 \\
 0 & 0 & 0 & -1 \\
 0 & 0 & 1 & -1 \\
\end{array}
\right).
\end{equation}
For $u=0$, the monodromy around the AD point $(u,v)=(0,1)$ can be calculated from the Picard-Fuchs solution,
\begin{equation}\label{M01}
\CM_{(0,1)}=\left(
\begin{array}{cccc}
 2 & 0 & 1 & -2 \\
 -2 & 1 & -2 & 4 \\
 -1 & -1 &1 & 0 \\
 0 & -1 & 1 & -1 \\
\end{array}
\right)=\CM_{\nu_1}\CM_{\nu_3}=\CM_{\nu_3}\CM_{\nu_5}.
\end{equation}
The remaining monodromy is fixed by the global consistency $\CM_{(0,\infty)}=\CM_{(0,1)}\CM_{(0,-1)}$.
This gives us
\begin{equation}
\CM_{(0,-1)}=\left(
\begin{array}{cccc}
0 & -1 & 1 & 1 \\
-1 & 0 & 1 & 1 \\
-1 & -1 & 2 & 1 \\
0 & -1 & 0 & 1 \\ 
\end{array}
\right)=\rho_v^{-1}\CM_{(0,1)}\rho_v=\CM_{\nu_2}\CM_{\nu_4}=\CM_{\nu_4}
\CM_{\nu_6},
\end{equation}
All of the above matrices are in $Sp(4,\mathbb Z)$. Due to the
relations \eqref{Ttauu=0} among $\tau_{11}$, $\tau_{12}$ and $\tau_{22}$, they act on $\tau_-=\tau_{11}-\tau_{12}$ as
\begin{equation}\begin{aligned}
M^-_{(0,1)}&=\left(
\begin{array}{cc}
 -4 & -7 \\
 1 & 1\\
\end{array}
\right), \\
 M^-_{(0,-1)}&=\left(
\begin{array}{cc}
 1 & 1 \\
 -1 & 2 \\
\end{array}
\right), \\
  M^-_{(0,\infty)}&= \left(
\begin{array}{cc}
 1 & -6 \\
 0 & 1 \\
\end{array}
\right).
\end{aligned}\end{equation}
We conjugate with $\left(\begin{smallmatrix}1&-1\\0&1\end{smallmatrix}\right)$, to match
with the coupling $\tau=\tau_-+1$ for (\ref{monstersol}). This
reproduces precisely the matrices (\ref{gengamma62}), which leave $v$ invariant. 

Similarly to the above, we can consider the action of the matrices
$\CM_{(0,\infty)}$ and $\CM_{(0,\pm 1)}$ on the parameter
$\tau_+=\tau_{11}+\tau_{12}$. This gives
\begin{equation}\label{key}
\begin{aligned}
M^+_{(0,1)}&=\begin{pmatrix}0&1\\-3&3\end{pmatrix},\\
M^+_{(0,-1)}&=\begin{pmatrix}-3&7\\-3&6\end{pmatrix},\\
M^+_{(0,\infty)}&=T^{-2},
\end{aligned}
\end{equation}
with again $M^+_{(0,1)}M^+_{(0,-1)}=M^+_{(0,\infty)}$ up to
normalisation. These matrices agree with what we found in Section
\ref{sectionu=0}, below  (\ref{pa1pu}).

\subsection{BPS quiver and origin of $\CU$}\label{periodsorigin}
A potential application of the previous sections is to 
interpolate between weak and strong coupling. One may follow the BPS
spectrum along such a trajectory using the connection to BPS
quivers \cite{Alim:2011kw, Chuang:2013wt, Galakhov:2013oja}. We briefly address this connection in this
subsection, and leave a more detailed analysis for future work.
 
Let us consider the origin of the moduli 
space, $(u,v)=(0,0)$. At this point, the two elliptic loci, $\CE_u$ and
$\CE_v$, touch. It is a perfectly regular point, since
$\Delta =729\Lambda^{18}$ does not vanish. We can compute the coupling matrix at the
origin of the moduli space $\CU$ starting from large $u$, and find
\begin{equation}\label{Omega_0}
	\Omega^u(0,0)= \begin{pmatrix} 1+\frac{\sqrt 3}{2}i& -\frac 12\\-\frac 12& 1+\frac{\sqrt 3}{2}i\end{pmatrix}.
\end{equation}
The above matrix can be obtained by expanding the periods to first
order in $v$ but exact in $u$, computing the coupling matrix, setting
$v=0$ and taking the limit $u\to 0$ for $u<0$. This is consistent
with the argument given in \cite{Alim:2011kw} that the origin should
be approached on the negative real $u$-line, as it avoids the
singularity $\underline u=1$ where the periods  pick up a monodromy. 

Analytically continuing the solutions for large $v$ \eqref{ajlargev}, we find that the coupling at the origin $(0,0)\in \CU$ is given by 
\begin{equation}\label{origincouplingT}
	\Omega^v(0,0)=\begin{pmatrix} \frac{2i}{\sqrt3}& 1-\frac{i}{\sqrt3}\\ 1-\frac{i}{\sqrt3}& -1+\frac{2i}{\sqrt3}\end{pmatrix}
\end{equation}
The two different matrices \eqref{Omega_0} and \eqref{origincouplingT} are related through the action  \eqref{symplectictransf} as
\be
\CT_\theta\, (\CM^{(r_2)})^{-1} \CM_{\nu_2}: \Omega^u(0,0)\mapsto \Omega^v(0,0),
\ee
with $\CT_\theta$ as below (\ref{Identu0}). The two effective
couplings at the origin $\Omega^{u,v}(0,0)$ are therefore related by a
monodromy up to $\CT_\theta$. This is expected, since $\CT_\theta$ transforms (\ref{Identu0}) to (\ref{lambdaC}).
 
 As shown in \cite{Alim:2011kw}, the
central charge configuration at the origin can be obtained from the
one for large $u$ by following the negative real axis on the $v=0$
plane from large $u$ to 0. At this point, the full $\mathbb
Z_6$-symmetry is restored and none of the central charges are zero. We
find that, for example, $Z_{\nu_1}=Z_{\nu_2}=e^{\frac{9\pi i}{6}}=-i$, $Z_{\nu_3}=Z_{\nu_4}=e^{\frac{5\pi
    i}{6}}$ and $Z_{\nu_5}=Z_{\nu_6}=e^{\frac{\pi i}{6}}$ in the normalisation
of Table \ref{tableperiods}. Together
with their charge conjugates, they all map into each other by
$\frac{2\pi}{6}$ rotations. In fact, the symmetry group is larger than
$\mathbb{Z}_6$. Since the symmetry group for the central charges of
$(\nu_j,\nu_{j+1},-\nu_j,-\nu_{j+1})$ for $j=1,3,5$ is $D_4$, and the
symmetry group of the equilateral triangle is $D_6$, the
total symmetry group becomes $D_4\rtimes D_6$. This group is known to be isomorphic to the group
$\mathbb{Z}_3\rtimes D_{8}$, which is the automorphism group of this genus 2 curve \cite{shaska2001}. 
Moreover, this group is isomorphic to $D_{12}\rtimes \mathbb{Z}_2$,
such that the automorphism group $D_4$ of $\CE_u$, and $D_{12}$ of
$\CE_v$ are both subgroups of the automorphism group at the origin.

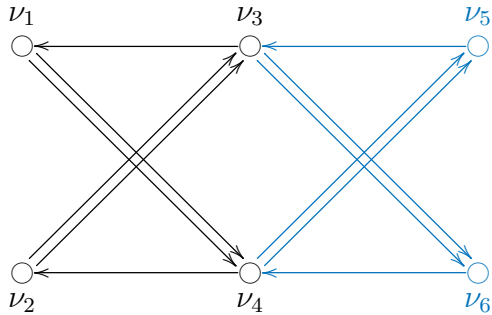
\begin{figure}
	\centering
	\raisebox{10mm}{
		\begin{xy}
			(0,-4)*{\nu_2}; (0,0)*{\Circle}="b1";
			(30,-4)*{\nu_4}; (30,0)*{\Circle}="b2";
			(30,34)*{\nu_3}; (30,30)*{\Circle}="t2";
			(0,34)*{\nu_1}; (0,30)*{\Circle}="t1";
			(60,-4)*{\color{RoyalBlue}\nu_6};(60,0)*{\color{RoyalBlue}\Circle}="b3";
			(60,34)*{\color{RoyalBlue}\nu_5};(60,30)*{\color{RoyalBlue}\Circle}="t3";
			{\ar@<0.37ex> "t1";"b2"};{\ar@<-0.37ex> "t1";"b2"};{\ar "t2";"t1"};{\ar@<0.37ex> "b1";"t2"};{\ar@<-0.37ex> "b1";"t2"};{\ar "b2";"b1"};{\ar@<0.37ex>@[RoyalBlue] "b2";"t3"};{\ar@<-0.37ex>@[RoyalBlue] "b2";"t3"};{\ar@<0.37ex>@[RoyalBlue] "t2";"b3"};{\ar@<-0.37ex>@[RoyalBlue] "t2";"b3"};{\ar@[RoyalBlue] "t3";"t2"};{\ar@[RoyalBlue] "b3";"b2"};
	\end{xy}}
	\caption{The mutation algorithm produces a finite spectrum consisting of 6 particles at strong coupling \cite{Alim:2011kw}. The generating matrix $\bar\sigma_v=\alpha\sigma_v$ maps the charges to the right. The coloured part does not belong to the $SU(3)$ quiver,  it merely highlights how all the charges at strong coupling can be obtained from $\bar \sigma_v$.} \label{fig:quiver2}
\end{figure}

The BPS quiver for strong coupling \cite{Alim:2011kw} is presented in Figure
\ref{fig:quiver2}. Every charge vector in the basis is represented by
a node. The number of arrows is determined by the symplectic inner
product between a pair of charges. The global $\mathbb{Z}_2$ symmetry
$\sigma_v$ acts in the picture
to the right as $ \nu_k\mapsto \nu_{k+2\mod 6}$.

\section{Discussion}\label{discussion}
In this paper we  discussed the modular properties of pure $\CN=2$
Yang-Mills theory in four dimensions with gauge group $SU(3)$. On the
two loci $\CE_u$ and $\CE_v$, where $v=0$ and $u=0$ respectively, we
express the parameters $u$ and $v$ of the moduli space as modular
functions for discrete subgroups of $SL(2,\mathbb R)$. See
\eqref{largeuv=0} and \eqref{monstersol}.  
To this end, we formulate the genus two $SU(3)$ SW curve in Rosenhain form in terms of Siegel theta series. The parameters of the  theory are then found by relating
the Rosenhain form to the PF solution of \cite{Klemm:1995wp}. 
We provide an explicit fundamental domain for the effective coupling on the two elliptic loci $\CE_u$ and $\CE_v$. The relation between cross-ratios of the curve and
theta constants suggests that the full moduli space can be
parametrized by higher genus modular forms. It would be interesting to
find a general solution to \eqref{generaluv} by expressing $u$ and $v$
as algebraic functions of theta constants.

On $\CE_u$, we established a nice
generalisation of the structure appearing in the $SU(2)$ case. In 
rank one, the  parameter $u$ is a weakly holomorphic modular function for the congruence subgroup
$\Gamma^0(4)$. For $SU(3)$, we instead found that on $\CE_u$ the
parameter $u$ is a weakly holomorphic modular function of $\tau_-$ for the group $\Gamma^0(9)\subset
SL(2,\BZ)$. The structure of the moduli space near the special points
of this locus also seems to generalize the rank one picture: We find
that $u$ maps the $\mathbb Z_2$ singularities to the cusps of its
fundamental domain. Furthermore, the duality group is generated by the nontrivial
monodromies on $\CE_u$. For the other choice of modular parameter
$\tau_+=\tau_{11}+\tau_{12}$, we find that $u$ is not invariant under
a congruence subgroup, but is rather a {\it sextic modular
function}, which is the same function as appears for rank 1 
$N_f=1$ SQCD. Nevertheless, we are able to show that the monodromies
can be viewed as paths in a new fundamental region, which we propose. 
 
On the other locus $\CE_v$ where $u=0$, we find that
$v$ can be expressed as a modular function for a subgroup $\Gamma_v\subset SL(2,\mathbb R)$ of Atkin-Lehner type. The AD points are
mapped to the elliptic fixed points of the quotient $ \Gamma_v\backslash
\mathbb H$. The group $ \Gamma_v$ includes a Fricke involution, which can be viewed as a manifestation of
$S$-duality \cite{Minahan:1995er,Argyres:2007cn,Minahan:1996ws}. We derive it from the monodromy group on $\CE_v$. On the locus $\CE_v$, the genus two hyperelliptic curve splits into two \emph{elliptic} curves with complex structures $\tau_\pm=\tau_{11}\pm\tau_{12}$. The appearance of the Fricke involution is a consequence of the two families of elliptic curves being isogenous \cite{Alim:2013eja,Zhou:2014rvr}. 
Fricke dualities also appear in  String theory,
where  they have been shown to play an important r\^{o}le
in the web of dualities of CHL models, i.e. orbifolds of heterotic
string theory on $T^6$ or type II on $K3\times T^2$ 
\cite{Persson:2015jka, Paquette:2016xoo}. They are also the natural generalisation of $S$-duality in the context of Olive-Montonen duality in $\CN=4$
super-Yang-Mills theory for non-simply laced gauge groups
\cite{Argyres:2006qr, Dorey:1996hx} and the geometric Langlands
program \cite{Kapustin:2006pk}. Moreover,
Fricke involutions are  familiar in topological string theory where they act on  higher genus amplitudes, which are described by quasi modular forms. They exchange the large complex structure of the Calabi-Yau threefold with the conifold loci, which gives an analogue of the action of electric-magnetic duality or $\CN=2$ $S$-duality  in topological string theory \cite{Alim:2013eja,Zhou:2014rvr}.

It would be interesting to extend this work to other theories,
such as those with gauge group $SU(N)$, including matter multiplets,
theories of class $S$ \cite{Gaiotto:2009hg}, or gravitational
couplings to these theories \cite{Huang:2009md, Huang:2011qx}.
For theories with $SU(N>2)$, one can for example consider to 
turn on only the bottom Casimir $u_2$ and setting  $u_3,\dots,u_{N}$ to zero. Our analysis naively suggests that it should be parametrized by a modular function for $\Gamma^0(N^2)$. The discriminant of the $SU(N)$ curve \cite{Klemm:1994qs}
\begin{equation}
y^2=\left(x^N-\sum_{j=2}^Nu_jx^{N-j}\right)^2-1
\end{equation}
intersects with this locus  in $u_2^N=N^N(N-2)^{2-N}/4$,
confirming that there are $N$ singularities at strong
coupling. However, it is easy to show that  $\Gamma^0(N^2)$ has $N$
cusps aside from $i\infty$ if and only if $N$ is prime. Note that this
worked for $N=2,3$.  It is furthermore not obvious how the modular
parameter would relate to the coupling matrix, and the map to elliptic
subcovers is more subtle in the higher rank case \cite{shaska2006}.  

We would like to finish by mentioning a few potential applications and
directions for further research: 
\begin{itemize}
\item We observe that the functions
parametrising the $SU(2)$ and $SU(3)$ moduli spaces are all
\emph{replicable} \cite{Conway:1979qga, Ferenbaugh1993,ford1994,
  Alexander:1992} modular functions. The $SU(2)$ order parameter $u$ is of
class 4C, $u_-$ of class 9B, and $v$ of class 6a. It would be interesting to
explore whether there is an underlying reason for the functions to 
have this property.

\item This work motivates exploring subloci of Coulomb branches for
  theories with other gauge
  groups and including matter multiplets. This could provide a
  better understanding of the modularity of these theories. Moreover,
  it would be interesting to understand whether the solution of the
  theory on a sublocus is equivalent to the solution of another
  theory, such as we found for $\CE_u$ and the massless $N_f=1$,
  $SU(2)$ theory for example.

\item The elliptic loci we consider are somewhat analogous to the
  special K\"ahler strata of Coulomb
branches being studied in the recent work
\cite{Argyres:2018zay,Martone:2020nsy,Argyres:2020wmq}. The latter aims to
classify higher rank $\CN=2$ SCFTs by decomposing the singular locus
into a nested series of one-dimensional building blocks.  It would be
interesting to see if our methods find applications in this
programme.

\item The last application which we would like to mention, is topological
  quantum field theory \cite{Witten:1988ze}. Evaluation of the path integral or
  correlation functions for a compact four-manifold $X$ involves the integration over the Coulomb branch (the so-called $u$-plane
integral) of the theory \cite{Moore:1997pc, Marino:1998bm,
  Shapere:2008zf}. For gauge group $SU(2)$, the integral becomes an integral over the modular fundamental domain
$\Gamma^0(4)\backslash\mathbb{H}$ \cite{Moore:1997pc, Malmendier:2008db, Korpas:2017qdo,
  Korpas:2019cwg}. A better understanding of the modularity of $SU(N >
2)$ Seiberg-Witten theory could possibly allow further progress in
this direction for theories with $N > 2$. 
\end{itemize}
   
\acknowledgments
We are happy to thank Philip Argyres, Yoshiaki Goto, Ling Long, Mario Martone, Saiei-Jaeyeong Matsubara-Heo, 
Gregory Moore and Ken Ono for correspondence and discussions. JA and JM  are supported by the Laureate Award 15175 “Modularity in Quantum Field
Theory and Gravity” of the Irish Research Council. EF is supported by the TCD
Provost's PhD Project Award.

\appendix 
\section{Automorphic forms}\label{sec:modularforms}
In this appendix we collect  examples of modular forms that are
used in the text above and discuss some general structures related to these. For further reading see \cite{Bruinier08,
  Rosenhain:1851, Previato:2013, ono2004, gordon1993, freitag1983}. 
\subsection{Elliptic modular forms}\label{appendA}

\subsubsection*{Modular groups and fundamental domains}
We first recall the notion of the congruence subgroups $\Gamma_0(n)$ and $\Gamma^0(n)$  of $ SL(2,\mathbb Z)$. They are defined as 
\be\begin{aligned}
\Gamma_0(n) = \left\{\begin{pmatrix}a&b\\c&d\end{pmatrix}\in SL(2,\mathbb Z)\big| \, c\equiv0 \; \mod n\right\},\\
\Gamma^0(n) = \left\{\begin{pmatrix}a&b\\c&d\end{pmatrix}\in SL(2,\mathbb Z)\big| \, b\equiv0 \; \mod n\right\},
\end{aligned}\ee
and are related by conjugation with the matrix $\text{diag}(n,1)$. We furthermore define $\Gamma(n)$ as the subgroup of $SL(2,\mathbb Z)\ni A$ with $A\equiv\mathbbm 1\mod n$.

The modular groups of $n|h$-type are defined in the following way \cite{Ferenbaugh1993}. Consider matrices of the form
\begin{equation}\label{ALelement}
\begin{pmatrix}ae& b/h\\ cn& de\end{pmatrix}
\end{equation}
with determinant $e$, where $a,b,c,d,e,h,n\in \mathbb Z$, and $h$ is the largest integer for which $h^2|N$ and $h|24$ with $n=N/h$. These matrices are also referred to as \emph{Atkin-Lehner involutions}.

In the case that $n$ is a positive integer and $h|n$, we define $\Gamma_0(n|h)$ as the set of above matrices with $e=1$. For any positive integer $e$ which satisfies $e|n/h$ and $(e,n/eh)=1$ ($e$ is called an \emph{exact divisor} of $n/h$), one can include also matrices of the above form with $e>1$, forming  a group denoted by $\Gamma_0(n|h)+e$. In fact, this construction works for any choice $\{e_1,e_2,\dots\}$ of exact divisors of $n/h$, resulting in the group $\Gamma_0(n|h)+e_1,e_2,\dots$. If $h=1$, the $|h$ is omitted in the notation, and in case that all the possible $e_i$ are included, the group is simply denoted by $\Gamma_0(n|h)+$.

In the $\Gamma^0$ convention the notation simplifies, since $\Gamma^0(n|h)=\Gamma^0(\tfrac nh)$. This can be checked by conjugating \eqref{ALelement} with $\text{diag}(n,1)$. The extension by non-unity determinant matrices follows by analogy.

A key concept of the theory of modular forms is the \emph{fundamental
  domain}. A fundamental domain for a group $\Gamma\subset
SL(2,\mathbb R)$ is an open subset $\CF\subset \mathbb H$ with the
property that no two distinct points of $\CF$ are equivalent under the
action of $\Gamma$ and every point in $\mathbb H$ is mapped to some
point in the closure of $\CF$ by the action of an element in
$\Gamma$. The quotient $\Gamma\backslash \mathbb H$ can be
compactified by adding finitely many points called \emph{cusps}. Cusps
are $\Gamma$-equivalence classes of $\mathbb Q\cup\{ i
\infty\}$. Special points in the fundamental domain are the
\emph{elliptic fixed points}, which are points in $\mathbb H$ that
have a non-trivial $\Gamma$-stabiliser. There,  the quotient
$\Gamma\backslash \mathbb H$ becomes singular. Elliptic points can
always be mapped to the boundary of the fundamental domain. They
furthermore contribute non-trivially to the order of vanishing, which
determines the dimension of the spaces of modular forms for fixed
weight.

\subsubsection*{Examples of modular forms}
The Eisenstein series $E_k:\mathbb{H}\to \mathbb{C}$ for even $k\geq 2$ are defined as the $q$-series 
\be
\label{Ek}
E_{k}(\tau)=1-\frac{2k}{B_k}\sum_{n=1}^\infty \sigma_{k-1}(n)\,q^n, \quad q=e^{2\pi i \tau},
\ee
with  $B_k$ the Bernoulli numbers and  $\sigma_k(n)=\sum_{d|n} d^k$ the divisor sum. For $k\geq 4$ even, $E_{k}$ is a modular form  of weight $k$ for
$\operatorname{SL}(2,\mathbb{Z})$. With this normalisation, the $j$-invariant can be written as 
\begin{equation}
\label{jfunction}
j=1728\frac{E_4^3}{E_4^3-E_6^2}.
\ee
The Jacobi theta functions $\vartheta_j:\mathbb{H}\to \mathbb{C}$,
$j=2,3,4$, are defined as
\be
\label{Jacobitheta}
\begin{split}
&\vartheta_2(\tau)= \sum_{r\in
  \mathbb{Z}+\frac12}q^{r^2/2},\\
&\vartheta_3(\tau)= \sum_{n\in
  \mathbb{Z}}q^{n^2/2},\\
&\vartheta_4(\tau)= \sum_{n\in 
  \mathbb{Z}} (-1)^nq^{n^2/2},
\end{split}
\ee
with $q=e^{2\pi i\tau}$. These functions transform under the generators $T$ and $S$ of $SL(2,\mathbb Z)$ as
\begin{alignat}{3}\nonumber
S:\quad& \vartheta_2(-1/\tau)=\sqrt{-i\tau}\vartheta_4(\tau),\quad&&\vartheta_3(-1/\tau)=\sqrt{-i\tau}\vartheta_3(\tau),\quad&&\vartheta_4(-1/\tau)=\sqrt{-i\tau}\vartheta_2(\tau)\\
T:\quad&\vartheta_2(\tau+1)=e^{\frac{\pi i}{4}}\vartheta_2(\tau),\quad &&\vartheta_3(\tau+1)=\vartheta_4(\tau),&&\vartheta_4(\tau+1)=\vartheta_3(\tau).
\end{alignat}
Some special values that we use are
\begin{equation}\label{specialvaluesjacobi}
\vartheta_2(i)=\vartheta_4(i)= \sqrt[4]{\tfrac\pi 2}/\Gamma(\tfrac34), \qquad \vartheta_3(i) =  \sqrt[4]{\pi}/\Gamma(\tfrac34).
\end{equation}
The Dedekind eta function $\eta: \mathbb H\to \mathbb C$ is defined as the infinite product
\begin{equation}
\label{etaf}
\eta(\tau)=q^{\frac{1}{24}}\prod_{n=1}^{\infty}(1-q^n), \quad q=e^{2\pi i\tau}.
\end{equation}
It transforms under the generators of $SL(2,\mathbb Z)$ as 
\be\begin{aligned}
S: \quad& \eta(-1/\tau)=\sqrt{-i\tau }\, \eta(\tau),\\
T: \quad& \eta(\tau+1)=e^{\frac{\pi i}{12}}\, \eta(\tau).
\end{aligned}\ee

Quotients of $\eta$ functions are frequently used to generate bases for the spaces of modular forms for congruence subgroups of $SL(2,\mathbb Z)$. We use the following\\

\textsc{Theorem} 1 \cite{ono2004,gordon1993}: Let $f(\tau)=\prod_{\delta|N}\eta(\delta\tau)^{r_\delta}$ be an eta-quotient with $k=\frac 12\sum_{\delta|N}r_\delta\in \mathbb Z$ and $\sum_{\delta|N}\delta r_\delta\equiv \sum_{\delta|N}\frac{N}{\delta} r_\delta\equiv  0\mod 24$. Then, $f$ is a weakly holomorphic modular form for $\Gamma_0(N)$ with weight $k$. In particular, $f$ transforms  as $f(\tau|_\gamma)=\chi(d)(c\tau+d)^k f(\tau)$ under $\gamma=\begin{pmatrix}a&b\\c&d\end{pmatrix}\in\Gamma_0(N)$ with character $\chi(d)=\left(\frac{(-1)^k s}{d}\right)$, where $s=\prod_{\delta|N}\delta^{r_\delta}$. \\

\subsection{Siegel modular forms}\label{sec:SiModForms}
Ordinary modular forms are constructed by the action of an
$SL(2,\mathbb Z)$ M{\"o}bius transformation on the upper half-plane
$\mathbb H$. Siegel modular forms \cite{Bruinier08,freitag1983}
generalize this notion by introducing an action of $Sp(2g,\mathbb Z)$
on the so-called Siegel upper half-plane $\mathbb{H}_g$, which works
for any \emph{genus} $g\in\mathbb N$. 

Define the Siegel modular group of genus $g$ as
\be\label{sp2gzdef}
Sp(2g,\mathbb Z) = \{M\in \text{Mat}(2g;\mathbb Z)\,|\, M^TJ M =J\} \quad\text{with } J = \left(\begin{smallmatrix} 0& \mathbbm 1_g\\ -\mathbbm 1_g&0\end{smallmatrix}\right).
\ee
The group $Sp(4,\BZ)$ can be generated \cite{Bruinier08} by the
elements $J$ and  $T=\left(\begin{smallmatrix} \mathbbm{1}_g & s \\0 &
    \mathbbm{1}_g \end{smallmatrix}\right)$ with $s=s^T$. 
The Siegel upper half-plane
\be 
\mathbb H_g =\{\Omega\in\text{Mat}(g;\mathbb C)\,|\, \Omega^T=\Omega, \, \im \Omega >0\}
\ee
consists of complex symmetric $g\times g$ matrices whose (componentwise) imaginary part is positive definite. This generalizes the ordinary upper half-plane $\mathbb H=\mathbb{H}_1$. For example, for $g=2$  this means that 
\begin{equation}
\Omega=\begin{pmatrix}\tau_{11}&\tau_{12}\\ \tau_{12}&\tau_{22}\end{pmatrix},\quad \im \tau_{11}>0, \quad \im\tau_{11}\im \tau_{22}-(\im \tau_{12})^2>0.
\end{equation}
 An element $\gamma=\left(\begin{smallmatrix} A&B\\C&D\end{smallmatrix}\right)\in Sp(2g,\mathbb Z)$ acts on the Siegel upper half-plane by 
\be\label{symplectictransf}
\Omega\longmapsto\gamma(\Omega)=(A\Omega+B)(C\Omega+D)^{-1}.
\ee
A (classical) Siegel modular form of weight $k$ and genus $g$ is then a holomorphic function $f:\mathbb{H}_g\to\mathbb C$ satisfying
\be f(\gamma(\Omega))=\det(C\Omega+D)^kf(\Omega) \qquad \forall \gamma =\begin{pmatrix}A&B\\C&D\end{pmatrix}\in Sp(2g,\mathbb Z), \ee
where for $g=1$ holomorphicity at $i\infty$ is required in addition.

\label{app:thetaconstants}
Theta series provide an explicit class of classical Siegel modular forms. For $a$, $b \in \BQ^2$ and $\Omega\in \mathbb{H}_2$, define
\be\label{definitionstheta}
\Theta\begin{bmatrix}a\\b\end{bmatrix}(\Omega) = \sum_{k\in \BZ^2}\exp\left(\pi i(k+a)^T\Omega(k+a)+2\pi i(k+a)^T\, b  \right).
\ee
We are especially interested in the case where the entries of these column vectors take values in the set $\{0,\frac{1}{2} \}$. The corresponding theta functions are usually referred to as the theta characteristics. We call $\gamma = \left[\begin{smallmatrix}a\\b\end{smallmatrix}\right]$ an even (odd) characteristic if $4a^T b$ is even (odd). In the case of genus two there are  ten even  theta constants \cite{Previato:2013},
\be
\begin{aligned}
&\Theta_{1} = \Theta\begin{bmatrix}0&0\\0&0\end{bmatrix},\hspace{5pt} \Theta_{2} = \Theta\begin{bmatrix}0&0\\\frac{1}{2}&\frac{1}{2}\end{bmatrix}, \hspace{5pt} \Theta_{3} = \Theta\begin{bmatrix}0&0\\\frac{1}{2}&0\end{bmatrix}, \hspace{5pt} \Theta_{4} = \Theta\begin{bmatrix}0&0\\0&\frac{1}{2}\end{bmatrix}, \hspace{5pt} \Theta_{5} = \Theta\begin{bmatrix}\frac{1}{2}&0\\0&0\end{bmatrix}, \\
&\Theta_{6} = \Theta\begin{bmatrix}\frac{1}{2}&0\\0&\frac{1}{2}\end{bmatrix},\hspace{5pt}\Theta_{7} = \Theta\begin{bmatrix}0&\frac{1}{2}\\0&0\end{bmatrix},\hspace{5pt}\Theta_{8} = \Theta\begin{bmatrix}\frac{1}{2}&\frac{1}{2}\\0&0\end{bmatrix},\hspace{5pt}\Theta_{9} = \Theta\begin{bmatrix}0&\frac{1}{2}\\\frac{1}{2}&0\end{bmatrix},\hspace{5pt}\Theta_{10} = \Theta\begin{bmatrix}\frac{1}{2}&\frac{1}{2}\\\frac{1}{2}&\frac{1}{2}\end{bmatrix}.
\end{aligned}
\ee
All even theta constants can be related through algebraic identities
to four \emph{fundamental} ones,  $\Theta_1$, $\Theta_2$,
$\Theta_3$, $\Theta_4$  \cite{Previato:2013}. 

The above theta functions are weight $\frac 12$ Siegel modular forms
for a subgroup of $Sp(4,\mathbb Z)$. Their transformation properties
under the Siegel modular group can be found in \cite{freitag1983}.

\section{Picard-Fuchs solution}\label{sec:pfsolutionappend}
In the limit of large $u$ and small $v$, reference \cite{Klemm:1995wp} determines the $a_I$ and
$a_{D,I}$ non-perturbatively in terms of the fourth Appell
hypergeometric function $F_4(a,b,c,d;x,y)$. For 
$\sqrt{|x|}+\sqrt{|y|}<1$, this function is given by
\be 
F_4(a,b,c,d;x,y)=\sum_{m,n\geq 0}\frac{(a)_{m+n}\,(b)_{m+n}}{m!\,n!\,(c)_m(d)_n}\,x^m\,y^n,
\ee
where $(a)_m=\frac{\Gamma(a+m)}{\Gamma(a)}$ is the Pochhammer
symbol. We will also need expansions of $F_4$ for large $y$, which can
be achieved by replacing the sum over $n$ by the hypergeometric series $_2F_1$, 
\be\label{F42F1}
F_4(a,b,c,d;x,y)=\sum_{m\geq 0}\frac{(a)_{m}\,(b)_{m}}{m!\,(c)_m}\,
{_2F_1}(a+m,b+m,d;y)\, x^m.
\ee
While analytic continuations are known for $_2F_1$, they are not well
established for $F_4$.

\subsection{Classical roots}\label{sec:roots}
In order to match the Picard-Fuchs solutions with the periods, we need to expand the periods around the classical solutions in \eqref{aIcubic}. We therefore need to find the roots of these two cubics.

The general formula for the roots of a depressed cubic equation, $ax^3+bx+c=0$, is given by
\begin{equation}\label{eq:gen_cubic}
\xi_k=-\frac{1}{3a}\left(\alpha^k C+\frac{\Delta_0}{\alpha^k C}\right), \hspace{10pt}k\in\{0,1,2 \},
\end{equation}
where $\alpha=e^{2\pi i/3}$, $C^3=\frac{\Delta_1\pm\sqrt{\Delta_1^2-4\Delta_0}}{2}$, $\Delta_0=-3ab$ and $\Delta_1=27a^2c$ \cite{Abramovitz:1964}. The choice of sign in front of the square root in $C$ is arbitrary, in the sense that it only corresponds to a permutation of the roots. 

It is however important to fix the ambiguities in taking the square and cubic root.
We fix the ambiguity in the square root by the following choice for the branch of the logarithm: For any complex number $z\in \mathbb{C}^*$, we set $\log(z)=\log\!|z|+i \mathrm{Arg}(z)$ with $-\pi<\mathrm{Arg(z)}\leq \pi$. The ambiquity in the cubic root of a complex number $z$ is fixed by demanding that the real part of $\sqrt[3]{z}$ has the largest absolute value among the three solutions to $\rho^3=z$. Thus $\sqrt[3]{1}=1$ and $\sqrt[3]{-1}=-1$. Two of the cube roots of $i$ and $-i$ have equal real parts. We fix the remaining ambiguity by setting $\sqrt[3]{i}=e^{\pi i/6}=\frac{\sqrt{3}}{2}+\frac{i}{2}$ and $\sqrt[3]{-i}=e^{-\pi i/6}=\frac{\sqrt{3}}{2}-\frac{i}{2}$. 

To list the roots of our two equations, we define
\be
s_\pm(a,b)=\sqrt[3]{\frac{b}{2}\pm \sqrt{\frac{b^2}{4}-\frac{a^3}{27}}}.
\ee
Using Eq. \eqref{eq:gen_cubic}, we then find that the roots of \eqref{aIcubic}  for $a_1$ are given by
\be\label{eq:class_sol}
\begin{split}
	\xi_{1}(u,v) & =s_+(u,v)+s_-(u,v),\\
	\xi_{2}(u,v) & =\alpha\,s_+(u, v)+\alpha^2\,s_-(u,v),\\
	\xi_{3}(u,v) & =\alpha^2\,s_+(u, v)+\alpha\,s_-(u, v),\\
\end{split} 
\ee 
and the roots for $a_2$ by $-\xi_j(u,v)$. This gives the $3\times 3=9$ solutions to the equations in (\ref{aIcubic}). However, (\ref{casimirs}) is supposed to have only
$2\times 3 =6$ solutions. Let us determine the 6 solutions in one of the
regimes of interest for $SU(3)$ Yang-Mills theory: we assume $u$ is
large and close to the positive axis: $u=\lambda -i\epsilon \lambda$ with $\lambda$ real
and very large and $0< \epsilon\ll 1$. Note that in this regime
\be
\begin{split}
	&s_\pm (u,v)=\sqrt[3]{\frac{v}{2}\pm i\sqrt{\frac{u^3}{27}-\frac{v^2}{4}}}.
\end{split}
\ee 
Furthermore, $s_+(u,v)\,s_-(u,v)=u/3$ and $s_{-}(u,-v)=e^{-\pi
	i/3}s_{+}(u,v)=-\alpha s_{+}(u,v)$ hold. 
For $v=0$, we have $s_+(u,0)=e^{\pi i/6} \sqrt{u/3}$ and
$s_-(u,0)=e^{-\pi i/6} \sqrt{u/3}$, and thus
\be
\begin{split}
	\xi_{1}(u,0) & =\sqrt{u}, \\
	\xi_{2}(u,0) & =-\sqrt{u}, \\
	\xi_{3}(u,0) & = 0.
\end{split}
\ee
This demonstrates that the solutions to \eqref{casimirs} for
$(a_1,a_2)$ are given by 
\begin{equation}\label{classroots}
(\xi_1,-\xi_2),\ (\xi_1,-\xi_3),\ (\xi_2,-\xi_1),\ (\xi_2,-\xi_3),\ (\xi_3,-\xi_1),\ (\xi_3,-\xi_2).
\end{equation}

\subsection{Picard-Fuchs system for large $u$}\label{pfuappendix}
To express $a_I$ and $a_{D,I}$ in terms of $u$ and $v$, we will start by working in the patch with large $u$ and small $v$, and use the variables
$x=\frac{27v^2}{4u^3}$ and $y=\frac{27\Lambda^6}{4u^3}$. In
\cite{Klemm:1995wp} the authors use the notation $P_3$ for this patch
and, similarly, $P_2$ for the patch where $v$ is large and $u$ is
small and we will adopt this notation in the following. We have four
solutions \cite[Eq. (6.1)]{Klemm:1995wp} to the Picard-Fuchs system \cite[Eq. (5.11)]{Klemm:1995wp} for 
$SU(3)$,
\be\label{PFP3}
\begin{aligned} 
	\omega_1^{P_3} & = \sqrt{3}\, 2^{\frac{2}{3}}\Lambda\,y^{-\frac{1}{6}}\,F_4\!\left(-\tfrac{1}{6},\tfrac{1}{6},\tfrac{1}{2},1;x,y\right), \\ 
	\omega_2^{P_3} & = \frac{2^{\frac{2}{3}}\Lambda}{3} \sqrt{x}\,y^{-\frac{1}{6}}\,F_4\!\left(\tfrac{1}{3},\tfrac{2}{3},\tfrac{3}{2},1;x,y\right), \\ 
	\Omega_1^{P_3} & = 36\pi\,e^{-\pi i/6}\, 2^{2/3}\Lambda\,\frac{\Gamma(\tfrac{1}{3})}{\Gamma(\tfrac{1}{6})^2}\,F_4\!\left(-\tfrac{1}{6},-\tfrac{1}{6},\tfrac{1}{2},\tfrac{2}{3};\tfrac{x}{y},\tfrac{1}{y}\right) + \beta_1^{P_3}\,\omega_1^{P_3},  \\
	\Omega_2^{P_3} & = - e^{\frac{\pi i}{3}} \frac{2^{\frac{2}{3}}\Lambda}{\sqrt{3}\,2\pi}\,\Gamma(\tfrac{1}{3})^3\,\sqrt{\frac{x}{y}}\,F_4\!\left(\tfrac{1}{3},\tfrac{1}{3},\tfrac{3}{2},\tfrac{2}{3};\tfrac{x}{y},\tfrac{1}{y}\right) +\beta_2^{P_3}\,\omega_2^{P_3},
\end{aligned}
\ee
where $\beta_1^{P_3}=(i-\sqrt{3})\pi+4\log(2)+3\log(3)-5$ and
$\beta_2^{P_3}=1+(i+\frac{1}{\sqrt{3}})\pi+3\log(3)$.\footnote{We corrected the power of
$\Gamma(\frac{1}{3})$ in the expression for $\Omega_2^{P_3}$ compared to \cite{Klemm:1995wp}, and removed the factor of $\sqrt3 \Lambda$ from the second terms of $\Omega_i^{P_3}$ which have been placed incorrectly in \cite{Klemm:1995wp} as they are already included in $\omega_1^{P_3}$ and $\omega_2^{P_3}$.} The $a_I$ and
$a_{D,I}$ are linear combinations of these periods found by comparing
the expansions of these solutions with the classical and
semi-classical solutions in the previous section for large $u$. Using
the classical solutions $(a_1,a_2)=(\xi_1,-\xi_2)$ one finds
\cite[Eq. 6.4]{Klemm:1995wp}, 
\be
\label{weakcouplingperiods}
\begin{split}
	a_{D,1}(u,v)&= -\frac{i}{4\pi}(\Omega_1^{P_3}+3\Omega_2^{P_3})-\frac{1}{\pi}(\alpha_1\omega_1^{P_3}-\alpha_2\omega_2^{P_3}) \\ 
	&=-\frac{i}{2\pi}
	\left(\sqrt{u}+\frac{3}{2}\frac{v}{u}\right)\log\! \left(
	\frac{27\Lambda^6}{4u^3} \right)-\frac{1}{\pi} \left(\frac{i}{2}+2\alpha_1\right)\sqrt{u}+O(u^{-1}),\\
	a_{D,2}(u,v) &=-\frac{i}{4\pi}(\Omega_1^{P_3}-3\Omega_2^{P_3})-\frac{1}{\pi}(\alpha_1\omega_1^{P_3}+\alpha_2\omega_2^{P_3})=a_{D,1}(u,-v)\\
	a_1(u,v) &=\frac{1}{2}(\omega_1^{P_3}+\omega_2^{P_3}) \sim \sqrt{u} +\frac{1}{2} \frac{v}{u}+\dots,\\
	a_2(u,v) &=\frac{1}{2}(\omega_1^{P_3}-\omega_2^{P_3}) \sim \sqrt{u} -\frac{1}{2} \frac{v}{u}+\dots,
\end{split}
\ee
with $\alpha_1 = \frac{5i}{4}-i\log(2)-\frac{3i}{4}\log(3)$ and
$\alpha_2=\frac{3i}{4}+\frac{9i}{4}\log(3)$. The chain rule then allows to compute the coupling matrix, 
\begin{equation}
\Omega(u,v)=\begin{pmatrix}\partial_ua_1 & \partial_u a_2\\\partial_va_1&\partial_va_2\end{pmatrix}^{-1}\, \begin{pmatrix}\partial_ua_{D,1}&\partial_ua_{D,2}\\ \partial_va_{D,1}&\partial_va_{D,2}\end{pmatrix}.
\end{equation}

\subsection{Picard-Fuchs system for large $v$}\label{pfderivation_vlarge}
We can run a similar analysis as in the previous section for the patch $P_2$, i.e., for large $v$ and small $u$. This is not done explicitly in \cite{Klemm:1995wp} but the authors hint at how it should be done. Here, we use the variables $x=\frac{4u^3}{27v^2}$ and $y=\frac{\Lambda^6}{v^2}$ to express the solutions of the Picard-Fuchs equations as
\begin{equation}\label{PFP2}
\begin{aligned}
\omega_1^{P_2}&=2y^{-1/6}F_4\left(-\frac{1}{6},\frac{1}{3},\frac{2}{3},1;x,y\right),\\
\omega_2^{P_2}&=2^{1/3}x^{1/3}y^{-1/6}F_4\left(\frac{1}{6},\frac{2}{3},\frac{4}{3},1;x,y\right),\\
\Omega^{P_2}_1 &= -\frac{\alpha^2}{2}\pi^{-3/2}\Gamma\left(-\tfrac{1}{6}\right)\Gamma\left(\tfrac{2}{3}\right)F_4\left(-\tfrac{1}{6},-\tfrac{1}{6},\tfrac{2}{3},\tfrac{1}{2};\tfrac{x}{y},\tfrac{1}{y}\right)+\beta_1^{P_2}\omega_1^{P_2},\\
\Omega_2^{P_2} &= -\frac{\alpha}{3}\pi^{-3/2}\sqrt[3]{\tfrac{x}{y}}\Gamma\left(-\tfrac{2}{3}\right)\Gamma\left(\tfrac{1}{6}\right)F_4\left(\tfrac{1}{6},\tfrac{1}{6},\tfrac{4}{3},\tfrac{1}{2};\tfrac{x}{y},\tfrac{1}{y}\right) +\beta_2^{P_2}\omega_2^{P_2},
\end{aligned}
\end{equation}
with 
\be
\begin{aligned}
	\beta_1^{P_2}&=-\frac{i}{4\pi}\left(2\log 2+3\log 3-6+\pi(i-2/\sqrt{3})\right), \\
	\beta_2^{P_2}&=-\frac{i}{2^{4/3}\pi}\left( 2\log 2+3\log 3+\pi(i+2/\sqrt{3})\right).
\end{aligned}
\ee

Comparing the expansions of these solutions with the asymptotic expansions of $a_{(D),I}$ for the semi-classical contributions fixes the coefficients. For this, one needs to match the $F_4$ expansions with  the leading coefficients of the (differentiated) prepotential \cite{Klemm:1994qs}
\be
\CF = \frac{\tau_0}{6}\sum_{i=1}^{3}Z_i^2+\CF_{1-\text{loop}}+\CF_{inst.},
\ee
where\footnote{We correct a typo in \cite[Eq. 6.8]{Klemm:1995wp}.}
\be\label{eq:tau_0}
\tau_0 =\frac{9-\log 4}{2\pi i}.
\ee
From this, one finds 
\be
\begin{aligned}\label{ajlargev}
	a_{D,1} &= -i\sqrt{3}\alpha\left(\Omega_1^{P_2}-2^{-2/3}\alpha\Omega_2^{P_2}\right)+\left(\alpha c_1-\tfrac{i\sqrt{3}}{2}\right)\omega_1^{P_2} +\left(\alpha^2c_2+\tfrac{i\sqrt{3}}{2}\right)\omega_2^{P_2}, \\
	a_{D,2} &=-i\sqrt{3}\left(\Omega_1^{P_2}-2^{-2/3}\Omega_2^{P_2}\right)+\left( c_1+\tfrac{i\sqrt{3}}{2}\right)\omega_1^{P_2} +\left(c_2-\tfrac{i\sqrt{3}}{2}\right)\omega_2^{P_2}, \\
	a_1 &= \frac{1}{2}\left(\omega^{P_2}_1+\omega^{P_2}_2\right), \\
	a_2 &= -\frac{\alpha}{2}\left(\omega^{P_2}_1+\alpha\omega^{P_2}_2\right),
\end{aligned}
\ee
where $c_1 = \frac{\sqrt{3}}{4\pi}\left(2\log 2+3\log 3+\frac{\pi}{\sqrt{3}}-6\right)$ and $c_2=-\frac{\sqrt{3}}{4\pi}\left(2\log 2+3\log 3-\frac{\pi}{\sqrt{3}}\right)$. We note that for $u=0$, we find $a_2=-\alpha a_1$.

\subsection{The $\BZ_2$ vacua and massless states}\label{relabelledroots}
In deriving the above results for the large $v$ regime we have used a different symplectic basis than what is used in for example \cite{Klemm:1995wp, Marino:1998bm}. In this subsection we briefly comment on how the two bases relate. The basis chosen in  \cite{Klemm:1995wp, Marino:1998bm} is more natural  to use when comparing such quantities as the strong coupling periods for the two different loci, and in this basis we also compute the periods for all the points of interest. The change of basis is done by interchanging the roots $\xi_2\leftrightarrow \xi_3$ as given in \eqref{eq:class_sol}. Quantum mechanically, the singular branch of the classical theory splits into two branches separated by the scale $\Lambda$. Therefore, we must also interchange $r_2\leftrightarrow r_3$ and $r_5\leftrightarrow r_6$. One finds that this symplectic change of basis is given by the semi-classical version of the second Weyl reflection of the $A_2$ root lattice,
\begin{equation}
	\CR_2=\begin{pmatrix}1&1&0&0\\0&-1&0&0\\0&0&1&0\\0&0&1&-1\end{pmatrix} \in Sp(4,\BZ).
\end{equation}
This merely changes some prefactors of the solution \eqref{ajlargev}. The change of roots modifies the cross-ratios in a trivial way, and they agree asymptotically with the theta quotients   \eqref{thetaconstants} computed from the new periods, as expected. One can show that the algebraic relations 
\eqref{eq:C_rel_v_large} for $u=0$ take the same form. However, on
this locus we now find 
\begin{equation}\label{u=0section}
\tau_{12} = \frac{1-\tau_{11}}{2}, \hspace{20pt}\tau_{22}=\tau_{11}-2,
\end{equation}
from which it follows that 
\be\begin{aligned}
	2i\sqrt{27}\,v&=-\alpha^2q^{-\frac16}+33\alpha q^{\frac16}+153 q^{\frac 12}+713\alpha^2q^{\frac56}+\CO(q^{\frac76})\\
	&=m\left(-\alpha q^{\frac16}\right)=m\left(\tfrac \tau6-\tfrac 16\right),
\end{aligned}\ee
which is identical to \eqref{vexpansion} up to phases. 

We can use the new solution to analyse the $\BZ_3$ symmetry $u\mapsto \alpha u$. This leads to the matrix
\begin{equation}\label{sigmav}
\tilde\sigma_v = \alpha^2\begin{pmatrix} 0&1&-1&2\\-1&-1&2&-1\\0&0&-1&1\\0&0&-1&0 \end{pmatrix}.
\end{equation} 
It can also be obtained from the previous result \eqref{tildesigmav} by conjugation with $\CR_2$. It satisfies $\tilde\sigma_v^3=\mathbbm{1}$ and we can use it to generate the charges of the states that become massless at the $\mathbb Z_2$ points. To this end, we introduce the purely integral matrix $U=\alpha^2\tilde\sigma_v^{-1}\in Sp(4,\BZ)$, which is the matrix used in \cite{Klemm:1995wp,Marino:1998bm}, and act with this on the monopole basis,
\begin{alignat}{3}\label{z2chargespp}\nonumber
\tilde\nu_1&=(1,0,0,0),\qquad && \tilde\nu_2 &&=\, (0,1,0,0), \\  
\tilde\nu_3 &=\tilde\nu_1 U=  (-1,-1,1,-2), \qquad && \tilde\nu_4 &&=\tilde\nu_2 U= \,(1,0,-2,1), \\ 
\tilde\nu_5 &=\tilde\nu_1 U^{-1}= (0,1,-1,2),\qquad &&\tilde\nu_6&&=\tilde\nu_2 U^{-1}= \,(-1,-1,2,-1).\nonumber
\end{alignat}
Using the periods from Table \ref{tableperiods} we can confirm that $\tilde\nu_{\{1,3,5\}}$ become massless at the AD point $(0,1)$ and $\tilde\nu_{\{2,3,6\}}$ at the AD point $(0,-1)$. Furthermore, the charges in row $k+1$ in \eqref{z2chargespp} become massless at the $\BZ_2$ point $(\underline u,v)=(\alpha^{k},0)$. It can be checked that the charges in each row are mutually local with respect to the symplectic inner product induced by $J$, given in \eqref{sp2gzdef}. The charges in both columns however are mutually non-local. This is a crucial observation that lead to the discovery of new superconformal theories \cite{Argyres:1995jj, ARGYRES199671, EGUCHI1996430}.

\begin{table}\begin{center}
		$\begin{tabu}{ c|c |c } 
		(\underline u,v)& \pi(\underline u,v)&\text{normalisation}\\
		\hline
		(0,1)&(0,-\sqrt3i,1,-\alpha^2)& \Gamma\left(\frac 13\right)\Gamma\left(\frac 76\right)/2^{1/3}\sqrt\pi\\
		(0,-1)&(-\sqrt3i,0,-\alpha,1)\\ \hline 
		(1,0)&(0,0,1,1)\\
		(\alpha,0)&(\alpha^2,\alpha^2,0,-\alpha^2)&\sqrt[3]{2}\, \pi/3\sqrt3\\
		(\alpha^2,0)&(-\alpha,-\alpha,-\alpha,0)\\ \hline
		(0,0)&(-i,-i,-\omega^5,\omega)&  2 \sqrt{\tfrac{\pi }{3}}\Gamma \left(\tfrac{7}{6}\right)/\Gamma \left(\tfrac{2}{3}\right)
		\end{tabu}$ 
		\caption{Periods at the $\mathbb Z_3$, $\mathbb Z_2$ points and the origin, computed from the analytic continuation of the large $v$ PF solution and appropriately normalized.}
		\label{tableperiods}\end{center}\end{table}

The matrix \eqref{sigmav} conjugates the strong coupling matrices \cite{Klemm:1995wp}  as well as the  semi-classical matrices according to
\begin{equation}
\tilde\sigma_v^{-1} M^{(r_1)}\tilde\sigma_v=M^{(r_2)}, \quad 
\tilde\sigma_v^{-1} M^{(r_2)}\tilde\sigma_v=M^{(r_3)},\quad 
\tilde\sigma_v^{-1} M^{(r_3)}\tilde\sigma_v=M^{(r_1)}.
\end{equation}
The same equations hold for the $\mathbb Z_2$ symmetry 
\begin{equation}
\tilde\rho_v=\left(
\begin{array}{cccc}
1 & 1 & -2 & 1 \\
-1 & 0 & 4 & -2 \\
0 & 0 & 0 & 1 \\
0 & 0 & -1 & 1 \\
\end{array}
\right),
\end{equation}
as is also the case for large $u$. As a consistency check, the  pair $(\tilde\sigma_v,\tilde\rho_v)$ again satisfies the relation \eqref{quantummonodromy}, and $\tilde\rho_v^2$ is a non-trivial monodromy. The matrix $\tilde\rho_v$ maps $\{\tilde\nu_2,\tilde\nu_4,\tilde\nu_6\}$ to $\{-\tilde\nu_1,-\tilde\nu_3,-\tilde\nu_5\}$  and therefore exchanges the AD points $v=\pm 1$.

The periods in Table \ref{tableperiods} obtain different values depending on the direction from which the various points are approached.\footnote{This is not only a problem involving monodromies. By computing coupling matrices at the origin from different directions we find that they generally  do not lie in the Siegel upper half-plane $\mathbb{H}_2$, even though it is a regular point of the curve. One cannot place them back in $\mathbb{H}_2$ by acting on them with monodromy matrices in $Sp(4,\mathbb Z)$.} On the locus $\CE_u$, where $v=0$, we have three singularities located at $\underline u=1,\, \alpha,\, \alpha^2$. Reference \cite{Alim:2011kw} argues that one finds consistent values if the points are approached from the negative real axis. In this way we can go from weak to strong coupling without crossing walls of the second kind.\footnote{Walls of the second kind are generally defined as hypersurfaces where a fixed quiver QM description of the BPS spectrum breaks down, and one needs to mutate the quiver to find the spectrum on the other side of the wall  \cite{Alim:2011kw}.} On $\CE_v$, with $u=0$, we instead have two singularities on the real line at $v=\pm 1$, analogous to the $u$-plane in the $SU(2)$ theory. There, we find a consistent picture by taking the limits from the lower half-plane in order to avoid the singular points (see discussion in \cite{Ferrari:1996sv}). 

The two patches with large $u$ and large $v$ (from this subsection) respectively are connected by a simple change of basis. It is given by 
\begin{equation}\label{Muv}
\CM= \CM_{\tilde \nu_2}=\left(
\begin{array}{cccc}
1 & 0 & 0 & 0 \\
0 & 1 & 0 & 0 \\
0 & 0 & 1 & 0 \\
0 & -1 & 0 & 1 \\
\end{array}
\right).
\end{equation}
This matrix is the strong coupling monodromy \eqref{strongmon} associated with the magnetic monopole  $\tilde \nu_2=(0,1,0,0)$. 

\subsection{The $\mathbb Z_3$ vacua}

With the explicit result \eqref{ADcouplingT} for the coupling matrix, the charges of the massless states \eqref{z2chargespp}  and the periods from Table \ref{tableperiods} we can revisit the results of \cite{Argyres:1995jj}. Starting from the three states $\tilde\nu_{\{1,3,5\}}$ which become massless at $(\underline u,v)=(0,1)$, we aim to find a symplectic projection such that the massless states are  charged only under the first $U(1)$ factor.   Following the logic of \cite{Argyres:1995jj,Marino:1998bm}, in this basis the coupling matrix becomes diagonal ($\tau_{12}=0$) and the curve splits into a small and a large torus, parametrized by $\tau_{11}$ and $\tau_{22}$, respectively. The modulus of the large torus is fixed by the $\mathbb Z_3$ symmetry to be $\tau_{22}=-\alpha^2$. The small torus $\tau_{11}=\tau(\rho)$ depends on the direction $\rho$ from which the AD point is approached, where $\delta v = 2\varepsilon^3$, $\delta u=3\varepsilon^3 \rho$. The small torus near the $\mathbb Z_3$ point takes the form $w^2=z^3-3\rho z-2$. This curve degenerates if $\rho^3=1$,  has a $\mathbb Z_2$ symmetry at $\rho^3=\infty$ and a $\mathbb Z_3$ symmetry at $\rho^3=0$. 

If we approach the AD point from the $\rho=0$ plane we find that $\tau_{11}=\alpha$. By an $Sp(4,\BZ)$ transformation we can go to a basis where the mutually non-local states $\tilde\nu_1$, $\tilde\nu_3$ and $\tilde\nu_5$ are mapped to an electron, a monopole and a dyon, all charged with respect to the first $U(1)$ factor only. This is done, for example, by the transformation
\begin{equation}
\CA = \left(
\begin{array}{cccc}
-1 & 0 & 0 & 0 \\
1 & 0 & 0 & -1 \\
1 & 1 & -1 & 2 \\
1 & 1 & 0 & 1 \\
\end{array}
\right) \in Sp(4,\BZ).
\end{equation}
This furthermore diagonalizes the coupling matrix 
\begin{equation}
\CA: \Omega(0,1)\mapsto  \begin{pmatrix}
\alpha & 0\\0 & -\alpha^2
\end{pmatrix},
\end{equation}
as anticipated. The periods $\pi(0,1)=(0,*,0,*)$ depend on the exact transformation, but the relations  $a_1=\alpha^2 a_{D,1}\to 0$ and $a_2=-\alpha^2 a_{D,2}$ are fixed.

\section{Proofs of modular identities}\label{sec:variouscomputations}
In this section we collect some rigorous proofs of exact statements made in the sections above.

\subsection{The origin of the moduli space}\label{zerosu2}
The zeros of $u(\tau)$ for both the $SU(2)$ and $SU(3)$ theory can be derived from the properties of the Jacobi theta functions.

\subsubsection*{The $SU(2)$ theory}
The moduli space of the pure $SU(2)$ theory is parametrized by the modular function \begin{equation}\label{urank1}
u(\tau) = \frac{\vartheta_2(\tau)^4+\vartheta_3(\tau)^4}{2\vartheta_2(\tau)^2 \vartheta_3(\tau)^2} = 1+\frac 18\left(\frac{\eta(\tfrac\tau4)}{\eta(\tau)}\right)^8.
\end{equation}
The Jacobi theta functions $\vartheta_j$ and their transformation properties are given in Appendix \ref{appendA}. The zeros of $u$ are given by the $\Gamma^0(4)$-orbit of $1+i$. To prove this, it suffices to observe that \begin{equation}
\vartheta_2(1+i)^4+\vartheta_3(1+i)^4= \left(e^{\frac{\pi i}{4}}\vartheta_2(i)\right)^4+\vartheta_4(i)^4=-\vartheta_2(i)^4+\vartheta_2(i)^4=0,
\end{equation}
where we have used the $T$-transformation in the first equation and the $S$-transformation of $\vartheta_4$ in the second equation. Using the result \eqref{specialvaluesjacobi}, we know that the denominator is nonzero. Therefore we have proven that $u(1+i)=0$.

\subsubsection*{The $SU(3)$ theory}\label{zerosu3}
Let us prove that \eqref{tauorigin} is a root of \eqref{largeuv=0}. Notice that
\begin{equation}
b_{3,0}(\tau) = \vartheta_3(2\tau)\vartheta_3(6\tau)+\vartheta_2(2\tau)\vartheta_2(6\tau).
\end{equation}
Without computing any of these sums, we can simplify the terms in $b_{3,0}\left(\frac{\tau_0}{3}\right)$ by making use of the transformation identities in Section \ref{appendA},
\be  \begin{aligned}
\vartheta_3(1+\tfrac{i}{\sqrt3})&= \vartheta_4(\tfrac{i}{\sqrt3}), \\
\vartheta_3(3+\sqrt3 i)&= \vartheta_4(\sqrt3 i), \\
\vartheta_2(1+\tfrac{i}{\sqrt3})&= e^{\frac{\pi i}{4}}\vartheta_2(\tfrac{i}{\sqrt3}) =\sqrt[4]{3}\, e^{\frac{\pi i}{4}}\vartheta_4(\sqrt3 i)\\
\vartheta_2(3+\sqrt3 i)&= e^{\frac{3\pi i}{4}}\vartheta_2(\sqrt3 i)=\tfrac{1}{\sqrt[4]{3}}\,e^{\frac{3\pi i}{4}} \vartheta_4(\tfrac{i}{\sqrt3}).
\end{aligned}\ee
We thus find that $b_{3,0}(\frac{\tau_0}{3})=b_{3,0}(\frac12+\frac{i}{2\sqrt3})=0$. The denominator 
\begin{equation}
b_{3,1}(\tau) = 3\, \frac{\eta(3\tau)^3}{\eta(\tau)}
\end{equation}
vanishes nowhere on $\mathbb H$, as $\eta^{24}(\tau)=\Delta(\tau)$ is a holomorphic cusp form of weight $12$ for $SL(2,\mathbb Z)$. This proves that indeed $u(\tau_0)=0$.

\subsection{The function $v$}\label{proofmo}
Since on the locus $\CE_v$ the  relations \eqref{Ttauu=0} among the $\tau_{IJ}$ are exact, it is possible to prove the step from \eqref{eq:v_exp_c1} to \eqref{monstersol} by computing the theta constants analytically instead of perturbatively (as done on $\CE_u$).
 First, note that $C_1=\lambda_3=\tfrac{\Theta_8^2}{\Theta_{10}^2}$, since $\Theta_1=\Theta_2$ due to  \eqref{Ttauu=0}. This lets us simplify, 
 \begin{equation}\label{vthetaapp}
v=- \frac{i}{\sqrt{27}}\frac{(\Theta_8^2-2\Theta_{10}^2)(\Theta_8^2+\Theta_{10}^2)(2\Theta_8^2-\Theta_{10}^2)}{\Theta_8^2 \Theta_{10}^2(\Theta_8^2-\Theta_{10}^2)}.
\end{equation}
Both sides of this equation are functions of $\Omega=\left(\begin{smallmatrix} \tau_{11} & -\tau_{11}/2+1 \\ \
      -\tau_{11}/2+1 & \tau_{11}-1 \end{smallmatrix}\right)$. We  have
  that $\tau_-=\frac32\tau_{11}-1$ and therefore
  $\tau=\tau_-+1=\frac32\tau_{11}$, as defined in Section
  \ref{sec:z3vacua}. In view of the claim $v\propto
  m(\tfrac\tau6)$, let us further define
  $\sigma=\tfrac\tau6=\frac{\tau_{11}}{4}$ to obtain integral powers
  of $\tq\coloneqq e^{2\pi i\sigma}$. This allows to compute the theta
  constants, 
      \be\begin{aligned}
\Theta_8(\Omega)&=e^{\frac{\pi i}{4}}\sum_{k,l\in\mathbb Z+\frac12}(-1)^{k+l}\tq^{2(k^2+k l+l^2)},\\
\Theta_{10}(\Omega)&=e^{\frac{\pi i}{4}}\sum_{k,l\in\mathbb Z+\frac12}(-1)^{2k}\tq^{2(k^2+k l+l^2)}.
\end{aligned}\ee
The $\tq$-series on the rhs should be interpreted as functions of $\sigma(\tau_{11})=\frac{\tau_{11}}{4}$, however it is convenient to consider them as functions of the new variable $\sigma\in \mathbb H$. On $\CE_v$, the theta constants $\Theta_8$ and $\Theta_{10}$  collapse to shifted theta functions of the $A_2$ root lattice, as defined in \eqref{b3j}.  In order to see this, define 
\begin{equation}
f(\tau)=\frac13(b_{3,0}(\tau)-b_{3,0}(4\tau))=\frac{2\eta(4\tau)^2\,\eta(12\tau)^2}{\eta(2\tau)\,\eta(6\tau)}=2(q+q^3+2\,q^7+\dots).
\end{equation}
According to Theorem 1 in Appendix \ref{appendA}, $f$ is a modular form of weight $1$ for $\Gamma_0(12)$.
By splitting the $b_{3,0}$ theta functions into even and odd exponents, it can be easily shown that
\begin{equation}
\Theta_{8}(\sigma)=-ie^{\frac{\pi i}{4}} f(\tfrac\sigma2+\tfrac14), \quad \Theta_{10}(\sigma)=-e^{\frac{\pi i}{4}}f(\tfrac\sigma2),
\end{equation}
with the abuse of notation $\Theta_j(\Omega)=\Theta_j(\sigma)$ with
$\Omega$ and $\sigma$ as below (\ref{vthetaapp}).
By modularity, it is enough to compare a finite number of coefficients between \eqref{vthetaapp} and \eqref{mo}, which proves $v=-\tfrac{i}{ 2\sqrt{27}}\,  m(\sigma)$.

\subsubsection*{Special points of $v$}\label{app:vtau=pm1}

The solutions to $v=1$  and  $v=-1$  are not straightforward to obtain. Let us start with the point $(\underline u,v)=(0,-1)$. In the following, all arguments are  those of $m$. Due to the prefactor in \eqref{monstersol}, $v=-1$ is in fact a quadratic equation with zero discriminant  and therefore satisfied if and only if 
\begin{equation}\label{27i}
\left(\frac{\eta(2  \tau )}{\eta(6 \tau)}\right)^6 = -\sqrt{27}\, i.
\end{equation}
A solution to this equation can be found to be 
\begin{equation}\label{tau+}
\tau_{-1}= \frac{\omega}{2\sqrt3}=\frac14+\frac{i}{4\sqrt 3}=\frac{\tau_{\text{AD},2}}{6}
\end{equation}
 with $\omega = e^{\pi i/6}$ as before and $\tau_{\text{AD},2}$ the argument of $v$ in \eqref{tildetauad}. The other AD point can be found using the symmetry of $m$, and it is given by
 \begin{equation}\label{tau-}
\tau_{+1}= \frac{\omega^5}{2\sqrt3}=-\frac14+\frac{i}{4\sqrt3}=\frac{\tau_{\text{AD},1}}{6}.
\end{equation}
The zero of $m$ (and therefore of $v$) is given by 
\begin{equation}\label{zerom}
\tau_0 = \frac{i}{2\sqrt3}.
\end{equation}
Note that all these numbers have the same absolute value $\frac{1}{2\sqrt3}$.

Let us prove \eqref{tau+} first: In order to compute both the numerator and the denominator, we can resort to the $S$- and $T$-transformations of $\eta$ as given in \ref{appendA},
\be\begin{aligned}
	\eta(2\tau_{-1})& \overset{S}{=} 3^{\frac{1}{4}}\omega e^{-\frac{\pi i}{12}}\eta(-\tfrac{1}{2}+\tfrac{\sqrt3}{2}i) \overset{T}{=} 3^{\frac14} e^{\frac{\pi i}{12}}\eta(\alpha)\\
	\eta(6\tau_{-1})&=\eta(\tfrac32+\tfrac{\sqrt3}{2}i) \overset{T}{=} e^{\frac{\pi i}{6}}\eta(\alpha).
\end{aligned}\ee
Equation \eqref{27i} follows immediately. \\
In order to find the point where $v=+1$, we can make the observation that $m(-\frac{1}{\tau})=-m(\frac{\tau}{12})$. This implies that under the Fricke involution $\left(\begin{smallmatrix}0&-1\\12&0\end{smallmatrix}\right)$, the solution receives a minus sign, 
\begin{equation}\label{mo12}
m\left(-\frac{1}{12\tau}\right)=-\,m(\tau).
\end{equation}
Using the $T$-transformation of $\eta$, one also finds that $m\left(\tau\pm\frac 12\right)=-m(\tau)$. We can use either of those maps, $\tau_{+1}=\tau_{-1}-\tfrac12=-\tfrac{1}{12\tau_{-1}}$ to obtain \eqref{tau-}.

We can also study the zeros of $v$. Every root of $m(\tau)$ is given by the equation $\eta(2\tau)^{12}=27\, \eta(6\tau)^{12}$. A solution to this equation is \eqref{zerom}, which we can prove: Using the $S$-transformation, we find 
\begin{equation}
\eta(2\tau_0) = \eta(\tfrac{i}{\sqrt3})=3^{\frac14}\eta(\sqrt3 i)=3^{\frac14}\eta(6\tau_0).
\end{equation}
The result follows immediately. Another proof follows simply from the fact that  $\tau_0$ is the fixed point under \eqref{mo12}.

\providecommand{\href}[2]{#2}\begingroup\raggedright\endgroup


\begin{thebibliography}{10}

\bibitem{Seiberg:1994rs}
N.~Seiberg and E.~Witten, \emph{{Electric - magnetic duality, monopole
  condensation, and confinement in N=2 supersymmetric Yang-Mills theory}},
  \href{http://dx.doi.org/10.1016/0550-3213(94)90124-4,
  10.1016/0550-3213(94)00449-8}{\emph{Nucl. Phys.} {\bf B426} (1994) 19--52},
  [\href{https://arxiv.org/abs/hep-th/9407087}{{\tt hep-th/9407087}}].

\bibitem{Seiberg:1994aj}
N.~Seiberg and E.~Witten, \emph{{Monopoles, duality and chiral symmetry
  breaking in N=2 supersymmetric QCD}},
  \href{http://dx.doi.org/10.1016/0550-3213(94)90214-3}{\emph{Nucl. Phys.} {\bf
  B431} (1994) 484--550}, [\href{https://arxiv.org/abs/hep-th/9408099}{{\tt
  hep-th/9408099}}].

\bibitem{Vafa:1994tf}
C.~Vafa and E.~Witten, \emph{{A Strong coupling test of S duality}},
  \href{http://dx.doi.org/10.1016/0550-3213(94)90097-3}{\emph{Nucl. Phys.} {\bf
  B431} (1994) 3--77}, [\href{https://arxiv.org/abs/hep-th/9408074}{{\tt
  hep-th/9408074}}].

\bibitem{Ne}
N.~A. Nekrasov, \emph{{Seiberg-Witten prepotential from instanton counting}},
  \href{http://dx.doi.org/10.4310/ATMP.2003.v7.n5.a4}{\emph{Adv. Theor. Math.
  Phys.} {\bf 7} (2003) 831--864},
  [\href{https://arxiv.org/abs/hep-th/0206161}{{\tt hep-th/0206161}}].

\bibitem{Nahm:1996di}
W.~Nahm, \emph{{On the Seiberg-Witten approach to electric - magnetic
  duality}},  \href{https://arxiv.org/abs/hep-th/9608121}{{\tt
  hep-th/9608121}}.

\bibitem{Moore:1997pc}
G.~W. Moore and E.~Witten, \emph{{Integration over the u plane in Donaldson
  theory}}, {\emph{Adv. Theor. Math. Phys.} {\bf 1} (1997) 298--387},
  [\href{https://arxiv.org/abs/hep-th/9709193}{{\tt hep-th/9709193}}].

\bibitem{Losev:1997tp}
A.~Losev, N.~Nekrasov and S.~L.~Shatashvili,
\emph{{Issues in topological gauge theory}},
Nucl. Phys. B \textbf{534}, 549-611 (1998)
doi:10.1016/S0550-3213(98)00628-2
[arXiv:hep-th/9711108 [hep-th]].

\bibitem{Aganagic:2006wq}
M.~Aganagic, V.~Bouchard and A.~Klemm, \emph{{Topological Strings and (Almost)
  Modular Forms}},
  \href{http://dx.doi.org/10.1007/s00220-007-0383-3}{\emph{Commun. Math. Phys.}
  {\bf 277} (2008) 771--819}, [\href{https://arxiv.org/abs/hep-th/0607100}{{\tt
  hep-th/0607100}}].

\bibitem{Huang:2009md}
M.-x. Huang and A.~Klemm, \emph{{Holomorphicity and Modularity in
  Seiberg-Witten Theories with Matter}},
  \href{http://dx.doi.org/10.1007/JHEP07(2010)083}{\emph{JHEP} {\bf 07} (2010)
  083}, [\href{https://arxiv.org/abs/0902.1325}{{\tt 0902.1325}}].

\bibitem{Huang:2011qx}
M.~x.~Huang, A.~K.~Kashani-Poor and A.~Klemm,
\emph{{The $\Omega$ deformed B-model for rigid $\mathcal{N}=2$ theories}},
Annales Henri Poincare \textbf{14}, 425-497 (2013)
doi:10.1007/s00023-012-0192-x
[arXiv:1109.5728 [hep-th]].

\bibitem{Klemm:1994qs}
A.~Klemm, W.~Lerche, S.~Yankielowicz and S.~Theisen, \emph{{Simple
  singularities and N=2 supersymmetric Yang-Mills theory}},
  \href{http://dx.doi.org/10.1016/0370-2693(94)01516-F}{\emph{Phys. Lett.} {\bf
  B344} (1995) 169--175}, [\href{https://arxiv.org/abs/hep-th/9411048}{{\tt
  hep-th/9411048}}].

\bibitem{Klemm:1994qj}
A.~Klemm, W.~Lerche, S.~Yankielowicz and S.~Theisen, \emph{On the monodromies
  of {$N=2$} supersymmetric yang-mills theory},
  \href{https://arxiv.org/abs/hep-th/9412158}{{\tt hep-th/9412158}}.

\bibitem{Klemm:1995wp}
A.~Klemm, W.~Lerche and S.~Theisen, \emph{{Nonperturbative effective actions of
  N=2 supersymmetric gauge theories}},
  \href{http://dx.doi.org/10.1142/S0217751X96001000}{\emph{Int. J. Mod. Phys.}
  {\bf A11} (1996) 1929--1974},
  [\href{https://arxiv.org/abs/hep-th/9505150}{{\tt hep-th/9505150}}].

\bibitem{Danielsson:1995is}
U.~H. Danielsson and B.~Sundborg, \emph{{The Moduli space and monodromies of
  N=2 supersymmetric SO(2r+1) Yang-Mills theory}},
  \href{http://dx.doi.org/10.1016/0370-2693(95)01010-N}{\emph{Phys. Lett.} {\bf
  B358} (1995) 273--280}, [\href{https://arxiv.org/abs/hep-th/9504102}{{\tt
  hep-th/9504102}}].

\bibitem{Masuda:1997nv}
T.~Masuda, T.~Sasaki and H.~Suzuki, \emph{{Seiberg-Witten theory of rank two
  gauge groups and hypergeometric series}},
  \href{http://dx.doi.org/10.1142/S0217751X98001542}{\emph{Int. J. Mod. Phys.}
  {\bf A13} (1998) 3121--3144},
  [\href{https://arxiv.org/abs/hep-th/9705166}{{\tt hep-th/9705166}}].

\bibitem{suzuki1}
T.~Masuda and H.~Suzuki, \emph{{Periods and prepotential of N=2 SU(2)
  supersymmetric Yang-Mills theory with massive hypermultiplets}},
  \href{http://dx.doi.org/10.1142/S0217751X97001791}{\emph{Int. J. Mod. Phys.}
  {\bf A12} (1997) 3413--3431},
  [\href{https://arxiv.org/abs/hep-th/9609066}{{\tt hep-th/9609066}}].

\bibitem{suzuki2}
T.~Masuda and H.~Suzuki, \emph{On explicit evaluations around the conformal
  point in n=2 supersymmetric yang-mills theories},
  \href{https://arxiv.org/abs/arXiv:hep-th/9612240}{{\tt
  arXiv:hep-th/9612240}}.

\bibitem{Minahan:1995er}
J.~A. Minahan and D.~Nemeschansky, \emph{{Hyperelliptic curves for
  supersymmetric Yang-Mills}},
  \href{http://dx.doi.org/10.1016/0550-3213(95)00672-9}{\emph{Nucl. Phys.} {\bf
  B464} (1996) 3--17}, [\href{https://arxiv.org/abs/hep-th/9507032}{{\tt
  hep-th/9507032}}].

\bibitem{Minahan:1996ws}
J.~A. Minahan and D.~Nemeschansky, \emph{{N=2 superYang-Mills and subgroups of
  SL(2,Z)}}, \href{http://dx.doi.org/10.1016/0550-3213(96)00167-8}{\emph{Nucl.
  Phys.} {\bf B468} (1996) 72--84},
  [\href{https://arxiv.org/abs/hep-th/9601059}{{\tt hep-th/9601059}}].

\bibitem{Argyres:1998bn}
P.~C. Argyres and A.~Buchel, \emph{{The Nonperturbative gauge coupling of N=2
  supersymmetric theories}},
  \href{http://dx.doi.org/10.1016/S0370-2693(98)01235-0}{\emph{Phys. Lett.}
  {\bf B442} (1998) 180--184},
  [\href{https://arxiv.org/abs/hep-th/9806234}{{\tt hep-th/9806234}}].

\bibitem{Ashok:2015cba}
S.~K. Ashok, M.~Bill{\`o}, E.~Dell'Aquila, M.~Frau, A.~Lerda and M.~Raman,
  \emph{{Modular anomaly equations and S-duality in $ \mathcal{N}=2 $ conformal
  SQCD}}, \href{http://dx.doi.org/10.1007/JHEP10(2015)091}{\emph{JHEP} {\bf 10}
  (2015) 091}, [\href{https://arxiv.org/abs/1507.07476}{{\tt 1507.07476}}].

\bibitem{Ashok:2016oyh}
S.~K. Ashok, E.~Dell'Aquila, A.~Lerda and M.~Raman, \emph{{S-duality, triangle
  groups and modular anomalies in $ \mathcal{N}=2 $ SQCD}},
  \href{http://dx.doi.org/10.1007/JHEP04(2016)118}{\emph{JHEP} {\bf 04} (2016)
  118}, [\href{https://arxiv.org/abs/1601.01827}{{\tt 1601.01827}}].

\bibitem{Argyres:1995jj}
P.~C. Argyres and M.~R. Douglas, \emph{{New phenomena in SU(3) supersymmetric
  gauge theory}},
  \href{http://dx.doi.org/10.1016/0550-3213(95)00281-V}{\emph{Nucl. Phys.} {\bf
  B448} (1995) 93--126}, [\href{https://arxiv.org/abs/hep-th/9505062}{{\tt
  hep-th/9505062}}].

\bibitem{ARGYRES199671}
P.~C. Argyres, M.~R. Plesser, N.~Seiberg and E.~Witten, \emph{New n=2
  superconformal field theories in four dimensions},
  \href{http://dx.doi.org/https://doi.org/10.1016/0550-3213(95)00671-0}{\emph{Nuclear
  Physics B} {\bf 461} (1996) 71 -- 84}.

\bibitem{EGUCHI1996430}
T.~Eguchi, K.~Hori, K.~Ito and S.-K. Yang, \emph{Study of n = 2 superconformal
  field theories in 4 dimensions},
  \href{http://dx.doi.org/https://doi.org/10.1016/0550-3213(96)00188-5}{\emph{Nuclear
  Physics B} {\bf 471} (1996) 430 -- 442}.

\bibitem{Galakhov:2013oja}
D.~Galakhov, P.~Longhi, T.~Mainiero, G.~W. Moore and A.~Neitzke, \emph{{Wild
  Wall Crossing and BPS Giants}},
  \href{http://dx.doi.org/10.1007/JHEP11(2013)046}{\emph{JHEP} {\bf 11} (2013)
  046}, [\href{https://arxiv.org/abs/1305.5454}{{\tt 1305.5454}}].

\bibitem{wang:2019}
Q.~{Wang}, \emph{{Wall Crossing Structures and Application to SU(3)
  Seiberg-Witten Integrable system}},
  \href{https://arxiv.org/abs/1903.10169}{{\tt 1903.10169}}.

\bibitem{shaska2001}
T.~{Shaska} and H.~{Voelklein}, \emph{{Elliptic subfields and automorphisms of
  genus 2 function fields}}, {\emph{arXiv Mathematics e-prints} (July, 2001)
  math/0107142}, [\href{https://arxiv.org/abs/math/0107142}{{\tt
  math/0107142}}].

\bibitem{shaska2012genus}
T.~Shaska, \emph{Genus two curves covering elliptic curves: a computational
  approach},  \href{https://arxiv.org/abs/1209.3187}{{\tt 1209.3187}}.

\bibitem{shaska2006}
J.~{Gutierrez} and T.~{Shaska}, \emph{{Hyperelliptic curves with extra
  involutions}}, {\emph{arXiv Mathematics e-prints} (Jan., 2006) math/0601456},
  [\href{https://arxiv.org/abs/math/0601456}{{\tt math/0601456}}].

\bibitem{gutierrez2012}
J.~{Gutierrez}, D.~{Sevilla} and T.~{Shaska}, \emph{{Hyperelliptic curves of
  genus 3 with prescribed automorphism group}}, {\emph{arXiv e-prints} (Sept.,
  2012) arXiv:1209.2938}, [\href{https://arxiv.org/abs/1209.2938}{{\tt
  1209.2938}}].

\bibitem{AlvarezGaume}
L.~Alvarez-Gaume and S.~F. Hassan, \emph{{Introduction to S duality in N=2
  supersymmetric gauge theories: A Pedagogical review of the work of Seiberg
  and Witten}}, \href{http://dx.doi.org/10.1002/prop.2190450302}{\emph{Fortsch.
  Phys.} {\bf 45} (1997) 159--236},
  [\href{https://arxiv.org/abs/hep-th/9701069}{{\tt hep-th/9701069}}].

\bibitem{Tachikawa13}
Y.~Tachikawa, \emph{{N=2 supersymmetric dynamics for pedestrians}}, vol.~890.
\newblock 2014,
  \href{http://dx.doi.org/10.1007/978-3-319-08822-8}{10.1007/978-3-319-08822-8}.

\bibitem{Bilal:1995hc}
A.~Bilal, \emph{{Duality in N=2 SUSY SU(2) Yang-Mills theory: A Pedagogical
  introduction to the work of Seiberg and Witten}},  in \emph{{Quantum fields
  and quantum space time. Proceedings, NATO Advanced Study Institute, Cargese,
  France, July 22-August 3, 1996}}, pp.~21--43, 1997.
\newblock \href{https://arxiv.org/abs/hep-th/9601007}{{\tt hep-th/9601007}}.

\bibitem{Laba05}
J.~Labastida and M.~Marino, \emph{{Topological quantum field theory and four
  manifolds}}.
\newblock 2005.

\bibitem{Klemm:1997gg}
A.~Klemm, \emph{{On the geometry behind N=2 supersymmetric effective actions in
  four-dimensions}},  in \emph{{33rd Karpacz Winter School of Theoretical
  Physics: Duality - Strings and Fields}}, 5, 1997.
\newblock \href{https://arxiv.org/abs/hep-th/9705131}{{\tt hep-th/9705131}}.

\bibitem{Diamond}
F.~Diamond and J.~Shurman, \emph{A First Course in Modular Forms}, vol.~228 of
  \emph{Graduate Texts in Mathematics}.
\newblock Springer-Verlag New York, 1~ed., 2005.

\bibitem{Conway:1979qga}
J.~H. Conway and S.~P. Norton, \emph{{Monstrous Moonshine}},
  \href{http://dx.doi.org/10.1112/blms/11.3.308}{\emph{Bull. London Math. Soc.}
  {\bf 11} (1979) 308--339}.

\bibitem{Alexander:1992}
D.~Alexander, C.~Cummins, J.~McKay and C.~Simons, \emph{\emph{"Completely
  replicable functions" in} {G}roups, {C}ombinatorics, and {G}eometry, {D}urham
  {S}ymposium, \emph{1990, {L}ondon {M}ath. {S}oc. {L}ecture {N}ote {S}er.}
  \textbf{165}, \emph{{C}ambridge {U}niv. {P}ress, {C}ambridge}},  1992.

\bibitem{ford1994}
D.~Ford, J.~McKay and S.~Norton, \emph{More on replicable functions},
  \href{http://dx.doi.org/10.1080/00927879408825127}{\emph{Communications in
  Algebra} {\bf 22} (1994) 5175--5193}.

\bibitem{Ferenbaugh1993}
C.~R. Ferenbaugh, \emph{The genus-zero problem for $n \vert h$ -type groups},
  \href{http://dx.doi.org/10.1215/S0012-7094-93-07202-X}{\emph{Duke Math. J.}
  {\bf 72} (10, 1993) 31--63}.

\bibitem{Ferrari:1996sv}
F.~Ferrari and A.~Bilal, \emph{{The Strong coupling spectrum of the
  Seiberg-Witten theory}},
  \href{http://dx.doi.org/10.1016/0550-3213(96)00150-2}{\emph{Nucl. Phys.} {\bf
  B469} (1996) 387--402}, [\href{https://arxiv.org/abs/hep-th/9602082}{{\tt
  hep-th/9602082}}].

\bibitem{Argyres:1994xh}
P.~C. Argyres and A.~E. Faraggi, \emph{{The vacuum structure and spectrum of
  N=2 supersymmetric SU(n) gauge theory}},
  \href{http://dx.doi.org/10.1103/PhysRevLett.74.3931}{\emph{Phys. Rev. Lett.}
  {\bf 74} (1995) 3931--3934},
  [\href{https://arxiv.org/abs/hep-th/9411057}{{\tt hep-th/9411057}}].

\bibitem{Matone:1995rx}
M.~Matone, \emph{{Instantons and recursion relations in N=2 SUSY gauge
  theory}}, \href{http://dx.doi.org/10.1016/0370-2693(95)00920-G}{\emph{Phys.
  Lett.} {\bf B357} (1995) 342--348},
  [\href{https://arxiv.org/abs/hep-th/9506102}{{\tt hep-th/9506102}}].

\bibitem{Eguchi:1995jh}
T.~Eguchi and S.-K. Yang, \emph{{Prepotentials of N=2 supersymmetric gauge
  theories and soliton equations}},
  \href{http://dx.doi.org/10.1142/S0217732396000151}{\emph{Mod. Phys. Lett.}
  {\bf A11} (1996) 131--138}, [\href{https://arxiv.org/abs/hep-th/9510183}{{\tt
  hep-th/9510183}}].

\bibitem{Rosenhain:1851}
G.~Rosenhain, \emph{Abhandlung \"uber die functionen zweier variablen mit vier
  perioden welche die inversion sind der ultra-elliptische integrale erster
  klasse \textnormal{(1851)}}, {\emph{\textnormal{Translation to German from
  Latin manuscript published in:} Klassiker der Exacten Wissenschaften} {\bf
  65} (1895) 1 -- 96}.

\bibitem{igusa1967}
J.-I. Igusa, \emph{Modular forms and projective invariants}, {\emph{American
  Journal of Mathematics} {\bf 89} (1967) 817--855}.

\bibitem{igusa1962}
J.-I. Igusa, \emph{On siegel modular forms of genus two}, {\emph{American
  Journal of Mathematics} {\bf 84} (1962) 175--200}.

\bibitem{Douglas:1995nw}
M.~R. Douglas and S.~H. Shenker, \emph{{Dynamics of SU(N) supersymmetric gauge
  theory}}, \href{http://dx.doi.org/10.1016/0550-3213(95)00258-T}{\emph{Nucl.
  Phys.} {\bf B447} (1995) 271--296},
  [\href{https://arxiv.org/abs/hep-th/9503163}{{\tt hep-th/9503163}}].

\bibitem{Eilers:2017}
K.~Eilers, \emph{Rosenhain-{T}homae formulae for higher genera hyperelliptic
  curves}, \href{http://dx.doi.org/10.1080/14029251.2018.1440744}{\emph{Journal
  of Nonlinear Mathematical Physics} {\bf 25} (07, 2017) }.

 
\bibitem{AFMtoappear}
J.~Aspman, E.~Furrer and J.~Manschot, \emph{To appear}, .

\bibitem{Magro:1997qs}
M.~Magro, L.~O'Raifeartaigh and I.~Sachs,
\emph{On the uniqueness of the effective Lagrangian for N = 2 SQCD},'
Nucl. Phys. B \textbf{508} (1997), 433-448
doi:10.1016/S0550-3213(97)00626-3
[arXiv:hep-th/9704027 [hep-th]].

\bibitem{Lian:1995js}
B.~H. Lian and S.-T. Yau, \emph{{Mirror maps, modular relations and
  hypergeometric series 1}},  \href{https://arxiv.org/abs/hep-th/9507151}{{\tt
  hep-th/9507151}}.

\bibitem{Alim:2013eja}
M.~Alim, E.~Scheidegger, S.-T. Yau and J.~Zhou, \emph{{Special Polynomial
  Rings, Quasi Modular Forms and Duality of Topological Strings}},
  \href{http://dx.doi.org/10.4310/ATMP.2014.v18.n2.a4}{\emph{Adv. Theor. Math.
  Phys.} {\bf 18} (2014) 401--467},
  [\href{https://arxiv.org/abs/1306.0002}{{\tt 1306.0002}}].

\bibitem{gordon1993}
K.~H. B.~Gordon, \emph{Multiplicative properties of $eta$-products}, vol.~143
  of \emph{Contemp. Math.}
\newblock Amer. Math. Soc., 1993.

\bibitem{ono2004}
K.~Ono, \emph{The Web of Modularity: Arithmetic of the Coefficients of Modular
  Forms and q-series}, vol.~102.
\newblock American Mathematical Society, cbms regional conference series in
  mathematics~ed., 2004.

\bibitem{Argyres:2015ffa}
P.~Argyres, M.~Lotito, Y.~L{\"u} and M.~Martone, \emph{{Geometric constraints
  on the space of $ \mathcal{N} $ = 2 SCFTs. Part I: physical constraints on
  relevant deformations}},
  \href{http://dx.doi.org/10.1007/JHEP02(2018)001}{\emph{JHEP} {\bf 02} (2018)
  001}, [\href{https://arxiv.org/abs/1505.04814}{{\tt 1505.04814}}].

\bibitem{igusa1960}
J.-I. Igusa, \emph{Arithmetic variety of moduli for genus two}, {\emph{Annals
  of Mathematics} {\bf 72} (1960) 612--649}.

\bibitem{Klemm:2015iya}
A.~Klemm, M.~Poretschkin, T.~Schimannek and M.~Westerholt-Raum, \emph{{Direct
  Integration for Mirror Curves of Genus Two and an Almost Meromorphic Siegel
  Modular Form}},  \href{https://arxiv.org/abs/1502.00557}{{\tt 1502.00557}}.

\bibitem{shaska2003}
T.~{Shaska}, \emph{{Determining the automorphism group of a hyperelliptic
  curve}}, {\emph{arXiv Mathematics e-prints} (Dec., 2003) math/0312284},
  [\href{https://arxiv.org/abs/math/0312284}{{\tt math/0312284}}].

\bibitem{Shaska:2009}
T.~Shaska and S.~Wijesiri, \emph{Theta functions and algebraic curves with
  automorphisms}, {\emph{NATO Science for Peace and Security Series D:
  Information and Communication Security} {\bf 24} (06, 2009) 193--237}.

\bibitem{Alim:2011kw}
M.~Alim, S.~Cecotti, C.~Cordova, S.~Espahbodi, A.~Rastogi and C.~Vafa,
  \emph{{$\mathcal{N} = 2$ quantum field theories and their BPS quivers}},
  \href{http://dx.doi.org/10.4310/ATMP.2014.v18.n1.a2}{\emph{Adv. Theor. Math.
  Phys.} {\bf 18} (2014) 27--127}, [\href{https://arxiv.org/abs/1112.3984}{{\tt
  1112.3984}}].

\bibitem{Chuang:2013wt}
W.-y. Chuang, D.-E. Diaconescu, J.~Manschot, G.~W. Moore and Y.~Soibelman,
  \emph{{Geometric engineering of (framed) BPS states}},
  \href{http://dx.doi.org/10.4310/ATMP.2014.v18.n5.a3}{\emph{Adv. Theor. Math.
  Phys.} {\bf 18} (2014) 1063--1231},
  [\href{https://arxiv.org/abs/1301.3065}{{\tt 1301.3065}}].

\bibitem{Shapere:2008zf}
A.~D. Shapere and Y.~Tachikawa, \emph{{Central charges of N=2 superconformal
  field theories in four dimensions}},
  \href{http://dx.doi.org/10.1088/1126-6708/2008/09/109}{\emph{JHEP} {\bf 09}
  (2008) 109}, [\href{https://arxiv.org/abs/0804.1957}{{\tt 0804.1957}}].

\bibitem{Argyres:2007cn}
P.~C. Argyres and N.~Seiberg, \emph{{S-duality in N=2 supersymmetric gauge
  theories}},
  \href{http://dx.doi.org/10.1088/1126-6708/2007/12/088}{\emph{JHEP} {\bf 12}
  (2007) 088}, [\href{https://arxiv.org/abs/0711.0054}{{\tt 0711.0054}}].

\bibitem{Zhou:2014rvr}
J.~Zhou, \emph{{Arithmetic Properties of Moduli Spaces and Topological String
  Partition Functions of Some Calabi-Yau Threefolds}}.
\newblock PhD thesis, Harvard U. (main), 2014.

\bibitem{Persson:2015jka}
D.~Persson and R.~Volpato, \emph{{Fricke S-duality in CHL models}},
  \href{http://dx.doi.org/10.1007/JHEP12(2015)156}{\emph{JHEP} {\bf 12} (2015)
  156}, [\href{https://arxiv.org/abs/1504.07260}{{\tt 1504.07260}}].

\bibitem{Paquette:2016xoo}
N.~M.~Paquette, D.~Persson and R.~Volpato,
\emph{Monstrous BPS-Algebras and the Superstring Origin of Moonshine,}
Commun. Num. Theor. Phys. \textbf{10} (2016), 433-526
doi:10.4310/CNTP.2016.v10.n3.a2
[arXiv:1601.05412 [hep-th]].

\bibitem{Argyres:2006qr}
P.~C. Argyres, A.~Kapustin and N.~Seiberg, \emph{{On S-duality for
  non-simply-laced gauge groups}},
  \href{http://dx.doi.org/10.1088/1126-6708/2006/06/043}{\emph{JHEP} {\bf 06}
  (2006) 043}, [\href{https://arxiv.org/abs/hep-th/0603048}{{\tt
  hep-th/0603048}}].

\bibitem{Dorey:1996hx}
N.~Dorey, C.~Fraser, T.~J. Hollowood and M.~A.~C. Kneipp, \emph{{S duality in
  N=4 supersymmetric gauge theories with arbitrary gauge group}},
  \href{http://dx.doi.org/10.1016/0370-2693(96)00773-3}{\emph{Phys. Lett.} {\bf
  B383} (1996) 422--428}, [\href{https://arxiv.org/abs/hep-th/9605069}{{\tt
  hep-th/9605069}}].

\bibitem{Kapustin:2006pk}
A.~Kapustin and E.~Witten, \emph{{Electric-Magnetic Duality And The Geometric
  Langlands Program}},
  \href{http://dx.doi.org/10.4310/CNTP.2007.v1.n1.a1}{\emph{Commun. Num. Theor.
  Phys.} {\bf 1} (2007) 1--236},
  [\href{https://arxiv.org/abs/hep-th/0604151}{{\tt hep-th/0604151}}].


\bibitem{Gaiotto:2009hg}
D.~Gaiotto, G.~W.~Moore and A.~Neitzke,
\emph{{Wall-crossing, Hitchin Systems, and the WKB Approximation}},
[arXiv:0907.3987 [hep-th]].

\bibitem{He:2012kw}
Y.-H. He and J.~McKay, \emph{{N=2 Gauge Theories: Congruence Subgroups, Coset
  Graphs and Modular Surfaces}},
  \href{http://dx.doi.org/10.1063/1.4772976}{\emph{J. Math. Phys.} {\bf 54}
  (2013) 012301}, [\href{https://arxiv.org/abs/1201.3633}{{\tt 1201.3633}}].

\bibitem{liwalker2013}
Z.~K. Li and A.~W. Walker, \emph{Arithmetic properties of picard-fuchs
  equations and holonomic recurrences},  2013.

\bibitem{Argyres:2018zay}
P.~C. Argyres, C.~Long and M.~Martone, \emph{{The Singularity Structure of
  Scale-Invariant Rank-2 Coulomb Branches}},
  \href{http://dx.doi.org/10.1007/JHEP05(2018)086}{\emph{JHEP} {\bf 05} (2018)
  086}, [\href{https://arxiv.org/abs/1801.01122}{{\tt 1801.01122}}].

\bibitem{Martone:2020nsy}
M.~Martone, \emph{{Towards the classification of rank-$r$ $\mathcal{N}=2$
  SCFTs. Part I: twisted partition function and central charge formulae}},
  \href{https://arxiv.org/abs/2006.16255}{{\tt 2006.16255}}.

\bibitem{Argyres:2020wmq}
P.~C. Argyres and M.~Martone, \emph{{Towards a classification of rank $r$
  $\mathcal{N}=2$ SCFTs Part II: special Kahler stratification of the Coulomb
  branch}},  \href{https://arxiv.org/abs/2007.00012}{{\tt 2007.00012}}.

\bibitem{Witten:1988ze}
E.~Witten, \emph{{Topological Quantum Field Theory}},
  \href{http://dx.doi.org/10.1007/BF01223371}{\emph{Commun. Math. Phys.} {\bf
  117} (1988) 353}.

\bibitem{Marino:1998bm}
M.~Marino and G.~W. Moore, \emph{{The Donaldson-Witten function for gauge
  groups of rank larger than one}},
  \href{http://dx.doi.org/10.1007/s002200050494}{\emph{Commun. Math. Phys.}
  {\bf 199} (1998) 25--69}, [\href{https://arxiv.org/abs/hep-th/9802185}{{\tt
  hep-th/9802185}}].

\bibitem{Malmendier:2008db}
A.~Malmendier and K.~Ono, \emph{{SO(3)-Donaldson invariants of $\mathbb{P}^2$
  and Mock Theta Functions}},
  \href{http://dx.doi.org/10.2140/gt.2012.16.1767}{\emph{Geom. Topol.} {\bf 16}
  (2012) 1767--1833}, [\href{https://arxiv.org/abs/0808.1442}{{\tt
  0808.1442}}].

\bibitem{Korpas:2017qdo}
G.~Korpas and J.~Manschot, \emph{{Donaldson-Witten theory and indefinite theta
  functions}}, \href{http://dx.doi.org/10.1007/JHEP11(2017)083}{\emph{JHEP}
  {\bf 11} (2017) 083}, [\href{https://arxiv.org/abs/1707.06235}{{\tt
  1707.06235}}].

\bibitem{Korpas:2019cwg}
G.~Korpas, J.~Manschot, G.~W. Moore and I.~Nidaiev, \emph{{Mocking the
  $u$-plane integral}},  \href{https://arxiv.org/abs/1910.13410}{{\tt
  1910.13410}}.

\bibitem{Bruinier08}
G.~H. J.H.~Bruinier, G. van der~Geer and D.~Zagier, \emph{The 1-2-3 of Modular
  Forms}.
\newblock Springer-Verlag Berlin Heidelberg, 2008,
  \href{http://dx.doi.org/10.1007/978-3-540-74119-0}{10.1007/978-3-540-74119-0}.

\bibitem{Previato:2013}
E.~{Previato}, T.~{Shaska} and G.~S. {Wijesiri}, \emph{{Thetanulls of cyclic
  curves of small genus}}, {\emph{arXiv e-prints} (Jan, 2013) arXiv:1301.4595},
  [\href{https://arxiv.org/abs/1301.4595}{{\tt 1301.4595}}].

\bibitem{freitag1983}
E.~Freitag, \emph{Siegelsche Modulfunktionen}, vol.~254 of \emph{Grundlehren
  der mathematischen Wissenschaften}.
\newblock Springer-Verlag Berlin Heidelberg, 1~ed., 1983.

\bibitem{Abramovitz:1964}
M.~Abramowitz and I.~A. Stegun, \emph{Handbook of Mathematical Functions with
  Formulas, Graphs, and Mathematical Tables}.
\newblock Dover, New York, ninth dover printing, tenth gpo printing~ed., 1964.

\end{thebibliography}
\end{document}